\documentclass[fleqn,usenatbib]{mnras}

\usepackage{newtxtext,newtxmath}

\usepackage[T1]{fontenc}

\DeclareRobustCommand{\VAN}[3]{#2}
\let\VANthebibliography\thebibliography
\def\thebibliography{\DeclareRobustCommand{\VAN}[3]{##3}\VANthebibliography}

\usepackage[draft]{graphicx}	
\usepackage{amsmath}	
\usepackage{CJKutf8}

\defcitealias{2010A&A...518A..10V}{VCV}
\defcitealias{2000A&AS..143....9W}{SMB}

\title[Viewing angle in AGN SED models]{The viewing angle in AGN SED models, a data-driven analysis}

\author[Andr\'{e}s F. Ramos P.\ et al.]{Andr\'{e}s F. Ramos Padilla$^{1,2}$\thanks{ramos@astro.rug.nl (AFRP)},
Lingyu Wang$^{1,2}$,
Katarzyna Ma{\l}ek$^{3,4}$,
Andreas Efstathiou$^{5}$
\newauthor{and
Guang Yang\begin{CJK*}{UTF8}{gbsn} (杨光)\end{CJK*}$^{6,7}$}
\\
$^{1}$Kapteyn Astronomical Institute, University of Groningen, Landleven 12, 9747 AD Groningen, The Netherlands\\
$^{2}$SRON Netherlands Institute for Space Research, Landleven 12, 9747 AD Groningen, The Netherlands\\
$^{3}$ National Centre for Nuclear Research, Pasteura 7, 02-093 Warsaw, Poland \\
$^{4}$ Aix Marseille Univ. CNRS, CNES, LAM, Marseille, France \\
$^{5}$ School of Sciences, European University Cyprus, Diogenes Street, Engomi, 1516 Nicosia, Cyprus\\
$^{6}$ Department of Physics and Astronomy, Texas A\&M University, College Station, TX 77843-4242, USA\\
$^{7}$ George P. and Cynthia Woods Mitchell Institute for Fundamental Physics and Astronomy, Texas A\&M University, College Station, TX 77843-4242, USA
}

\date{Accepted XXX. Received YYY; in original form ZZZ}

\pubyear{2020}

\begin{document}
\label{firstpage}
\pagerange{\pageref{firstpage}--\pageref{lastpage}}
\maketitle

\begin{abstract}
The validity of the unified active galactic nuclei (AGN) model has been challenged in the last decade, especially when different types of AGNs are considered to only differ in the viewing angle to the torus.
We aim to assess the importance of the viewing angle in classifying different types of Seyfert galaxies in spectral energy distribution (SED) modelling. 
We retrieve photometric data from publicly available astronomical databases: CDS and NED, to model SEDs with \textsc{X-CIGALE} in a sample of 13\,173 Seyfert galaxies located at redshift range from $z=0$ to $z=3.5$, with a median redshift of $z\approx0.2$. We assess whether the estimated viewing angle from the SED models reflects different Seyfert classifications. Two AGN models with either a smooth or clumpy torus structure are adopted in this paper. 
We find that the viewing angle in Type-1 AGNs is better constrained than in Type-2 AGNs. Limiting the viewing angles representing these two types of AGNs do not affect the physical parameter estimates such as star-formation rate (SFR) or AGN fractional contribution ($f_{\rm{AGN}}$). In addition, the viewing angle is not the most discriminating physical parameter to differentiate Seyfert types. 
We suggest that the observed and intrinsic AGN disc luminosity can: i) be used in $z<0.5$ studies to distinguish between Type-1 and Type-2 AGNs, and ii) explain the probable evolutionary path between these AGN types. Finally, we propose the use of \textsc{X-CIGALE} for AGN galaxy classification tasks. All data from the 13\,173 SED fits are available at Zenodo\footnotemark.
\end{abstract}


\footnotetext{\url{https://doi.org/10.5281/zenodo.5221764}}



\section{Introduction}

The presence of an optically thick structure in a Seyfert galaxy \citep{1985ApJ...297..621A} led to the creation of the unified model of active galactic nuclei (AGN), where an obscuring torus explains the variety of AGN types due to the orientation with respect to the line of sight \citep[e.g.] [] {1993ARA&A..31..473A,1995PASP..107..803U}. This simple model uses the viewing angle to separate AGN galaxies into two types: unobscured (Type-1 AGN) and obscured (Type-2 AGN). Type-1 AGNs have broad emission lines, in terms of the full width at half maximum (FWHM), while Type 2 AGNs do not, due to differences in the viewing angle $\,i\,$. We observe narrow-line regions (NLR, with FWHM\,$\lesssim 1000\,\rm{km}\,\rm{s}^{-1}$) or broad line regions (BLR, with FWHM\,$\gtrsim 1000\,\rm{km}\,\rm{s}^{-1}$) depending on $\,i\,$. In Type-1 AGN, the BLR and NLR are viewed directly because the galaxies are viewed at small angles with respect to the line of sight ($i\approx0-30\degr$), while for Type-2 AGN, only the NLR is visible because the galaxies are viewed at high angles with respect to the line of sight  ($i\approx70-90\degr$) and obscuration hides the BLR \citep[e.g.][]{1993ARA&A..31..473A,2003MNRAS.346.1055K}. However, in the last decade, the ``zoo'' of AGNs has become more complex and difficult to explain with this simple toroidal structure model \citep[][]{2017A&ARv..25....2P} and the viewing angle has not been easy to estimate \citep[e.g][]{2016MNRAS.460.3679M}. Besides, obscuration is not static and may depend on different physical conditions that may vary \citep[e.g.][]{2017ApJ...838L..20H, 2018ARA&A..56..625H}. In addition, the ``changing look'' AGNs cannot be explained with the unified model, but rather with the accretion state of the AGN \citep{2015ApJ...800..144L,2014MNRAS.438.3340E}. Therefore, updates to the unified model of AGNs have been proposed describing new AGN scenarios such as clumpy structures \citep[e.g.][]{1988ApJ...329..702K,2002ApJ...570L...9N,2005A&A...436...47D}, radiation-pressure modes \citep[e.g.][]{2008MNRAS.385L..43F,2017Natur.549..488R, 2015ApJ...812...82W}, polar dust \citep[e.g.][]{1993ApJ...409L...5B,1993ApJ...419..136C, 1995MNRAS.277.1134E,2006MNRAS.371L..70E} and disc winds \citep[e.g.][]{1992ApJ...385..460E,2015ARA&A..53..365N,2006ApJ...648L.101E,2019ApJ...884..171H}. 

The study of Seyfert galaxies can help to understand the nature of the AGNs in these scenarios. Seyferts are moderate luminosity AGN galaxies that possess high excitation emission lines  \citep{2017A&ARv..25....2P} which can be used to classify these galaxies in Type-1 AGN (Seyfert 1, hereafter Sy1) and Type-2 AGN (Seyfert 2, hereafter Sy2) in catalogues \citep[e.g.][]{2010A&A...518A..10V}. In addition, Seyfert sub-classes, like the narrow line Sy1 \citep[NLSy1,][]{1985ApJ...297..166O,2017ApJS..229...39R} or the intermediate Seyfert types \citep[][]{ 1981ApJ...249..462O,1992MNRAS.257..677W}, could be ideal to understand the new AGN scenarios \citep{2014MNRAS.438.3340E}. Nevertheless, large samples of spectroscopically classified Seyfert galaxies are mainly limited to $z<1$ \citep[e.g.][]{2010A&A...518A..10V,2017ApJ...850...74K}. 

One solution to increase the number of Seyfert galaxies at higher redsfhits is to identify AGNs through colour selections in IR broad-bands \citep[a compilation of these selection criteria is presented by][table 2]{2017A&ARv..25....2P} and then observe their spectrum in optical wavelengths for the classifications. However, these photometric broad-bands can also be used in spectral energy distribution (SED) analysis, which allows us to obtain a more reliable estimation of the contribution of the AGN than using only the IR colours \citep{2015A&A...576A..10C,2018MNRAS.480.3562D,2020MNRAS.499.4325R,2021A&A...646A..29M,2020MNRAS.495.1853P}. The contribution of the AGN in SED models comes from AGN templates \citep[e.g.][]{2011MNRAS.414.1082M,2021MNRAS.503.2598B} or AGN models \citep[e.g.][]{1992ApJ...401...99P,1994MNRAS.268..235G,1995MNRAS.273..649E,2006MNRAS.366..767F,2008ApJ...685..160N,2012MNRAS.420.2756S,2016MNRAS.458.2288S,2015A&A...583A.120S,2019ApJ...877...95T}, which are fitted together with dust emissions and stellar populations \citep[e.g.][]{ 2016ApJ...833...98C,2018ApJ...854...62L,2019A&A...622A.103B} in different configurations \citep[check][for an overview of the most popular SED fitting codes]{2021MNRAS.505..540T,2021A&ARv..29....2P}. SED models without an AGN contribution do not provide the adequate physical properties of AGNs \citep[e.g.][]{2018ApJ...854...62L, 2018MNRAS.480.3562D}, which could lead to over-estimations in star-formation rate (SFR) and stellar masses, especially in X-ray selected AGNs \citep{2020MNRAS.497.3273F}. 

When the AGN is included in the SED modelling, it is shown to be possible to identify Type-1 and Type-2 AGN \citep[e.g.][]{2016ApJ...833...98C,2020MNRAS.499.4325R}. These AGN types tend to show differences not only in the spectrum, but also in colours from photometric bands. For example, Type-1 AGNs tend to have typically bluer colours than Type-2 because of their higher brightness and lower extinction in the UV and optical bands \citep[][]{2017A&ARv..25....2P}. In addition, in Type-1 the contribution from the AGN seem to be more dominant in UV and NIR--MIR bands, while for Type-2 this contribution is dominant in MIR--FIR bands \citep[][]{2015A&A...576A..10C}, which can explain why it is possible to differentiate Type-1 and Type-2 AGNs according to their fractional contribution of the AGN to the IR \citep[][]{2006MNRAS.366..767F}. Therefore it should also be possible to identify Sy1 and Sy2 galaxies with SED analysis when observing broad-band emissions.

In this work, we aim to assess the importance of the estimated viewing angle in classifying AGN galaxies, and highlight its implications in high-redshift studies. Particularly, we gather a sample of Seyfert galaxies with available photometry in astronomical databases to develop a data-driven approach with easily accessible data. We use \textsc{X-CIGALE} \citep{2020MNRAS.491..740Y}, a modified version of \textsc{CIGALE} \citep{2019A&A...622A.103B}, one of the most popular SED tools to obtain physical parameters in host galaxies and AGN itself. \textsc{X-CIGALE} has several AGN-related improvements compared to \textsc{CIGALE}, ideal for this work. In addition, we test the two different AGN models inside \textsc{X-CIGALE} to see how the classifications depend on the selected model. We compare two popular machine learning techniques, random forest \citep{2001MachL..45....5B} and gradient boosting \citep{2016arXiv160302754C}, with individual physical parameters when classifying unclassified and discrepant cases.   

We present the Seyfert sample selection, the description of the SED models and the verification of the estimations with a similar model in Sect.~\ref{sec:methods}. Then, we select our main physical parameters, compare the estimated galaxy physical parameters from different AGN SED setups, and we compare different classifications in Sect.~\ref{sec:results}. After that, we present the discussions about the role of the viewing angle, its implications and possible bias of these results (Sect.~\ref{sec:disc}). Finally, we present our conclusions (Sect.~\ref{sec:conclu}).

\section{Data and Analysis} \label{sec:methods}

\subsection{Seyfert Sample} 

\begin{figure}
	\includegraphics[width=\columnwidth]{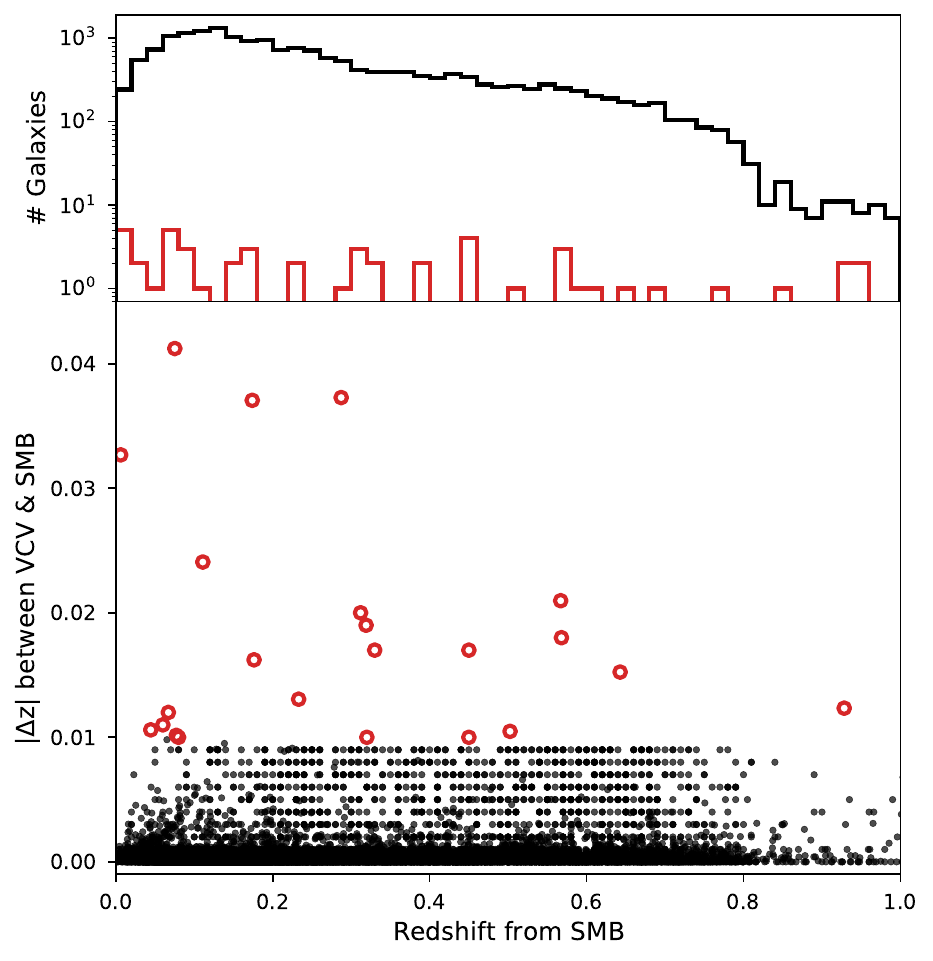}
   \caption{Redshift distribution for the matches between \citetalias{2000A&AS..143....9W} and \citetalias{2010A&A...518A..10V} catalogues. \textit{Upper-panel}: Histogram of the redshift distribution for galaxies where the difference in redshift between the catalogues was below (black line) or above (red line) the threshold at 0.01. Only a few galaxies were discarded using this threshold. \textit{Bottom-panel}: Absolute difference in redshift ($|\Delta z|$) between the catalogues with respect to the \citetalias{2000A&AS..143....9W} redshift. Galaxies with redshifts above 1, or with a large difference in redshift in the catalogues are not shown. Discarded galaxies are indicated as red circles.}
   \label{fig:F1}
\end{figure}

Seyfert galaxies are a good starting point to differentiate between Type-1 and Type-2 AGN, as described by the AGN unification model. Thus, we selected a sample of Seyfert galaxies by combining the \textit{SIMBAD} astronomical database \citep[][hereafter \citetalias{2000A&AS..143....9W}]{2000A&AS..143....9W}\footnote{Data and classification types of the galaxies were retrieved on 2020 December 3.} and the dedicated catalogue of AGNs by \citet[][hereafter \citetalias{2010A&A...518A..10V}]{2010A&A...518A..10V}. \citetalias{2000A&AS..143....9W} is widely used for retrieving basic information of galaxies in an homogeneous manner, while \citetalias{2010A&A...518A..10V} is one of the most popular catalogues for AGN studies. From \citetalias{2000A&AS..143....9W}, we picked galaxies whose main type was Seyfert, including: Seyfert 1 (Sy1), Seyfert 2 (Sy2) and, unclassified Seyfert galaxies. From \citetalias{2010A&A...518A..10V}, we selected all Seyfert types galaxies, including all intermediate numerical classifications (e.g. Sy1.5). We cross-matched \citetalias{2000A&AS..143....9W} and \citetalias{2010A&A...518A..10V} samples using a cross-matching radius of 2\arcsec. We removed galaxies where the difference in redshifts between the catalogues ($|\Delta z|$) was higher than 0.01, which is the limit of the reported numerical accuracy between the catalogues, as shown in Figure~\ref{fig:F1}. This decision help us to avoid misidentification and uncertain redshifts in the sample of Seyfert galaxies.

\subsubsection{Classification type}

We used the classification type from both \citetalias{2000A&AS..143....9W} and \citetalias{2010A&A...518A..10V} samples. Classifications types in \citetalias{2010A&A...518A..10V} come from spectroscopic measurements with SDSS data \citep{ 2009ApJS..182..543A}, while classification types in \citetalias{2000A&AS..143....9W} are a compendium of the literature. The information gathered in \textit{SIMBAD} was manually added by documentalist till the 90's, and now is done semi-automatically with \textit{COSIM} \citep{ 2018EPJWC.18602004B}. Unfortunately, the source of the classifications was not recorded until 2006, therefore almost half of the Seyfert classifications in \citetalias{2000A&AS..143....9W} are marked as coming from \textit{SIMBAD}. A small fraction of our Seyfert galaxies ($\sim 5\%$) still have an unknown source, as the object type classification is still under development \citep{2020ASPC..522..105O}\footnote{We use the 2018 August 2 classification version.}. Therefore, classifications in \citetalias{2000A&AS..143....9W} should be taken with caution. If the classification source is unknown and the main Seyfert classification in \citetalias{2000A&AS..143....9W} matches \citetalias{2010A&A...518A..10V}, we assume that the classification source is \citetalias{2010A&A...518A..10V}. If the main Seyfert classification source is unknown and the classification in \citetalias{2010A&A...518A..10V} is Seyfert 3 (also known as LINERs) we remove the galaxies from the sample. We re-classified the remaining unknown sources, 49 galaxies, as unclassified Seyfert to study them further. These decisions led us to a sample of 18\,921 Seyfert galaxies. 

For the classifications in \citetalias{2000A&AS..143....9W}, we found: i) Almost half of the classifications (45\%) came from the basic data of the galaxy (asigned by the astronomical database); ii) \citet{2014ApJ...788...45T} work contributed to 21\% of the Sy1 and Sy2 classifications iii) \citet{2006ApJS..166..128Z}, \citet{2015ApJS..219....1O}, and \citet{2017ApJS..229...39R} together contributed to 25\% of \citetalias{2000A&AS..143....9W} classifications, all of them in Sy1 galaxies. The \citetalias{2000A&AS..143....9W} sample contains in total: 13\,760 Sy1, 5\,040 Sy2, and 121 unclassified Seyfert galaxies. 

In \citetalias{2010A&A...518A..10V}, we found 17 different Seyfert type classifications. In this work, we focus on the typical Sy1, Sy2 and unclassified Seyferts which all together account for 71\% of the sample. We added the narrow-line Sy1 (NLSy1) galaxies \citep[e.g.][]{2006ApJS..166..128Z,2017ApJS..229...39R} to the Sy1 classification because most of the NLSy1  galaxies in \citetalias{2010A&A...518A..10V} are classified as Sy1 in \citetalias{2000A&AS..143....9W}. However, some differences in the estimates may indicate that the total accretion power in  NLSy1 galaxies is higher than in normal Sy1 galaxies, as we verified in Appendix~\ref{App:S1n}. Three of the NLSy1 galaxies (2MASX J10194946+3322041, 2MASS J09455439+4238399 and 2MASX J23383708-0028105) were classified as Sy2 in \citetalias{2000A&AS..143....9W}, so we reclassified them as unclassified Seyfert for further study. In addition, we checked the subgroups between Sy1 and Sy2 as divided by \citet{1977ApJ...215..733O,1981ApJ...249..462O} and \citet{1992MNRAS.257..677W} which account for 5\% of the sample. We denoted a small fraction of the galaxies from \citetalias{2010A&A...518A..10V} ($\sim$1\%), which do not fall in the classifications described before (e.g. LINERS, NLSy1.2 and polarised classifications), as alternative Seyfert galaxies. \citetalias{2010A&A...518A..10V} sample contains in total: 13\,180 Sy1, 4\,567 Sy2, 84 unclassified Seyfert galaxies, 920 in the intermediate numerical subgroups between Sy1 and Sy2, and 170 alternative Seyfert galaxies. 

\subsubsection{Photometry}

We used 31 bands in the UV-FIR wavelength range to get a well-sampled SED for our sample of Seyfert galaxies. We list the selected bands for the SED modelling in Table~\ref{tab:Bands} with their respective effective wavelength and number of galaxies detected in that band. We retrieved photometric values of these bands available in CDS\footnote{\url{http://cdsportal.u-strasbg.fr/}} and NED\footnote{The NASA/IPAC Extragalactic Database (NED) is funded by the National Aeronautics and Space Administration and operated by the California Institute of Technology.}. CDS and NED photometric data points are ideal for this data-driven work as they are published and curated by other researchers, saving time in the photometric reduction. However, we needed to make sure that the retrieved data were good enough for our purpose.

We keep in mind that the use of heterogeneous measurements may lead to some systematics in the analysis. For example, for galaxies in the local Universe, or where the instrument resolution is good enough to resolve the galaxies, measurements will come from specific regions within the galaxies, like their centres. In contrast, for galaxies at higher redshifts or instruments where the resolution is not high enough to resolve them spatially the measurements will correspond to the whole galaxy as we observe the galaxies as unresolved point sources. Fortunately, in terms of spatial resolution, most of the galaxies in this sample could be treated as point sources for most of the instruments operating at different wavelengths, therefore we expect measurements at different wavelengths to be consistent with each other. When this is not the case, discrepant apertures at different wavelengths will lead to unphysical jumps in the SED models, which will give us erroneous fittings that we can ignore before going further with the analysis.

Hence, we followed a series of steps to obtain a set of galaxies with useful photometry. First, we decided not to use upper or lower limits from published values in CDS or NED. Second, we remove duplicate data photometry values between the CDS and NED, keeping the value reported in CDS. Third, we used the mean value when more than one measurement was available per band. These measurements can also come from the same apertures but from different works or methods. Fourth, we selected photometric data points with a relative error (after propagating the initial reported errors) below 1/3. Finally, we accounted for the absolute calibration error for each band as in \citet{2020MNRAS.499.4325R}, where instrument-dependent uncertainties were added to the measurement uncertainties.

\begin{table}
\centering
\caption{Photometric bands used in the SEDs modelling. The last column shows the number of galaxies detected in a given band.}
\label{tab:Bands}
\begin{tabular}{llcc}
\hline
\hline
Mission or & Band & Effective & Number of \\
Survey& & Wavelength [$\mu$m] & galaxies \\
\hline
\textit{GALEX} & FUV & 0.152& 6456\\
& NUV & 0.227 & 9266\\
SDSS & u & 0.354 & 12024\\
& g & 0.477 & 12542\\
& r & 0.623 & 12326\\
& i & 0.762 & 12274\\
& z & 0.913 & 11604\\
2MASS & J & 1.25 & 7018\\
& H & 1.65 & 6566\\
& Ks & 2.17 & 8215\\
\textit{Spitzer} & IRAC-1 & 3.6 & 4063\\
& IRAC-2 & 4.5 & 4048\\
& IRAC-3 & 5.8 & 458\\
& IRAC-4 & 8.0 & 447\\
& MIPS1 & 24.0 & 809\\
& MIPS2 & 70.0 & 225\\
& MIPS3 & 160.0 & 110\\
\textit{WISE} & W1 & 3.4 & 13170\\
&W2 & 4.6 & 13165\\
&W3 & 12.0 & 12361\\
&W4 & 22.0 & 8295\\
\textit{IRAS} & IRAS-1 & 12.0 & 462\\
&IRAS-2 & 25.0 & 634\\
&IRAS-3 & 60.0 & 979\\
&IRAS-4 & 100.0 & 722\\
\textit{Herschel} & PACS-blue & 70.0 & 265\\
&PACS-green & 100.0 & 178\\
&PACS-red & 160.0 & 303\\
&SPIRE-PSW & 250.0 & 840\\
&SPIRE-PMW & 350.0 & 476\\
&SPIRE-PLW & 500.0 & 233\\
\hline
\end{tabular}
\end{table}

We constrained the galaxies to have good coverage over the optical and IR wavelengths. We only include sources satisfying both criteria: i) more than five photometric data points in wavelengths between $0.1-3\micron$ (\textit{GALEX}, SDSS and 2MASS), and ii) more than three photometric data points in wavelengths between $3-500\micron$ (\textit{Spitzer}, \textit{WISE}, \textit{IRAS} and \textit{Herschel}). With these criteria, we ended up with 13\,173 Seyfert galaxies for which we carry out the following SED modelling analysis.

We also looked for X-ray and radio photometric data points. However, the coverage at those wavelengths was not homogeneously tabulated in CDS or NED as in the selected bands in Table~\ref{tab:Bands}. We decided not to use X-ray and radio wavelengths as this will require more computational and time efforts for a few number of galaxies (only $\sim 0.01\%$ of the sample). In addition, currently \textsc{X-CIGALE} does not include a AGN radio component. We discuss the implications of not using X-ray data in Sect.~\ref{sec:DiscXray}. 

\subsection{SED Models}\label{subsec:SEDmodels}

\subsubsection{Parameter grids}

We modelled the SEDs of the Seyfert galaxies with \textsc{X-CIGALE} \citep{2020MNRAS.491..740Y}. \textsc{X-CIGALE} is a modified version of \textsc{CIGALE} \citep{2019A&A...622A.103B}, a SED fitting code based on an energy balance principle. The difference between \textsc{X-CIGALE} and \textsc{CIGALE} is the addition of i) an X-ray photometry module and, ii) a polar dust model in AGNs. These two enhancements help to connect the X-ray emission to the UV-to-IR SED, and account for dust extinction in the polar angles, respectively. The X-ray emission is helpful to constrain AGN intrinsic accretion power in the SED \citep{2018ApJ...866...92L,2021ApJ...912...91T}, and the polar dust follow observational results from MIR interferometry \citep{2016A&A...591A..47L,2017NatAs...1..679R}.

We included six modules which account different galactic emission processes to fit the SEDs. The first module defines the star-formation history (SFH). We used a delayed SFH model for our sample of galaxies because this has shown a good agreement in different types of galaxies with ongoing or recent starburst events \citep{2018MNRAS.480.3562D,2020MNRAS.499.4325R}, and can provide better estimates for physical parameters such as star-formation rate (SFR) and stellar mass \citep{2015A&A...576A..10C}. The second module defines the single-age stellar population (SSP). We selected the standard \citet{2003MNRAS.344.1000B} model taking into account the initial mass function (IMF) from \citet{2003PASP..115..763C} and a metallicity close to solar. The dust attenuation law from \citet{2000ApJ...533..682C} is our third module. This module helps us control the UV attenuation with the colour excess E(B-V), and also the power-law slope ($\delta$) that modifies the attenuation curve. The fourth module takes the dust emission in the SED into account. We modelled the dust emission following \citet{2014ApJ...784...83D}, implementing a modified blackbody spectrum with a power-law distribution of dust mass at each temperature, 
\begin{eqnarray}
dM \propto U^{-\alpha} dU, \label{eq:alpha}
\end{eqnarray}
where $U$ is the local heating intensity.  We also included the nebular emission module although we did not change the default parameters.

The sixth and most important module for this work is the module that describes the AGN SED. For our experiments setups, we selected the two AGN modules available in \textsc{X-CIGALE}: A simple smooth torus \citep{2006MNRAS.366..767F}, and a tho-phase (smooth and clumpy) torus \citep[][also known as SKIRTOR]{2016MNRAS.458.2288S}. For both models, we covered a larger sample of parameters for the viewing angle $\,i\,$ and the fraction of AGN contribution to the IR luminosity  $f_{\rm{AGN}}$ \citep[][eq. 1]{2015A&A...576A..10C}, 
\begin{eqnarray}
\rm{L}_{\rm{IR}}^{\rm{AGN}} = f_{\rm{AGN}} \times \rm{L}_{\rm{IR}}^{\rm{total}}, \label{eq:fagn}
\end{eqnarray}
to investigate the effect of $\,i\,$ in Seyfert galaxies. In addition, we set the extinction law of polar dust to the SMC values \citep{1984A&A...132..389P}, with a temperature of polar dust to 100\,K  \citep[][]{2021A&A...646A..29M,2021A&A...654A..93B} and the emissivity index of polar dust to 1.6 \citep[][]{2012MNRAS.425.3094C}. The values for the colour excess of polar dust go from no extinction (E$(B-V)=0$) to E$(B-V)=1.0$ because E$(B-V)$ cannot be well constrained only from the SED shape \citep{2020MNRAS.491..740Y}. However, adding E$(B-V)$ as a free parameter can improve the accuracy of the classification type \citep{2021A&A...646A..29M}.

The ratio of the outer to inner radii of the dust torus R$_{\textnormal{out}}$/R$_{\textnormal{in}}$ and the optical depth at 9.7 $\micron$ $\tau$ are parameters that both AGN models share. The selection of these values changes in studies similar to this one on AGN galaxies depending on the AGN model used. When using the Fritz model it is common to use  $\tau = 6.0$ and R$_{\textnormal{out}}$/R$_{\textnormal{in}} = 60$ \citep[e.g.][]{2017A&A...597A..51V,2018A&A...620A..50M,2020MNRAS.499.4068W}, while for SKIRTOR $\tau = 7.0$ and R$_{\textnormal{out}}$/R$_{\textnormal{in}} = 20$ are often used \citep[e.g.][]{2020MNRAS.491..740Y,2021A&A...646A..29M}. We adopted the same values as in the literature, even though the $\tau$ values can be considered large, with the difference that we used R$_{\textnormal{out}}$/R$_{\textnormal{in}} = 30$ for the Fritz model to make it more similar to SKIRTOR. We used the default geometrical parameters (power-law densities) in both models to focus on $\,i\,$, $f_{\rm{AGN}}$ and E$(B-V)$. Finally, we tested two angle configurations: i) with viewing angles between 0\degr and 90\degr, and ii) using typical viewing angles of Type-1 and Type-2 AGNs of 30\degr (unobscured) and 70\degr (obscured). This comparison helps us to understand how important the viewing angle input parameter is in \textsc{X-CIGALE}. 

In summary, we used the parameters and values given in Table~\ref{tab:Par_CIG} to define the grid of \textsc{X-CIGALE} SED models for the sample of Seyfert galaxies. For the remaining parameters not shown in Table~\ref{tab:Par_CIG}, we adopted the \textsc{X-CIGALE} default settings. We decided not to include the X-ray or radio modules due to the lack of homogeneous information for the selected sample of Seyfert galaxies (see Sect.~\ref{sec:DiscXray}). We assumed the redshifts from \citetalias{2000A&AS..143....9W} in the SED fits.

\begin{table*}
\centering
\caption{\textsc{X-CIGALE} grid parameter values adopted for the modelling described in Section~\ref{subsec:SEDmodels}}
\label{tab:Par_CIG}
\begin{tabular}{p{0.15\textwidth}p{0.25\textwidth}p{0.45\textwidth}}
\hline
\hline
Parameter & Values & Description \\
\hline
\multicolumn{3}{c}{Star formation history (SFH): Delayed}\\
$\tau_{\textnormal{main}}$&50, 500, 1000, 2500, 5000, 7500& e-folding time of the main stellar population model (Myr).\\
Age &500, 1000, 2000, 3000, 4000, 5000, 6000&Age of the oldest stars in the galaxy (Myr).\\
\hline
\multicolumn{3}{c}{Single-age stellar population (SSP): \citet{2003MNRAS.344.1000B}}\\
IMF& 1& Initial Mass Function from {\citet{2003PASP..115..763C}}.\\
Metallicity & 0.02 & Assuming solar metallicity. \\
\hline
\multicolumn{3}{c}{Dust attenuation: \citet{2000ApJ...533..682C}}\\
E$(B-V)$ &0.2, 0.4, 0.6, 0.8 & Color excess of the nebular light for the young and old population.\\
E$(B-V)_{\rm{factor}}$ &0.44 & Reduction factor for the  E$(B-V)$ to compute the stellar continuum attenuation.\\
Power-law slope ($\delta$) & -0.5, -0.25, 0.0, 0.25, 0.5& Slope delta of the power law modifying the attenuation curve.\\
\hline
\multicolumn{3}{c}{Dust emission: \citet{2014ApJ...784...83D}}\\
$\alpha$ & 1.0, 1.5, 2.0, 2.5, 3.0 & Alpha from the power-law distribution in Eq.~\ref{eq:alpha}. \\
\hline
\multicolumn{3}{c}{AGN models:}\\
$i$ & 0 -- 90$^{\rm a}$ & Viewing angle (face-on: $i=0$\degr, edge-on: $i=90$\degr). \\
$f_{\rm{AGN}}$ & 0.1 -- 0.9 in steps of 0.05 & Fraction of AGN torus contribution to the IR luminosity in Eq.~\ref{eq:fagn} \\
E$(B-V)_{\rm{polar}}$ & 0.0, 0.03, 0.1, 0.2, 0.3, 0.4, 0.6, 1.0 & E$(B-V)$ of polar dust (fig 4 of {\citet{2020MNRAS.491..740Y}}).\\
$\rm{T_{pd}}$ & 100 & Temperature of polar dust (eq. 10 of {\citet{2020MNRAS.491..740Y}})..\\
$\beta_{\rm{pd}}$& 1.6 & Emissivity index of polar dust (eq. 10 of {\citet{2020MNRAS.491..740Y}}).\\
\multicolumn{3}{c}{Fritz model (\citet{2006MNRAS.366..767F})}\\
R$_{\textnormal{out}}$/R$_{\textnormal{in}}$ & 30.0 & Ratio of the outer to inner radii of the dust torus. \\
$\tau$ & 6.0 & Optical depth at 9.7 $\micron$. \\
$\beta$ & $-$0.50 & Beta from the power-law density distribution for the radial component of the dust torus (eq. 3 of \citet{2006MNRAS.366..767F}).\\
$\gamma$ & 4.0 & Gamma from the power-law density distribution for the polar component of the dust torus (eq. 3 of \citet{2006MNRAS.366..767F}).\\
Opening Angle ($\theta$) & 100.0& Full opening angle of the dust torus (fig 1 of \citet{2006MNRAS.366..767F}). \\
\multicolumn{3}{c}{SKIRTOR model (\citet{2016MNRAS.458.2288S})}\\
R$_{\textnormal{out}}$/R$_{\textnormal{in}}$ & 20.0 & Ratio of the outer to inner radii of the dust torus.\\
$\tau$ & 7.0 & Optical depth at 9.7 $\micron$. \\
$p$ & 1.0 & Power-law exponent of the radial gradient of dust density (eq. 2 of {\citet{2012MNRAS.420.2756S}}).\\
$q$ & 1.0 & Angular parameter for the dust density (eq. 2 of {\citet{2012MNRAS.420.2756S}}).\\
$\Delta$ & 40 & Angle between the equatorial plane and edge of the torus (half opening angle).\\
\hline
\end{tabular}

\begin{flushleft}
$^{\rm a}$ We covered viewing angles between 0\degr and 90\degr in steps of 10\degr. We used the values closest to the predefined angle grid in \textsc{X-CIGALE}. For other setups, we used only 30\degr and 70\degr, taking into account that $i = 90- \psi$ (with $\psi$ the angle between equatorial axis and line of sight).
\end{flushleft}
\end{table*}

\subsubsection{Cleaning \textsc{X-CIGALE} fits}

We ran another setup of \textsc{X-CIGALE} without the AGN module (hereafter No-AGN) in addition to the \textsc{X-CIGALE} setups with AGN models described in Table~\ref{tab:Par_CIG}. The No-AGN setups helped us to identify bad-fittings and ambiguous cases where an AGN is not dominant in the SED, even though the galaxies are classified as Seyfert.

\begin{figure*}
	\includegraphics[width=\textwidth]{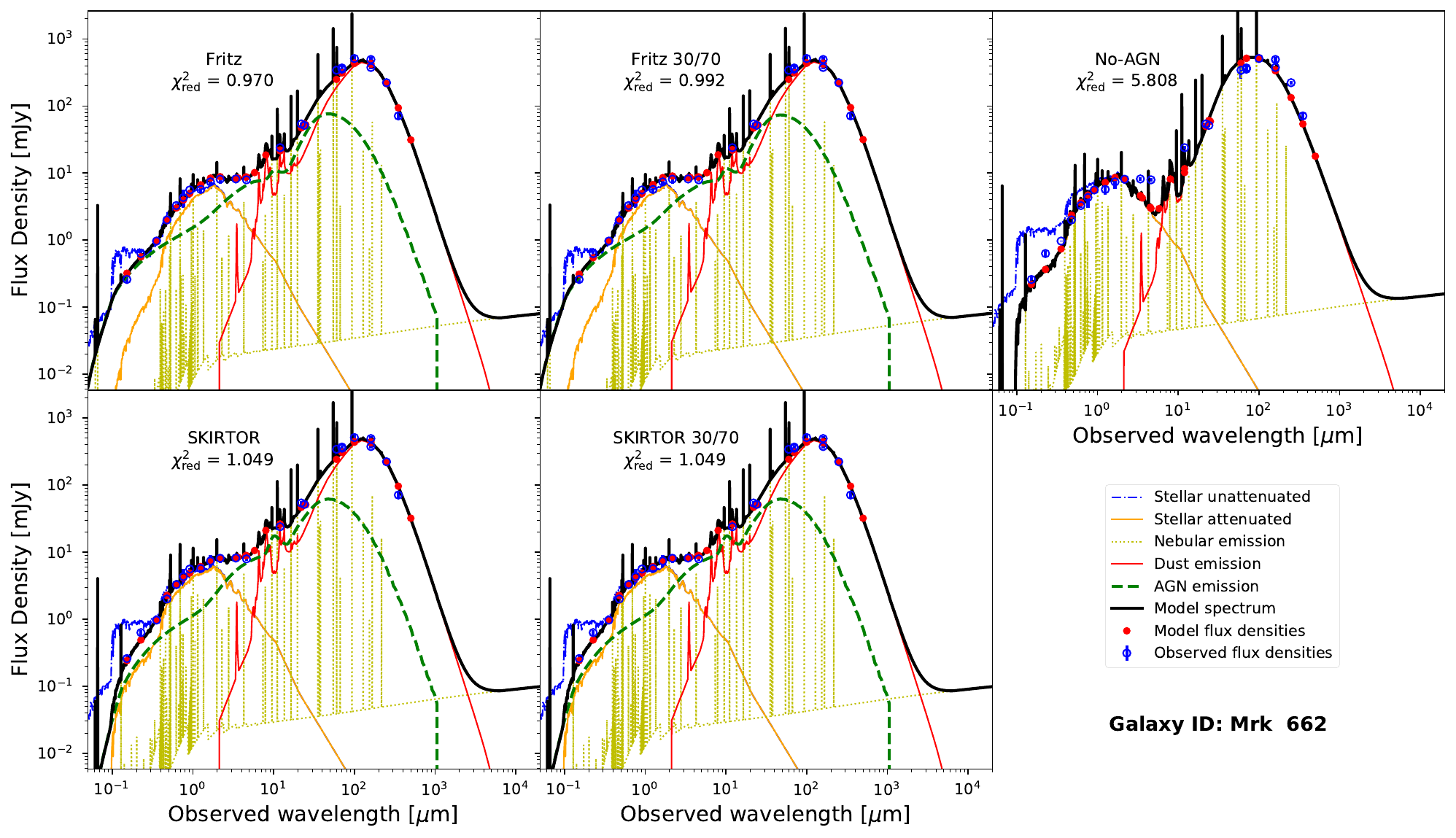}
   \caption{Example SEDs of the five SED fitting setups used in this work for the galaxy Mrk 662 at $z=0.05$. No-AGN setup (upper-right panel) usually have higher reduced $\chi^2$ ($\chi^2_{\rm{red}}$) values than AGN setups (other four panels). Each plot contains the contribution to the model spectrum (black line) of: nebular emission (gold dotted lines), attenuated (orange) and non-attenuated stellar emission (blue dot-dashed), dust emission (red solid), and AGN emission (green dashed). The red dots are the best model flux densities and the blue squares mark the observed flux densities with 1$\sigma$ error bars.}
   \label{fig:SEDExample}
\end{figure*}

\textsc{X-CIGALE} minimises the $\chi^2$ statistic and produces probability distribution functions for the grid parameters by assuming Gaussian measurement errors \citep{2005MNRAS.360.1413B,2009A&A...507.1793N,2011ApJ...740...22S}. In Fig.~\ref{fig:SEDExample}, we show an example of the SED fitting in one of the galaxies (Mrk 662) using the five different setups: smooth torus (Fritz setup from now on), smooth and clumpy torus (SKIRTOR setup from now on), smooth and clumpy torus with only two viewing angles (Fritz 30/70 and SKIRTOR 30/70 setups), and a model without AGN (No-AGN setup). The No-AGN setup (upper-right panel) shows a significant difference with the AGN setups in terms of reduced $\chi^2$ ($\chi^2_{\rm{red}}$), which is expected as an AGN model is needed for most of our Seyfert galaxies. However, in some cases, No-AGN setup have a lower or equal $\chi^2_{\rm{red}}$ than AGN setups, meaning a worse fit with the AGN setup and/or a non-dominant AGN. Another slight difference in these SEDs is the contribution from the AGN (green dashed line) at 10$\micron$, which varies depending on the best-fitting to a given setup. These differences can play a role in some physical parameters (e.g. attenuation), that may affect the classification type.

\begin{figure}
	\includegraphics[width=\columnwidth]{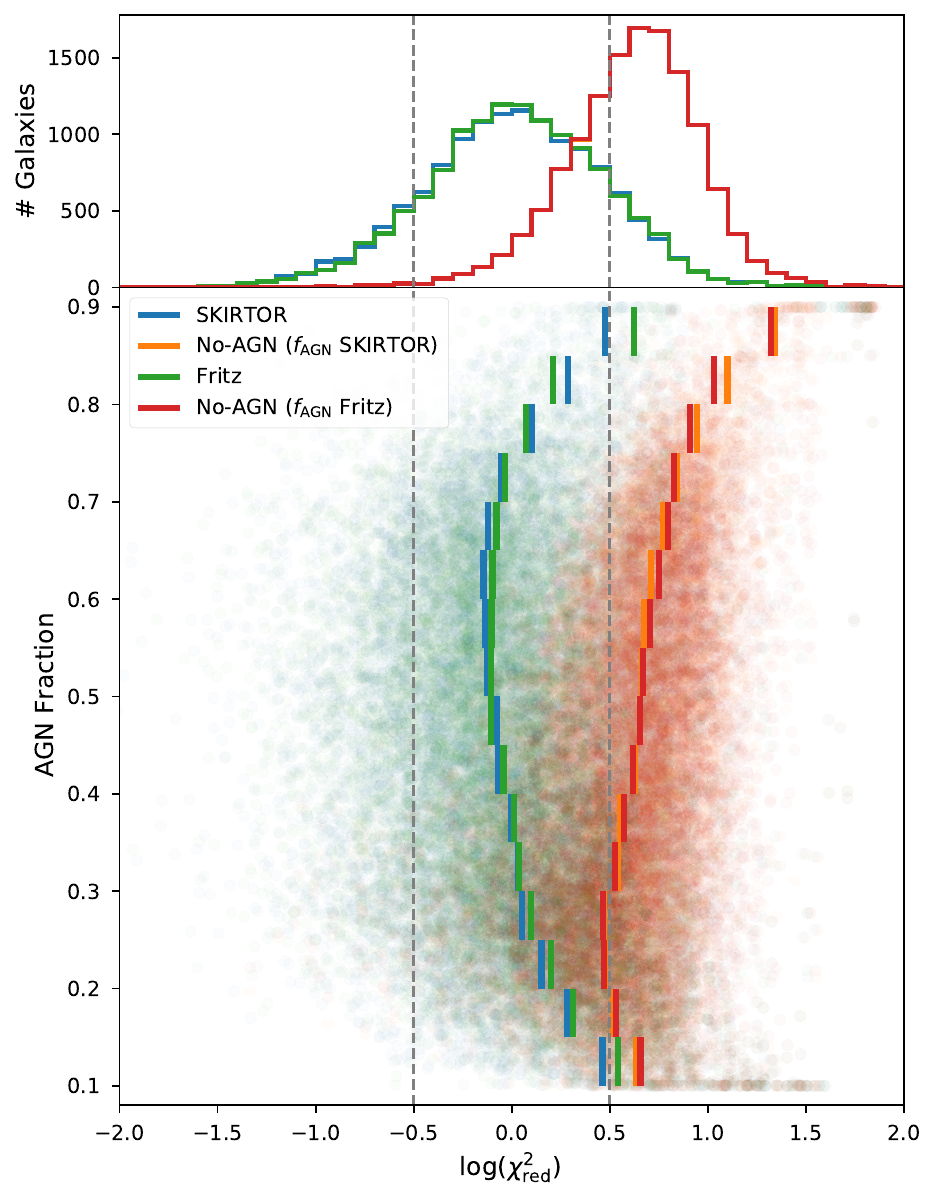}
   \caption{Reduced-chi-square ($\chi^2_{\rm{red}}$) distribution for the setups used in this work. \textit{Upper-panel}: histogram of the $\chi^2_{\rm{red}}$ distribution for SKIRTOR (blue), Fritz (green) and No-AGN (red) setups. \textit{Bottom-panel}: $\chi^2_{\rm{red}}$ against  the estimated AGN fraction ($f_{\rm{AGN}}$) for each galaxy (shaded dots) and the running median in bins of $f_{\rm{AGN}}$ values (solid lines). No-AGN setup ($f_{\rm{AGN}}=0$) is plotted assuming the $f_{\rm{AGN}}$ values of the AGN setups, thus showing slightly different median values with Fritz and SKIRTOR setups. We assumed that well fitted galaxies have a $\log(\chi^2_{\rm{red}})$ between -0.5 and 0.5 (grey dashed lines).}
   \label{fig:Chisq}
\end{figure}

We compare the $\log(\chi^2_{\rm{red}})$ distribution for the SED setups in Fig.~\ref{fig:Chisq}. There is a small difference in $\chi^2_{\rm{red}}$ between AGN setups (Fritz and SKIRTOR), with and average value of $\Delta \chi^2_{\rm{red}}=0.147$, which shows that both setups fit the data similarly. Besides, we found that AGN and No-AGN setups have an average difference in $\log(\chi^2_{\rm{red}})$ of $\sim$0.4 dex, favouring AGN setups. For the No-AGN setup, if we use the same $f_{\rm{AGN}}$ value as in the AGN setups (which by construction have a $f_{\rm{AGN}}=0$), then we can compare the difference in $\chi^2_{\rm{red}}$ when adding an AGN model in the SED. We found the smallest $\chi^2_{\rm{red}}$ differences at $f_{\rm{AGN}}$ below 0.2, while the largest differences are at $f_{\rm{AGN}}\sim0.7$, when comparing the setups with and without AGN. For the SED setups with AGN, we found that galaxies with $f_{\rm{AGN}}$ between 0.2 and 0.8 have $\chi^2_{\rm{red}}$ values close to one. Therefore, for most galaxies outside this $f_{\rm{AGN}}$ range have poorer fittings. This differences shows the importance of adding the AGN model in the SED fitting and how $\chi^2_{\rm{red}}$ changes for non-dominant AGNs ($f_{\rm{AGN}}<0.2$) and highly-dominant AGNs ($f_{\rm{AGN}}>0.85$).

To better compare the No-AGN and AGN setups, we use the Bayesian Inference Criterion (BIC) to see if the AGN module is preferred for the fits, as in other \textsc{CIGALE} works \citep[e.g.][]{2019A&A...632A..79B}. The BIC is defined as $\rm{BIC}= \chi^2 + k\times\ln(N)$, with $k$ the number of free parameters and $N$ the number of data points used for the fit \citep{2018A&A...615A..61C} and works as an approximation of the Bayes factor \citep{kass1995bayes}. Then, the difference between the setups can be calculated as $\Delta\rm{BIC} = \chi^2_{\rm{AGN}} - \chi^2_{\rm{No-AGN}} + \ln(N)$, as we are just fixing the $f_{\rm{AGN}}$ to zero in the No-AGN setup. We adopt a positive evidence criterion for No-AGN setup \citep{2016ApJ...827...20S}, meaning that galaxies with a $\Delta\rm{BIC} \geq 2$ will prefer the No-AGN setup.

We imposed some constraints in the \textsc{X-CIGALE} estimated values to clean the set of derived parameters for this work. First, we used galaxies with a $\log(\chi^2_{\rm{red}})$ between -0.5 and 0.5 (grey dashed lines in Fig.~\ref{fig:Chisq}) to avoid over and underestimations, respectively. Second, we selected galaxies where the AGN setups were preferred i.e. $\Delta\rm{BIC} < 2$. Finally, we selected galaxies where their estimated $1\sigma$ error in SFR was below one dex, to obtain reliable SFR estimations. Unfortunately, this last selection causes a bias against quiescent galaxies. These constraints led us to remove between 4\,757 and 5\,181 galaxies (depending on the AGN setup), from which: 69--75\% galaxies were over or underestimated fits, 9--15\% galaxies had a better fit with the No-AGN setup, and 31-38\% galaxies where SFR was not well constrained.

To summarise, we present in Table~\ref{tab:NumSey} the total number of galaxies of the original samples (\citetalias{2000A&AS..143....9W} and \citetalias{2010A&A...518A..10V}), samples with photometry that meet our criteria for the SED fitting procedure, and the well-constrained fits with the \textsc{X-CIGALE} AGN models with respect to their Seyfert classification. In Appendix~\ref{App:mock}, we verify the quality of the fits for the main parameters studied in this work and the parameter space used for the fitting procedure by mock analysis. The mock analysis is a standard procedure included inside the \textsc{CIGALE}. A detailed description of this process can be found in \citet[][]{2019A&A...622A.103B}.

\begin{table*}
\centering
\caption{Summary of the Seyfert samples used in this work. The original samples of Seyfert galaxies are in columns 2 (\citetalias{2010A&A...518A..10V}) and 3 (\citetalias{2000A&AS..143....9W}). Galaxies with photometry from the samples fulfilling our criteria are in column 4. The last four columns show the final counts for well-constrained SEDs in \textsc{X-CIGALE}. We show the counts from \citetalias{2010A&A...518A..10V} and \citetalias{2000A&AS..143....9W} classifications in columns 4-8 for Sy1, Sy2 and unclassified Seyfert rows. We show the counts for \citetalias{2010A&A...518A..10V} classification for intermediate or other Seyfert galaxies. The last row shows the total number of galaxies in each of the samples.}
\label{tab:NumSey}
\begin{tabular}{l|cc|c|cccc}
\hline
Seyfert & \multicolumn{2}{|c|}{Samples} &  With  & \multicolumn{4}{|c|}{\textsc{X-CIGALE} AGN models} \\
\cline{2-3} \cline{5-8}
Classification & VCV & SMB & Photometry & SKIRTOR & Fritz & SKIRTOR 30/70 & Fritz 30/70\\
\hline
Seyfert 1 & 13\,177 & 13\,760 & 8\,942 / 9\,421 & 5\,913 / 6\,328 & 6\,295 / 6\,683 & 6\,064 / 6\,453 & 6\,350 / 6\,723 \\
Seyfert 2 & 4\,567 & 5\,040 & 3\,284 / 3\,679 & 1\,473 / 1\,626 & 1\,535 / 1\,697 & 1\,390 / 1\,544 & 1\,361 / 1\,515\\
Unclassified Seyfert & 87 & 121 & 54/73 & 27 / 38 & 28 / 36 & 25 / 34 & 28 / 37\\
Intermediate Seyfert & 920 & $\cdots$ & 756 & 507 & 489 & 492 & 479\\
Alternative Seyfert& 170 & $\cdots$ & 137 & 72 & 69 & 60 & 57\\
\hline
Total galaxies & 18\,921 & 18\,921 & 13\,173 & 7\,992 & 8\,416 & 8\,031 & 8\,275\\
\hline
\end{tabular}
\end{table*}

\subsubsection{Verification with other estimates}\label{sec:verif}

We verified the estimates from our procedure by comparing with a similar study done with \textsc{CIGALE} by \citet{2017A&A...597A..51V}. \citet{2017A&A...597A..51V} uses a sample of 1\,146 galaxies selected from the CASSIS spectroscopic sample \citep{2015ApJS..218...21L} with good photometric coverage from UV to mid-IR in the redshift range $0<z<2.5$. As all these galaxies have been observed with \textit{Spitzer}/IRS, the sample is biased to significantly brighter mid-IR galaxies. There are two main differences between the estimated physical parameters from \citet{2017A&A...597A..51V} and this work. The first difference is the way the photometry was retrieved. \citet{2017A&A...597A..51V} used specific catalogues that contain broad-band photometry for their sample of galaxies, while in this work we use data available in CDS and NED. Thus, we include additional information as the databases collect more broad-band photometry. The second difference is the assumed grid values for the SED modelling. Although the numerical values in most of the input parameters are not the same, here we mention the three most important differences between the grids. First, the IMF in this work comes from \citet{2003PASP..115..763C}, while \citet{2017A&A...597A..51V} uses the IMF from \citet[][]{1955ApJ...121..161S}. The use of the Chabrier IMF will lead to lower stellar mass values in this work with respect to \citet{2017A&A...597A..51V}. Second, the parameter space for $f_{\rm{AGN}}$ in this work is finer sampled and more homogeneously distributed than the one from \citet{2017A&A...597A..51V}. And third, the selected viewing angles for the AGN in \citet{2017A&A...597A..51V} are only $i=0\degr$ and $i=90\degr$ for the Fritz AGN model.  

\begin{figure}
	\includegraphics[width=\columnwidth]{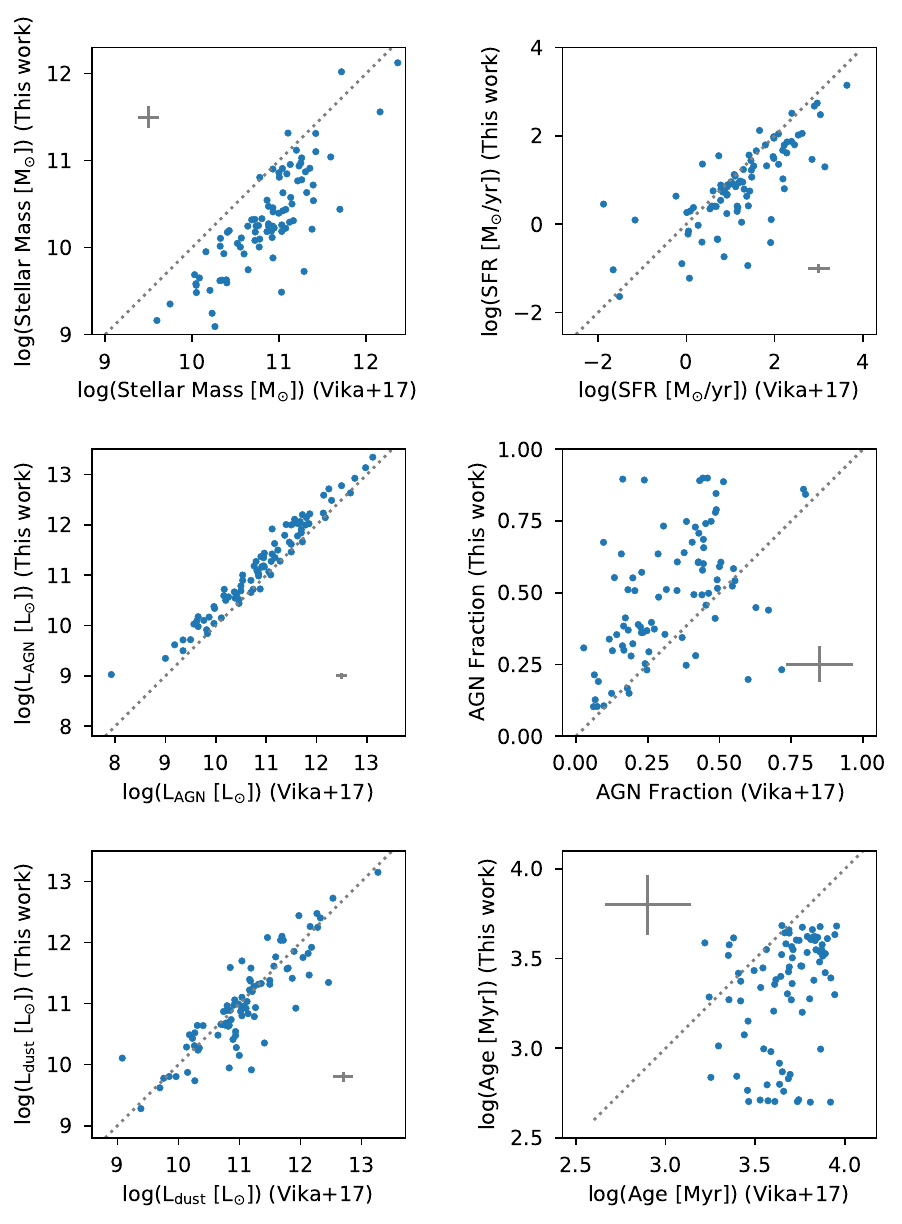}
   \caption{Comparison between the estimated physical parameters from this work and by \citet{2017A&A...597A..51V}. The pointed line represents the 1:1 relation. The grey crosses represent the median estimated error for each of the parameters. In general, the estimates from this work agree with the results presented by \citet{2017A&A...597A..51V}. Based on the median estimated errors, our results are better constrained than the ones from \citet{2017A&A...597A..51V}.}
   \label{fig:Verif}
\end{figure}

We selected the smooth torus model with two viewing angles (Fritz 30/70 setup) to compare the results from \citet{2017A&A...597A..51V}, as it is the most similar model in this work. We cross-matched the \citet{2017A&A...597A..51V} catalogue with the Fritz 30/70 setup between 3\arcsec and we found 87 galaxies for the comparison in the range of $0.02<z<1.4$, with a median of $z=0.13$. In Fig.~\ref{fig:Verif}, we present the comparison of six physical parameters between this work and \citet{2017A&A...597A..51V}. In terms of SFR, AGN luminosity and dust luminosity, there is no clear difference between the estimates, the small systematic offsets are related to the different assumed values of the grid. For example, in both works the dust model from \citet{2014ApJ...784...83D} is used to estimate the dust luminosity, which depends mainly on the power-law parameter $\alpha$ of the mass distribution (Eq.~\ref{eq:alpha}). In \citet{2017A&A...597A..51V}, they only use two values for $\alpha$, while in this work we use five. On the contrary, the estimates on stellar mass are lower in this work mainly due to the adopted IMF, but the estimated median errors are similar between both works.

Age and $f_{\rm{AGN}}$ are the only physical parameters that are very different from \citet{2017A&A...597A..51V}. In the case of $f_{\rm{AGN}}$, the estimates are constrained by the different grid selection. However, the estimated median error is lower in this work than in \citet{2017A&A...597A..51V}, as we use more photometric bands and a finer $f_{\rm{AGN}}$ grid. In the case of the age, \citet{2017A&A...597A..51V} noted that age estimates are not well constrained in \textsc{CIGALE}, besides the differences in the grid values. This problem is also obvious from our estimates, which have uncertainties similar to those presented by \citet{2017A&A...597A..51V}. In general, the physical parameters estimated in \citet{2017A&A...597A..51V} are similar to those presented in this work, validating our approach of obtaining data directly from astronomical databases.

\section{Results}\label{sec:results}

\subsection{Feature selection}\label{sec:FeatureSel}

Recent advances in algorithms and machine learning techniques are helping to classify very complicated physical systems \citep[][]{2019RvMP...91d5002C,2020NatMe..17..261V}. These classification tasks have covered the full range of galactic and extragalactic sources \citep[e.g.][]{2016ApJ...832L..22T,2018Ap&SS.363..197M,2018MNRAS.477.3145J,2020MNRAS.492...96B,2021A&A...645A..87B}. Nowadays, these methods are helping to classify astrophysical objects not only from reduced fluxes, but also from astronomical imaging surveys, where Type-1 AGN are separated from normal galaxies \citep{2021MNRAS.503.4136G}. Therefore, joining \textsc{X-CIGALE} physical parameters with these classification techniques could be useful to solve the AGN ``zoo'' of galaxies for different AGN scenarios. 

However, \textsc{X-CIGALE} estimates more than 60 physical parameters depending on the number of modules included in the SED fitting. Therefore, it is necessary to select a smaller number of physical parameters which are the most informative for a classification task. We use a set of 29\,737 estimates from the sum of the four \textsc{X-CIGALE} AGN setups (SKIRTOR, Fritz, and their respective 30/70 setups), where \citetalias{2010A&A...518A..10V} and \citetalias{2000A&AS..143....9W} share the same classification of Sy1 or Sy2. We split this set randomly into train and test subsets with a proportion of 80\% and 20\%, respectively. Then, we scale the subsets by subtracting the median and transforming according to the interquartile range. We perform this scaling to obtain classifications that are robust against outliers. We discard physical parameters directly related to inputs (e.g. redshift) and those dividing old and young stellar populations (e.g. old and young stellar masses).

We implement two machine learning ensemble techniques for the classification task: random forest and gradient boosting. Both ensemble techniques provide estimates using multiple estimators. The first technique is composed of a collection of trees randomly distributed where each one decides the most popular class \citep{2001MachL..45....5B}. While the second technique takes into account ``additive'' expansions in the gradient descent estimation to improve the selection of the class \citep{friedman2001greedy}. For the random forest we use the \texttt{scikit-learn} \citep{2012arXiv1201.0490P} classifier \texttt{RandomForestClassifier}. For the gradient boosting we use the \texttt{XGBoost} \citep{2016arXiv160302754C} classifier \texttt{XGBClassifier}. We tune the classifier's parameters using an estimator from \texttt{scikit-learn} that uses cross-validation in a grid-search \texttt{GridSearchCV}. The grid is defined with two parameters: the number of trees in the forest \texttt{n\_estimators}, and the maximum depth of the tree \texttt{max\_depth}. Values in the grid for \texttt{n\_estimators} cover the range between 100 and 500 in steps of 100, while for \texttt{max\_depth} the values cover the range between 10 and 40 in steps of 5. We use the F1-score to evaluate the predictions on the test set. The F1-score is the harmonic mean of precision and recall, where precision is the fraction of true positives over true and false positives and recall is the fraction of true positives over true positives and false negatives. With the \texttt{GridSearchCV} estimator, the best values for the \texttt{RandomForestClassifier} are $\texttt{n\_estimators}=200$ and $\texttt{max\_depth}=25$, while for \texttt{XGBClassifier} are $\texttt{n\_estimators}=300$ and $\texttt{max\_depth}=25$.

\begin{figure}
	\includegraphics[width=\columnwidth]{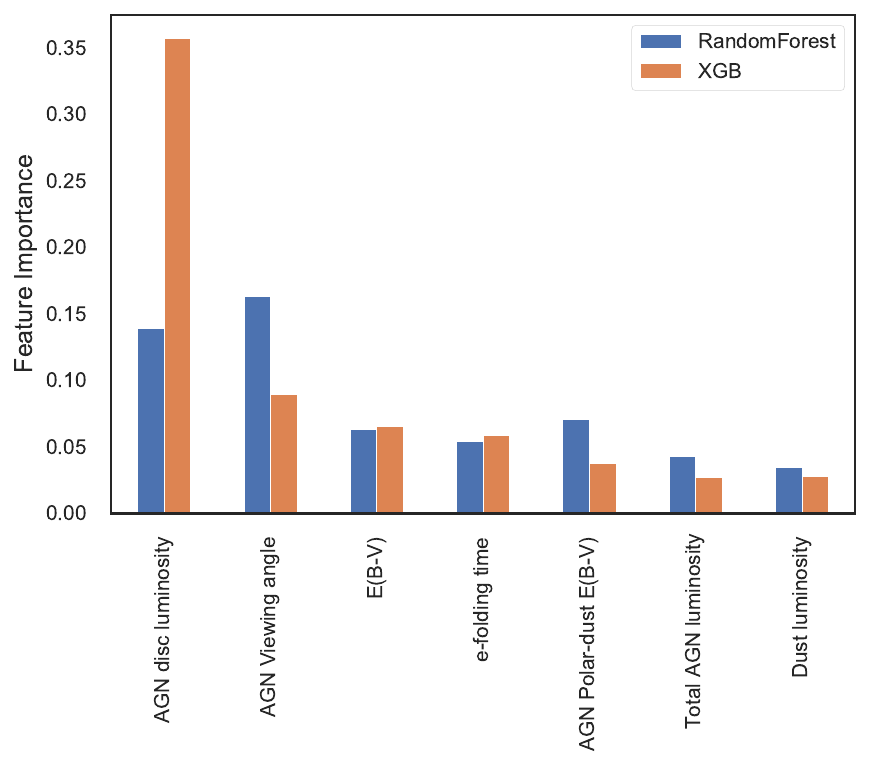}
   \caption{Feature importance scores for the seven most important physical parameters in both \texttt{RandomForestClassifier} and \texttt{XGBClassifier}. Parameters with scores below 0.03 (i.e. $<3\%$) are not presented. The observed AGN disc luminosity is the most important physical parameter to describe Seyfert types.}
   \label{fig:F5}
\end{figure}

We apply a recursive feature elimination and cross-validation selection (RFECV) to select the ideal number of physical parameters to study. We perform 10 k-fold cross-validations with \texttt{RandomForestClassifier} and \texttt{XGBClassifier}. In Figure~\ref{fig:F5}, we present the feature importance (the score of a feature in a predictive model) for both classifiers after the RFECV has been applied. We find that five physical parameters contribute the most in the classification task. These physical parameters are observed AGN disc luminosity, AGN viewing angle, AGN polar dust $E(B-V)$, e-folding time ($\tau_{\textnormal{main}}$), and the colour excess. Though we find that the feature importance is low for SFR and $f_{\rm{AGN}}$, the classification scores improve when we include these parameters, which are important when comparing to observational results. Thus, we decided to focus on these previous seven parameters to assess the impact of the viewing angles in the samples of Seyfert galaxies. In Sect.~\ref{sec:DiscView}, we describe the role that these physical parameters play in the classification task. 

It is necessary to clarify here that the $\rm{L}_{\rm{AGN}}^{\rm{disc}}$ is the observed total luminosity of the accretion disc in \textsc{X-CIGALE}. The AGN accretion power is the term we use to refer to the intrinsic luminosity of the accretion disc. 

\subsection{Comparison of \textsc{X-CIGALE} outputs from different SED fitting setups}\label{sec:Comparisons}

We compare the seven selected physical parameters using the density distribution of the selected setups and Seyfert type samples. We quantify the difference between the distributions using the two-sample Kolmogorov--Smirnov (KS) test. The KS test checks the null hypothesis that two distributions are drawn from the same underlying distribution. We reject the null hypothesis if $D$ (the distance between cumulative distributions) is higher than the critical value at a significance level of $\alpha=0.05$ (e.g., $P < 0.05$). We visualise the distributions using a bandwidth of the density estimator following the Scott's Rule \citep{2015mdet.book.....S}. The visualisation and the KS statistics $D$ tell us how different the samples are.

\subsubsection{AGN setup comparison}

\begin{figure*}
	\includegraphics[width=\textwidth]{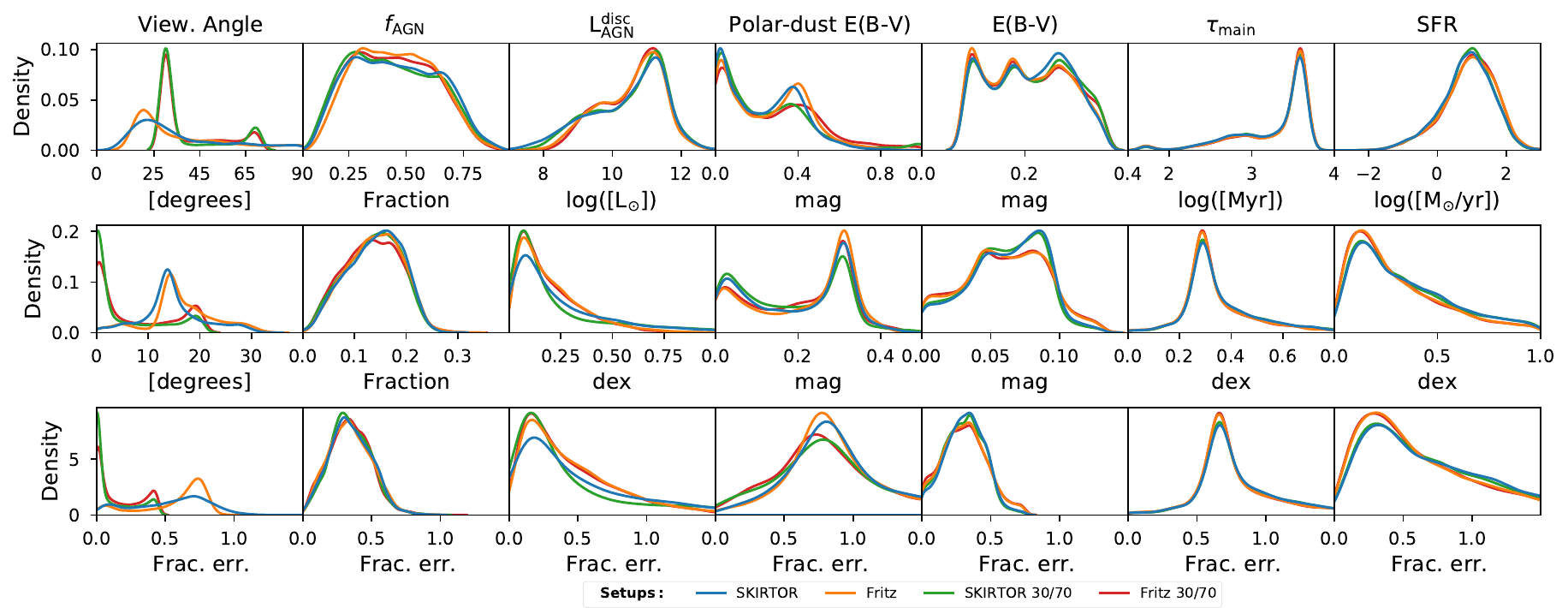}
   \caption{Probability density functions for the estimated parameters (upper panels), their respective errors (middle panels) and their relative fractional errors (lower panels). We compare the SKIRTOR (blue), SKIRTOR 30/70 (green), Fritz (orange), and Fritz 30/70 (red) setups. The difference between the setups that use the full range of viewing angles (SKIRTOR and Fritz) and those that use only two angles (30/70 setups) is significant only in the viewing angle, as expected.}
   \label{fig:F6}
\end{figure*}

We compare the AGN setups (SKIRTOR, Fritz, and their respective 30/70 setups) before comparing the results between different Seyfert types.  We present the probability distribution functions for the parameters and their errors in Fig.~\ref{fig:F6}.

First, we check the differences between the setups covering all viewing angles and setups with viewing angles of only 30\degr and 70\degr.  We do not observe a clear difference between these two sets of setups in most of the physical parameters. Their main discrepancy in the distributions is in the viewing angle, as expected. In setups with 30\degr and 70\degr, we find galaxies with viewing angles around 50\degr. These values are related to the bayesian nature of the estimations, as seen in the estimated error. Something different happens in setups with the full range of viewing angles used for the SED modelling. The distribution of the viewing angle peaks at around 20\degr while at larger angles ($\gtrsim 45$\degr) the distribution is almost flat.

Second, we check the differences between SKIRTOR and Fritz setups. In this case, small differences are noticeable in some physical parameters (e.g. observed AGN disc luminosity and polar dust). However, both kinds of setups follow similar trends in all parameters. If we observe estimated errors, all of them are well constrained, partly due to our $\chi^2_{\rm{red}}$ selection criteria. 

Finally, the null hypothesis in the KS test is almost always rejected in all the setups. The only cases where the null hypothesis cannot be rejected are at Fritz and Fritz 30/70 setups for the $\tau_{\rm{main}}$ and SFR parameters ($D\sim0.02$, $P\sim0.7$), and at SKIRTOR and SKIRTOR 30/70 setups for the SFR parameter ($D\sim0.02$, $P\sim0.1$). We find $D<0.1$ in most parameters when comparing all setups except for the viewing angle due to the way we designed the setups, as expected. Variations in the setups will only give different individual results, but in general, the setups will be similar when interpreting the physical results in Seyfert galaxies.

\subsubsection{Seyfert types 1 and 2 comparison}\label{subsec:Dichotomy}

\begin{figure*}
	\includegraphics[width=\textwidth]{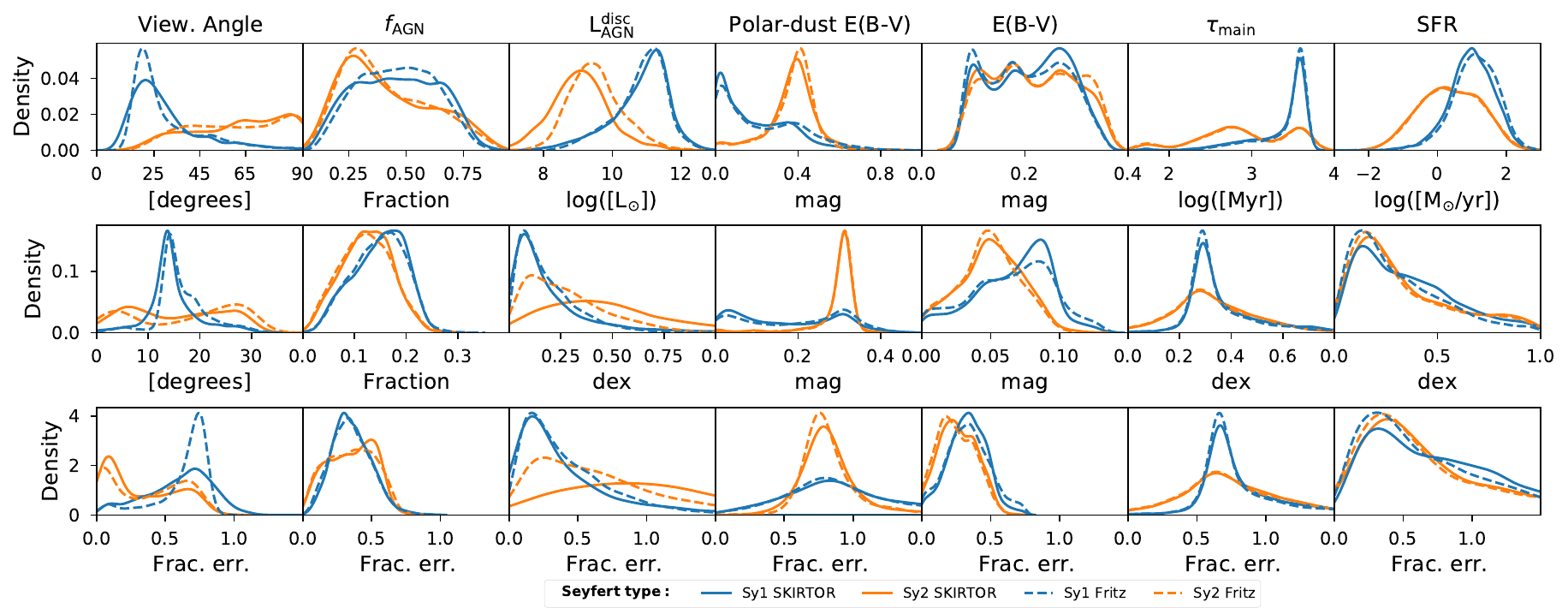}
    \caption{Probability density functions for the estimated parameters (upper panels), their respective errors (middle panels) and their relative fractional errors (lower panels). We compare the Seyfert types 1 (blue) and 2 (orange) for the SKIRTOR (solid) and Fritz (dashed) setups. A clear difference is observed in some of these parameters between Sy1 and Sy2 (e.g. viewing angle, polar dust $E(B-V)$ and observed AGN disc luminosity).}
    \label{fig:F7}
\end{figure*}

According to the AGN unified model, the viewing angle is the main physical parameter to classify Seyfert galaxies into Type-1 AGNs (face-on) and Type-2 AGNs (edge-on). In \textsc{X-CIGALE}, SED models which include AGN follow the AGN-unification scheme \citep{2020MNRAS.491..740Y}. The viewing angle estimated by the models should coincide with the classification scheme of Sy1 and Sy2 galaxies. For this analysis, we use galaxies where Seyfert classifications from \citetalias{2010A&A...518A..10V} and \citetalias{2000A&AS..143....9W} agree. In Fig.~\ref{fig:F7}, we present the probability distribution functions for the seven selected parameters now separating the galaxies in Sy1 and Sy2 for the SKIRTOR and Fritz setups. The distribution of viewing angles for Sy1 coincides with the expected low values of face-on galaxies. For Sy2 galaxies, the viewing angle distribution extends over a wide range of angles, but agrees with a viewing angle close to edge-on galaxies.

KS statistic values $D$ show: i) independent of the used AGN setup, the distributions of the parameter values are similar for the same Seyfert galaxy types, and ii) the different Seyfert types can be well separated as the distributions are different for both setups. The only exception is in terms of the $E(B-V)$, where $D$ values for Sy1 galaxies depend on the used setup. 

In terms of $f_{\rm{AGN}}$, Sy2 galaxies have mainly lower values than Sy1 galaxies. We observe something similar for the parameters of SFR and e-folding time ($\tau_{\rm{main}}$), with low values typically associated to Sy2 galaxies. Although for these two parameters, the difference with Sy1 could be due to the strong UV / optical emission that bias these parameters. Values of the polar-dust show that in Sy2 galaxies the emission is obscured, in contrast to the Sy1 where $E(B-V)=0$ is more common. 

The most interesting result in this comparison is the significant difference in the $\rm{L}_{\rm{AGN}}^{\rm{disc}}$ between the two Seyfert types. Most Sy2 have $\rm{L}_{\rm{AGN}}^{\rm{disc}}$ below $\sim 10^{10} \rm{L}_{\sun}$. The opposite happens for Sy1 galaxies where most $\rm{L}_{\rm{AGN}}^{\rm{disc}}$ are above $\sim 10^{10} \rm{L}_{\sun}$. This result is also verified with the KS statistic $D$, where higher $D$ values are found when comparing Sy1 and Sy2 samples for $\,i\,$ and $\rm{L}_{\rm{AGN}}^{\rm{disc}}$. In a classification task, this latter physical parameter might be more informative than others, as we have seen with the feature importance score (Sect~\ref{sec:FeatureSel} and Fig.~\ref{fig:F5}). We test this idea when predicting the classification type in unclassified and discrepant Seyfert galaxies (Sect.~\ref{sec:Predictions}).

We verify the impact of missing bands in the SED fitting in our sample of galaxies for the estimated parameters. The differences we found between the probability density functions with and without a given band can be explained by the way the Seyfert types are distributed in the sample. For example, in the 2MASS bands, a third of the galaxies detected in these bands are classified to be Seyfert 2, while in the cases where we do not have these bands most of the galaxies ($\sim$90\%) are of Seyfert 1 type. Most of the galaxies not detected by 2MASS follow the trends of Seyfert 1 galaxies, with high AGN disc luminosities, lower viewing angles and polar dust close to zero. This shows that even if some bands are missing, the SED fitting procedure still gives similar values for the physical parameters compared to other galaxies which have the same classification. In other words, the lack of some data does not necessarily impact the results of this work significantly, but it could be important in individual cases, which is out of the scope of this work.

\subsubsection{Intermediate Seyfert types comparison}\label{subsec:InterSey}

From the results presented in Figs.~\ref{fig:F6} and \ref{fig:F7}, we observe a small difference between the AGN setups. Thus, we select the SKIRTOR setup to visualise the difference between intermediate Seyfert types. These intermediate galaxy types are classified in subgroups following \citet{1977ApJ...215..733O,1981ApJ...249..462O} with the quantitative approach of \citet{1992MNRAS.257..677W}. The subgroups are divided using the ratio between H$\beta$ and [\ion{O}{III}] fluxes ($R$) and the spectral profiles of the Balmer lines (see also \citet{2010A&A...518A..10V}). In this classification scheme Seyfert galaxies are: Sy1.0 if $R > 5$, Sy1.2 if $2.0 < R < 5.0$, Sy1.5 if $0.33 <R< 2.0$, Sy1.8 if $R < 0.33$ with a broad component in H$\alpha$ and H$\beta$, and Sy1.9 if the broad component is visible in H$\alpha$ but not in H$\beta$.

\begin{figure*}
	\includegraphics[width=\textwidth]{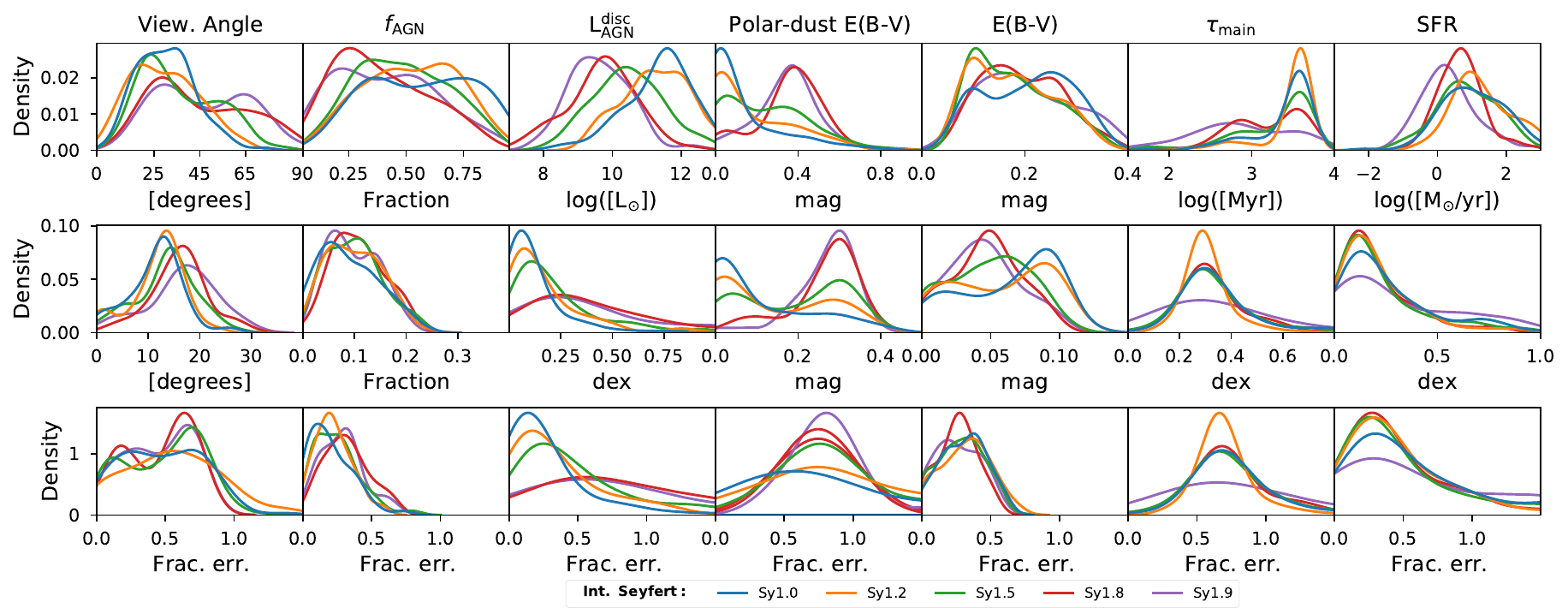}
    \caption{Probability density functions for the estimated parameters (upper panels), their respective errors (middle panels) and their relative fractional errors (lower panels). We compare the intermediate Seyfert types 1.0 (blue), 1.2 (orange), 1.5 (green), 1.8 (red), and 1.9 (purple) for the SKIRTOR setup. A transition between Sy1 and Sy2 galaxies is observed in the viewing angle, polar dust $E(B-V)$ and observed AGN disc luminosity.}
    \label{fig:F8}
\end{figure*}

We show in Fig.~\ref{fig:F8} the probability density functions of the physical parameters for the intermediate Seyfert types. Interestingly, these intermediate types coincide with the picture observed in Sy1 and Sy2 in most of the physical parameters. In some parameters, a possible numerical sequence from Sy1 to Sy2 can be observed (Sy1.0>Sy1.2>Sy1.5>Sy1.8>Sy1.9). Three parameters show this sequence in their KS statistic $D$ values: viewing angle, observed AGN disc luminosity and the polar dust $E(B-V)$. For the viewing angle $\,i\,$, Sy1.0, Sy1.2 and Sy1.5 tend to estimate values around 25\degr, while for Sy1.8 and Sy1.9 there are more $\,i\,$ values above 45\degr. For the $\rm{L}_{\rm{AGN}}^{\rm{disc}}$, the density functions for Sy1.8 and Sy1.9 peak below $10^{10}$ L$_{\sun}$, Sy1.5 peaks at $\sim10^{10}$ L$_{\sun}$, and Sy1.2 and Sy1.0 peak at values above $\sim10^{10}$ L$_{\sun}$. The polar dust in Sy1.0 and Sy1.2 shows in general no extinction, Sy1.5 shows mild values (0.1-0.3), and Sy1.8 and Sy1.9 peak at around 0.4. These results agree with the expected behaviours of the transition between Sy1 and Sy2 type galaxies.

Other physical parameters (e.g. $f_{\rm{AGN}}$ or $\tau_{\rm{main}}$) also show similarities between close intermediate Seyfert types, but not as clear as the parameters described before. This suggests that AGN SED models, as the ones available in \textsc{X-CIGALE}, can estimate the possible transitional phase between different Seyfert types. In Sect.~\ref{sec:DiscCrack}, we discuss a possible explanation for this transitional phase. In addition, in Sect.~\ref{sec:DiscClass}, we discuss the effect of using these AGN classifications when interpreting our results.

\subsection{Redshift behaviour/evolution}\label{subsec:Redshift}

\begin{figure}
	\includegraphics[width=\columnwidth]{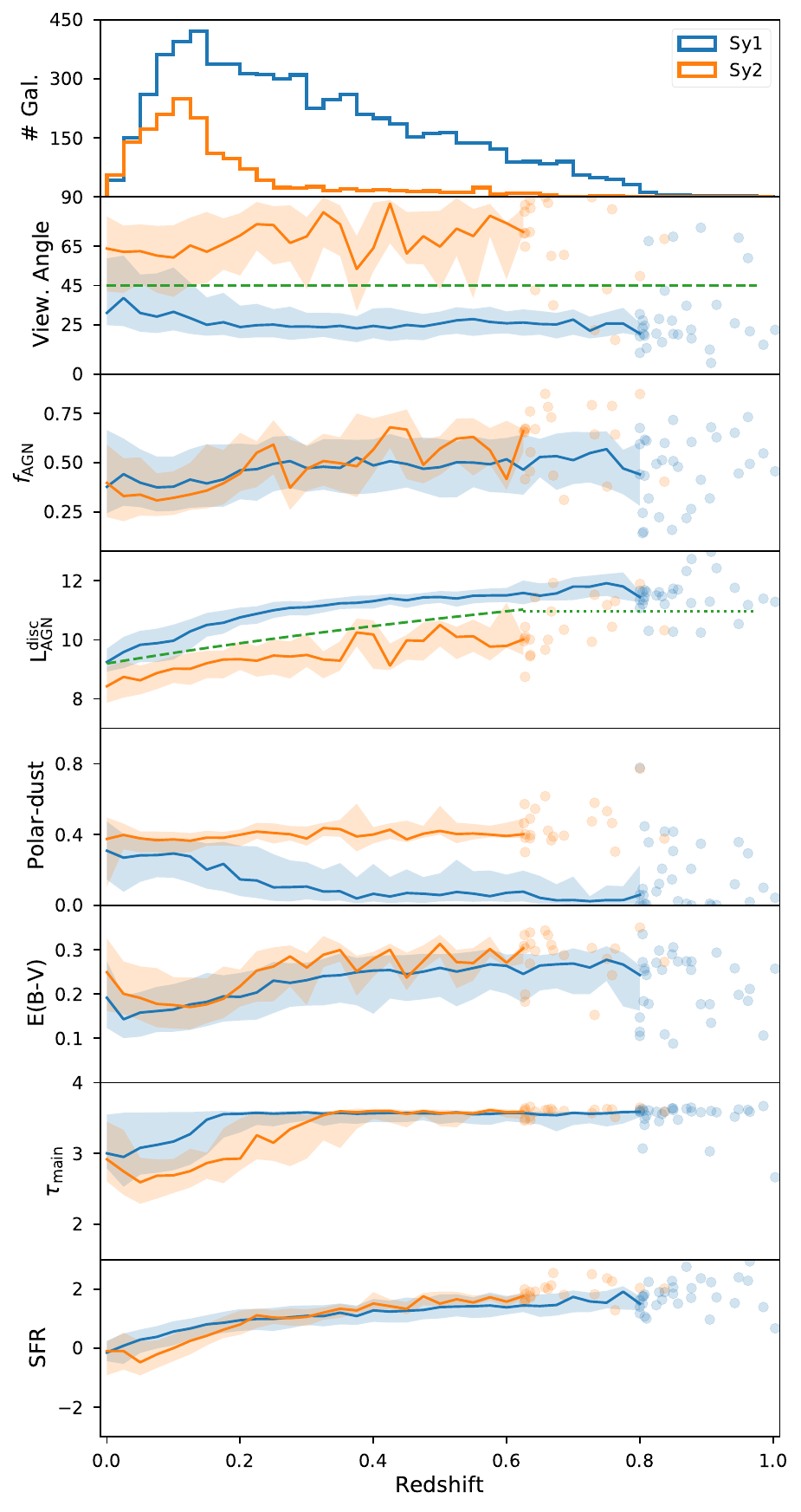}
    \caption{Redshift evolution of the physical parameters for the sample of Seyfert galaxies using the SKIRTOR setup. In the uppermost panel, we present the histogram of the Sy1 and Sy2 galaxies in terms of redshift. At redshift higher than 0.6, we find only a few Sy2 galaxies. At redshift above 0.8, the number of Sy1 galaxies reduces to tens of galaxies. In the rest of the panels, we show the running median of the physical parameters in Sy1 (blue) and 2 (orange), with the range between the 25th and 75th percentiles as shaded regions. We calculate the running median for bins with at least ten galaxies. For bins with fewer than ten galaxies, we plot the galaxies as scatter points. For the viewing angle we include the value of $\,i\,=45$\degr that separates the two Seyfert samples (green dashed line). We also include the separation limit for the two samples using $\rm{L}_{\rm{AGN}}^{\rm{disc}}$, assuming the linear relation described in Eq.~\ref{eq:LAGNsep} (green dashed line) and a constant of $\log\left(\rm{L}_{\rm{AGN}}^{\rm{disc}}\right) = 10.97$ at $z>0.6$ (green pointed line). Galaxies with $z>1$ are not shown.}
    \label{fig:F9}
\end{figure}

Separating Seyfert galaxies in Type-1 or Type-2 is very important to understand the nature of these types of galaxies. In previous section (Sect.~\ref{sec:Comparisons}), we notice that some physical parameters could be used to separate the two Seyfert types. Now, we verify if the separation between Seyfert types holds and/or evolves with redshift. In Fig.~\ref{fig:F9}, we present the evolution of the SED physical parameters as a function of redshift. For this figure, we use estimates from the SKIRTOR setup and separate the Seyfert types using the classifications from \citetalias{2000A&AS..143....9W}. 

In the upper panel of Fig.~\ref{fig:F9}, we notice that the number of classified Sy2 galaxies are almost always below the number of classified Sy1 galaxies. Most of these classifications come from redshifts below $z\sim0.5$, where spectroscopic information of the local Universe is more readily available, compared to high redshift galaxies. In the lower panels, the median of the physical parameters is similar between the two Seyfert types for $f_{\rm{AGN}}$, $E(B-V)$, $\tau_{\rm{main}}$ and SFR. However, the estimated values for the viewing angle, observed AGN disc luminosity and polar dust separate the two Seyfert types. 

The viewing angle does not evolve with redshift, as expected. In general, the viewing angle for Sy1 galaxies is mostly located at $\sim$25\degr while for Sy2 galaxies is at $\sim$65\degr. Thus, we can use the value of 45\degr to separate the two types of Seyfert galaxies. For $\rm{L}_{\rm{AGN}}^{\rm{disc}}$, the difference is always above 0.8 dex (at $z\sim0$) and can go up to $\sim2.3$ dex at $z\sim0.43$. We define a separation limit with the median values of the separation between Seyfert type as linear relation
\begin{eqnarray}
\log{\left(\rm{L}_{\rm{AGN}}^{\rm{disc}}\right)} = \left(9.20 \pm 0.08\right) + \left(8.67 \pm 0.61\right) \times \log{\left(1+z\right)},
\label{eq:LAGNsep}
\end{eqnarray}
where $z$ is redshift. Finally, for the polar dust, the minimum difference of $\sim$0.1 occurs at $z<0.1$ and increases with redshift because most of the Sy2 galaxies above $z=0.2$ have an estimate of 0.4 compared to values close to zero in Sy1, as expected. We do not define a separation for the polar dust due to the similarity of the estimates at $z<0.1$. 

These three parameters (viewing angle, observed AGN disc luminosity and polar dust) are tightly related in \textsc{X-CIGALE} \citep{2020MNRAS.491..740Y}. However, polar dust have larger uncertainties in the estimations, as show in Figs.~\ref{fig:F6}--\ref{fig:F8}. On the other side, $\rm{L}_{\rm{AGN}}^{\rm{disc}}$ could be a more robust parameter in classification tasks than the viewing angle. In Sy1 galaxies the median contribution from the observed disc luminosity to the total AGN luminosity is $\sim52$\%, while for Sy2 the median contribution is around $\sim4$\%. Furthermore, as we see in Figs.~\ref{fig:F7} and \ref{fig:F8}, the separation is more evident between Seyfert types in observed AGN disc luminosity than in the viewing angle. The extent of the viewing angles and their uncertainties for Sy2 galaxies calls into question the use of this parameter to assess type in SED models. 

\subsection{Classifiers in Seyfert galaxies}

We compare different classifiers to test if the physical parameters estimated by the SED modelling are useful to classify AGN galaxies. We define two different scenarios to do the classifications: i) using individual \textsc{X-CIGALE} physical parameters, ii) using ensemble methods with \textsc{X-CIGALE} selected physical parameters. For the first scenario, we use the two limits defined for $\,i\,$ and $\rm{L}_{\rm{AGN}}^{\rm{disc}}$, $i =45$\degr and Eq.~\ref{eq:LAGNsep} respectively, as illustrated in Fig.~\ref{fig:F9}. For the second scenario, we use two machine learning ensemble methods: \texttt{RandomForestClassifier} and \texttt{XGBoostClassifier}, to be consistent with the assumptions in the initial selection of physical parameters (Sect.~\ref{sec:FeatureSel}). In the second scenario, we randomly split the sample of galaxies in train and test sets with a contribution of 80\% and 20\%, respectively. We use three different metrics to compare and test the quality of these binary classifications: Matthews correlation coefficient \citep[MCC,][]{matthews1975comparison}, F1-score and accuracy. MCC is the correlation coefficient between observed and predicted classifications, the F1-score is the harmonic mean of precision and recall (as described in Sect.~\ref{sec:FeatureSel}), while accuracy is the fraction of samples correctly classified. These metrics are usually used in binary classifications, however they can be sensitive to imbalanced data \citep{tharwat2020classification}. Thus, we assess the scenarios by defining a baseline using the \texttt{DummyClassifier} from \texttt{scikit-learn}. This classifier respects the class distribution (i.e. stratified with Seyfert types) and generates random predictions, so it works as a sanity check. Therefore, we use five classifiers with three metrics to test the classification methods.

We use the subset of Seyfert galaxies where the classifications (of Type-1 and Type-2) were the same. We apply the different classifiers described previously for each AGN setup and present their metrics in Table~\ref{tab:metrics}. We notice that all classifiers outperform the baseline. In general, the best AGN setup in all the classifiers is SKIRTOR 30/70. In the scenario where we use individual \textsc{X-CIGALE} physical parameters, $\rm{L}_{\rm{AGN}}^{\rm{disc}}$ is in most cases a better discriminator than the viewing angle $\,i\,$. In the scenario where we use ensemble methods, both classifiers are similar but better than the $\rm{L}_{\rm{AGN}}^{\rm{disc}}$ and $\,i\,$, as expected.  Part of the success of the SKIRTOR 30/70 AGN setup with individual classifiers is the assumption of using two viewing angles that follows the edge-on and face-on geometrical configurations. However, if $\,i\,$ is the most robust parameter in the classification task, we would also expect higher numbers in the metrics for Fritz 30/70, which is not the case. For the classifier using $\rm{L}_{\rm{AGN}}^{\rm{disc}}$, we find that most of the metrics between SKIRTOR and SKIRTOR 30/70 are not far from each other. These results mean that $\rm{L}_{\rm{AGN}}^{\rm{disc}}$ in SKIRTOR setups can be used to assess Seyfert type of a galaxy. Nevertheless, this classifier will not achieve as accurate predictions as the ensemble methods used in this work. In Sect.~\ref{sec:DiscClass}, we discuss the main drawback of using these classifiers in AGNs with SED models. 

\begin{table}
\centering
\caption{Prediction metrics for galaxies with the same classification in \citetalias{2000A&AS..143....9W} and \citetalias{2010A&A...518A..10V}. The baseline is defined from a random classifier as mention in the text. In bold, we highlight the highest metric value for a  given classifier. In general, the SKIRTOR 30/70 setup is the one that best separates Sy1 and Sy2 galaxies, although SKIRTOR metrics are good too.}
\label{tab:metrics}
\begin{tabular}{llccc}
\hline
\hline
Classifier & Setup & \multicolumn{3}{c}{Metrics} \\
\cline{3-5}
&& MCC & F1 & Accu.\\
\hline
Viewing angle ($i$) &SKIRTOR & 0.541 &0.763& 0.827\\
&Fritz &0.492 &0.737 &0.808\\
&SKIRTOR 30/70 &\textbf{0.580} &\textbf{0.781} &\textbf{0.846}\\
&Fritz 30/70 &0.530 &0.746 &0.817\\
\hline
AGN disk luminosity &SKIRTOR & \textbf{0.635} & \textbf{0.808}& \textbf{0.860}\\
($\rm{L}_{\rm{AGN}}^{\rm{disc}}$) &Fritz &0.542 &0.769 &0.842\\
&SKIRTOR 30/70 & 0.621 & 0.801 &\textbf{0.860}\\
&Fritz 30/70 &0.535 &0.764 &0.848\\
\hline
Random Forest &SKIRTOR &0.717 &0.856 &0.907\\
&Fritz &0.732 & 0.865 &0.913\\
&SKIRTOR 30/70 & \textbf{0.737} & \textbf{0.866} &\textbf{0.914}\\
&Fritz 30/70 &0.646 &0.822 &0.900\\
\hline
XGBoost &SKIRTOR & 0.666 & 0.833 & 0.896\\
&Fritz &0.680 &0.840 &0.898\\
&SKIRTOR 30/70 & \textbf{0.716} & \textbf{0.857} & 0.910\\
&Fritz 30/70 &0.707 &0.853 &\textbf{0.912}\\
\hline
Baseline & All & -0.003 & 0.499 &0.691\\
\end{tabular}
\end{table}

\subsection{Predictions on unclassified and discrepant Seyfert galaxies}\label{sec:Predictions}

We test if the estimation from the $\rm{L}_{\rm{AGN}}^{\rm{disc}}$ and the machine learning classifiers are robust and useful to predict the unclassified and discrepant cases in Seyfert galaxies. We compare the Seyfert type predictions of these classifiers with the literature. We gather information about other activity type classifications, outside \citetalias{2010A&A...518A..10V} and \citetalias{2000A&AS..143....9W} where possible, and we present this information in Table~\ref{tab:Liter}. We discuss the classifications for the 59 galaxies (14 unclassified and 45 discrepant) in the following paragraphs. 

First, we focus on unclassified Seyfert galaxies by neither \citetalias{2010A&A...518A..10V} nor \citetalias{2000A&AS..143....9W}, as shown in Table~\ref{tab:Pred}. From the 14 unclassified Seyfert galaxies with the $\rm{L}_{\rm{AGN}}^{\rm{disc}}$ and ML methods: i) one galaxy (6dFGS gJ234635.0-205845) is classified for both methods as Sy2 galaxies, ii) seven are classified for both methods as Sy1 galaxies, and iii) six have mixed classifications in the two methods. From the literature classifications, two galaxies (LEDA 1485346 and MCG+00-11-002) coincides with the Seyfert classification. For MCG+03-45-003, we are not able to compare it with our classifiers because classifications change depending on the AGN setup used. In addition, we found a mixed blazar (2MASX J12140343-1921428), five quasar (CADIS 16-505716, LEDA 3095610, LEDA 3096762, QSO B1238+6232 and [HB93] 0248+011A), and a composite galaxy (2MASX J23032790+14434). We assume these seven galaxies cannot be Seyfert galaxies. However, we can compare the prediction's similarity with Type-1 AGN. For the mixed blazar and quasars (QSO), the $\rm{L}_{\rm{AGN}}^{\rm{disc}}$ and ML methods classify them as Sy1 types (except for QSO B1238+6232 with $\rm{L}_{\rm{AGN}}^{\rm{disc}}$), which are close to Type-1 AGNs in the unified model scheme. Thus, 8/14 of the galaxies in the unclassified subset have similar classifications to the AGN type. Therefore, it may be possible to distinguish the AGN type from the classifiers presented in this work. 

\begin{table}
\centering
\caption{Classification types in unclassified Seyfert galaxies. The second and third columns show the Seyfert type obtained using the $\rm{L}_{\rm{AGN}}^{\rm{disc}}$ and machine learning ensemble methods, respectively. The fourth column shows the assumed classification from the literature. Finally, the last column shows if the previous classifiers are similar with the AGN type in the literature. In case $\rm{L}_{\rm{AGN}}^{\rm{disc}}$ and ML classifications are different split the last column in two options.}
\label{tab:Pred}
\begin{tabular}{lcccc}
\hline
\hline
Object ID & \multicolumn{4}{c}{Classification type} \\
\cline{2-5}
 & $\rm{L}_{\rm{AGN}}^{\rm{disc}}$ & ML & Literature & Sim. \\
\hline
2MASX J12140343-1921428 & Sy1 & Sy1 & Mix. blazar & Y \\
2MASX J18121404+2153047 & Sy2 & Sy1 & --  & --\\
2MASX J21560047-2144325 & Sy2 & Sy1 & -- & -- \\
2MASX J23032790+1443491 & Sy1 & Sy1 & Composite & N \\        
6dFGS gJ234635.0-205845 & Sy2 & Sy2 & -- & --\\
CADIS 16-505716 & Sy1 & Sy1 & QSO & Y\\
ESO 373-13 & Sy2 & Sy1 & -- & --\\
LEDA 1485346 & Sy1 & Sy1 & Sy1 & Y\\
LEDA 3095610 & Sy1 & Sy1 & QSO & Y\\
LEDA 3096762 & Sy1 & Sy1 &  QSO & Y\\
MCG+00-11-002 & --$^{\rm a}$ & Sy1 &  Sy1 & --/Y\\
MCG+03-45-003 & --$^{\rm a}$ & --$^{\rm a}$ & Sy2 & --\\
QSO B1238+6232 & Sy2 & Sy1 & QSO & N/Y\\
{[HB93] 0248+011A} & Sy1 & Sy1 & QSO & Y\\
\hline
\end{tabular}
\begin{flushleft}
$^{\rm a}$ The classification changes depending on the AGN setup used.\\
\end{flushleft}
\end{table}

Second, we focus in the 45 discrepant galaxies that are unclassified in \citetalias{2010A&A...518A..10V} but classified in \citetalias{2000A&AS..143....9W}, and vice versa, shown in Table~\ref{tab:Pred2}. From this sample, we find that 36 of them (80\%) are mainly classified as Sy1 while only two are Sy2 with the $\rm{L}_{\rm{AGN}}^{\rm{disc}}$ and machine learning ensemble methods. For the other seven galaxies the classifications are still discrepant. We compare the similarity of the predictions of the $\rm{L}_{\rm{AGN}}^{\rm{disc}}$ and machine learning ensemble methods with the literature. We assume that: i) mixed blazar, BZQ, QSO, NLSy1, Sy1.5 and Sy1.2 galaxies are Type-1 AGN; ii) Sy1.8 are Type-2 AGN; and iii) LINERs cannot be classified as Type-1 nor Type-2 AGN. From galaxies with literature classifications (42 galaxies), we find that more than half of these galaxies ($\sim59\%$) are classified as QSO, and only 11 galaxies ($\sim26\%$) are classified as a Sy1. From the seven galaxies with still discrepant classifications five have other literature classifications. From those five, two galaxies have AGN types similar to those in the literature when $\rm{L}_{\rm{AGN}}^{\rm{disc}}$ is used as a classifier, and only one galaxy when we use machine learning ensemble methods. For the remaining galaxies with literature classifications (37 galaxies), we find that $\sim95\%$ of discrepant galaxies have AGN types similar to those in the literature when both classifiers coincide in the classification.

\begin{table}
\centering
\caption{Classification types in discrepant Seyfert galaxies. Columns are similar to Table~\ref{tab:Pred}.}
\label{tab:Pred2}
\begin{tabular}{lcccc}
\hline
\hline
Object ID & \multicolumn{4}{c}{Classification type} \\
\cline{2-5}
 & $\rm{L}_{\rm{AGN}}^{\rm{disc}}$ & ML & Literature & Sim. \\
\hline
2E 2294 & Sy1 & Sy1 & QSO & Y \\
2E 2628 & Sy1 & Sy1 & QSO & Y \\
2E 3786 & Sy1 & Sy1 & QSO & Y \\
2MASS J00423990+3017514 & Sy1 & Sy1 & Sy1 & Y \\        
2MASS J01341936+0146479 & Sy1 & Sy1 & QSO & Y \\
2MASS J02500703+0025251 & Sy1 & Sy1 & QSO & Y \\
2MASS J08171856+5201477 & Sy2 & Sy1 & Sy1 & N/Y \\
2MASS J09393182+5449092 & Sy1 & Sy1 & QSO & Y \\
2MASS J09455439+4238399 & Sy1 & Sy1 & NLSy1 & Y \\
2MASS J09470326+4640425 & Sy1 & Sy1 & QSO & Y \\
2MASS J09594856+5942505 & Sy1 & Sy1 & QSO & Y \\
2MASS J10102753+4132389 & Sy1 & Sy1 & QSO & Y \\
2MASS J10470514+5444060 & Sy1 & Sy1 & QSO & Y \\
2MASS J12002696+3317286 & Sy1 & Sy1 & QSO & Y \\
2MASS J15142051+4244453 & Sy1 & Sy1 & QSO & Y \\
2MASSI J0930176+470720 & Sy1 & Sy1 & QSO & Y \\
2MASX J02522087+0043307 & Sy1 & Sy1 & QSO & Y \\
2MASX J02593816+0042167 & Sy1 & --$^{\rm a}$ & QSO & Y/-- \\
2MASX J06374318-7538458 & Sy1 & Sy1 & Sy1 & Y \\
2MASX J09420770+0228053 & Sy1 & Sy1 & LINER$^{\rm b}$ & N \\
2MASX J09443702-2633554 & Sy1 & Sy1 & Sy1.5$^{\rm b}$ & Y \\
2MASX J09483841+4030436 & Sy2 & Sy2 & Sy1 & N \\
2MASX J10155660-2002268 & Sy1 & --$^{\rm a}$ & Sy1 & Y/-- \\
2MASX J10194946+3322041 &  Sy1 & Sy1 & NLSy1 & Y \\
2MASX J15085291+6814074 &  Sy1 & Sy1 & Sy1 & Y \\
2MASX J16383091-2055246 & --$^{\rm a}$ & Sy2 & NLSy1 & --/N \\
2MASX J21033788-0455396 & Sy1 & Sy1 & Sy1 & Y\\
2MASX J21512498-0757558 & Sy2 & Sy2 & -- & --\\
2MASX J22024516-1304538 & Sy1 & Sy1 & Sy1 & Y \\
2dFGRS TGN357Z241 & Sy1 & Sy1 & QSO & Y \\
3C 286 & Sy1 & Sy1 & Mix. blazar & Y \\
6dFGS gJ034205.4-370322 & Sy1 & Sy1 & Mix. blazar & Y \\
6dFGS gJ043944.9-454043 & Sy1 & Sy1 & QSO & Y \\
6dFGS gJ084628.7-121409 & Sy1 & Sy1 & Sy1 & Y \\
CTS 11 & Sy1 & Sy1 & NLSy1 & Y \\
HE 0226-4110 & Sy1 & Sy1 & QSO & Y \\
ICRF J025937.6+423549 & Sy1 & Sy1 & Mix. blazar & Y \\
ICRF J081100.6+571412 & Sy1 & Sy1 & QSO & Y \\
ICRF J100646.4-215920 & Sy1 & Sy1 & Mix. blazar & Y \\
ICRF J110153.4+624150 & Sy1 & Sy1 & BZQ & Y \\
ICRF J135704.4+191907 & Sy1 & Sy1 & BZQ & Y \\
IRAS 10295-1831 & Sy1 & Sy1 & Sy1 & Y \\
Mrk 1361 & --$^{\rm a}$ & Sy2 & -- & -- \\
PB 162 & --$^{\rm a}$ & Sy1 & -- & -- \\
UGC 10683 & Sy2 & --$^{\rm a}$ & Sy1 & N/-- \\
\hline
\end{tabular}
\begin{flushleft}
$^{\rm a}$ The classification changes depending on the AGN setup used.\\
$^{\rm b}$ Decision based on \citetalias{2010A&A...518A..10V}.\\
\end{flushleft}
\end{table}

However, most of the results come from Type-1 AGNs and less than half of these galaxies end up being Seyfert galaxies. The unclassified  and discrepant Seyfert type classifications, presented in Tables~\ref{tab:Pred} and ~\ref{tab:Pred2}, only show part of the predictability potential of these methods but support the usefulness of the classifiers in this work. These results, together with Table~\ref{tab:metrics}, show that we can classify AGN types correctly using the physical parameters estimated from broad-band SED fitting using \textsc{X-CIGALE}. In a future study, we expect to use a more complete set of intermediate numerical values for Seyfert galaxies to test their classification. 

\section{Discussions}\label{sec:disc}

In this section, we further examine the results of this work. First, we discuss the importance of the viewing angle in the AGN SED models (Sect.~\ref{sec:DiscView}) and how they affect current AGN studies (Sect.~\ref{sec:DiscCrack}). Then, we focus on the AGN types in terms of physical parameters as $f_{\rm{AGN}}$ and SFR (Sect.~\ref{sec:DiscFAGNSFR}), and 
their classifications (Sect.~\ref{sec:DiscClass}). Finally, we examine the effect of not including X-ray data in the SED AGN models (Sect.~\ref{sec:DiscXray}).

\subsection{The role of the viewing angle}\label{sec:DiscView}

In this work, we restricted our results to the seven most important, according to the machine learning techniques, physical parameters when classifying Sy1 and Sy2. Following the AGN unification model, the main difference between these two types of galaxies resides in the viewing angle. Therefore, by comparing these seven parameters, we could understand the role that the viewing angle plays in determining the AGN type. 

In Fig.~\ref{fig:F6}, we show how different the AGN setups are when assuming only two viewing angles instead of a full range. With the exception of the viewing angle, all the parameters show a similar behaviour between the different setups. Only small differences in the E(B-V) and observed AGN disc luminosity are observed when comparing SKIRTOR and Fritz setups. Interestingly, in setups with ten viewing angles the estimates give a frequent value of $\sim$25\degr. Therefore, it seems that the full range of viewing angles in the setup does not significantly affect the other estimated SED parameters.

By construction, the viewing angle in \textsc{X-CIGALE} can determine the AGN type when the angle is close to the face-on (0-30\degr) and edge-on (70-90\degr) scenarios \citep{2020MNRAS.491..740Y}. Using the Chandra COSMOS Legacy survey \citep{ 2016ApJ...817...34M} and \textsc{X-CIGALE}, \citet{2020MNRAS.491..740Y} estimated an accuracy of $\sim$71\% in spectroscopic Type-1 and Type-2 AGNs. In our case, the accuracy of the classifications is around 82-85\% when using only the viewing angle (Table~\ref{tab:metrics}), although with a different sample size (590 AGNs in \citet{2020MNRAS.491..740Y} while $\sim8\,000$ in this work). The highest value in the accuracy (and other metrics) is obtained when using a setup with only two angles. If we use a setup with the full range of viewing angles, the distributions for the viewing angle (Fig.~\ref{fig:F7}) are similar to what \citet{2021A&A...650A..75G} found, where Sy1 are located at values around 20-30\degr while Sy2 galaxies are more scattered in a wider range of viewing angles. However, the estimations in terms of redshift (Sect.~\ref{subsec:Redshift} and Fig.~\ref{fig:F9}) seems to favour values around 20-30\degr and 60-70\degr when looking at Sy1 and Sy2, respectively. Thus, AGNs classifications can be improved by forcing the viewing angle to two values that follow the Type-1 and Type-2 classifications. This result justifies the selection of two or even three viewing angles in similar studies using \textsc{CIGALE} or \textsc{X-CIGALE} in AGN galaxies \citep[e.g.][]{2017A&A...597A..51V,2019ApJ...878...11Z,2020MNRAS.495.1853P,2020MNRAS.499.4068W,2021A&A...646A..29M}.

This dichotomy in the viewing angle of AGN SED models could indicate that using the (IR) SED is not an adequate tool to estimate the viewing angle of AGNs as a continuous distribution, compared to spectroscopic measurements of the NLR \citep{2013ApJS..209....1F,2016MNRAS.460.3679M}. In Fig.~\ref{fig:F5}, we show that the viewing angle is not the most important physical parameter when classifying Sy1 and Sy2. The observed AGN disc luminosity has the highest importance score to classify Seyferts, although its UV-optical emission is (in theory) angle-dependent \citep{2015ARA&A..53..365N,2020MNRAS.491..740Y}. The power of the observed AGN disc luminosity in separating Sy1 and Sy2 galaxies is stunning when comparing with the viewing angle and other parameters in this work (Figs.~\ref{fig:F7}-\ref{fig:F9} and Table~\ref{tab:metrics}). Using supernovae host galaxies catalogues, \citet{2017ApJ...837..110V} shows that the AGN luminosity, together with the stellar age, could play an important role in the AGN unification model beyond the viewing angle estimated from the torus when counting the number of supernovae in Type-2 AGNs. This result is similar to what we find in this work, although the age in the SEDs is not well constrained (Sect.~\ref{sec:verif}). 

\begin{figure}
	\includegraphics[width=\columnwidth]{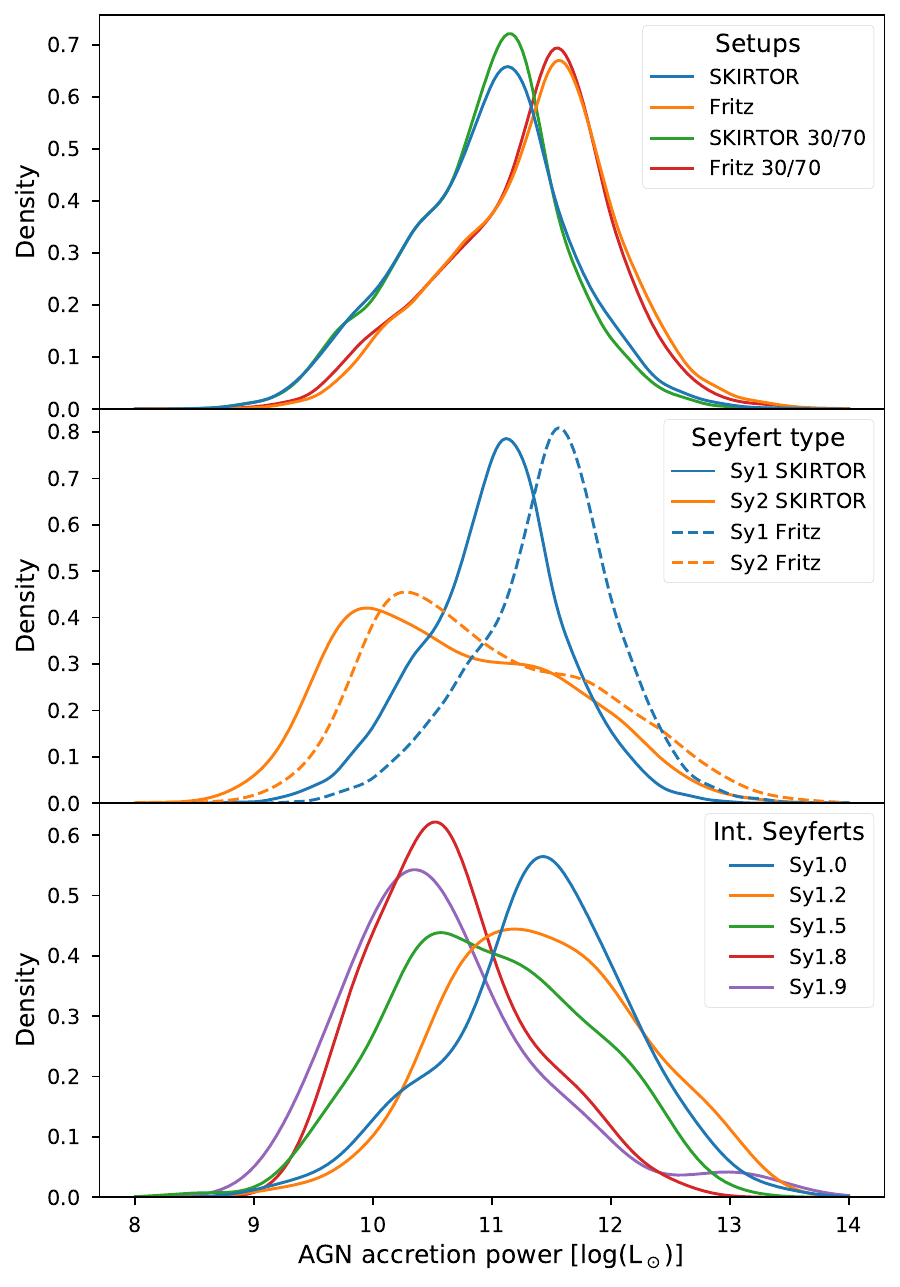}
    \caption{Probability density functions for the AGN accretion power of the four AGN setups (upper panel), Seyfert Type-1 and Type-2 for SKIRTOR and Fritz setups (middel panel), and intermediate Seyfert types for the SKIRTOR setup (lower panel). The AGN accretion power (intrinsic disc luminosity) depends on the selected AGN model, but the path from Type-2 to Type-1 AGNs by increasing the AGN accretion power.}
    \label{fig:F10}
\end{figure}

In Fig.~\ref{fig:F10}, we verify the estimations of the AGN accretion power for the different Seyfert types and AGN models. We notice that Fritz setups estimate higher accretion powers than SKIRTOR setups. This result is due to a different anisotropy correction applied in \textsc{X-CIGALE}, where the AGN accretion power does not depend on the viewing angle for the Fritz models. As a result of this correction, the difference between Seyfert types (middle panel) have different accretion powers depending on the AGN model. Therefore, $\rm{L}_{\rm{AGN}}^{\rm{disc}}$ cannot be easily transformed into bolometric luminosities due to its complex dependencies on the models. However, when we use only one of the AGN models in the intermediate Seyfert types, we notice similar distributions as with the $\rm{L}_{\rm{AGN}}^{\rm{disc}}$. Therefore, the more detailed spectroscopic classifications follow a path from Type-2 to Type-1 AGNs by increasing the observed and intrinsic AGN disc luminosity (Sect.~\ref{subsec:InterSey} and Figs.~\ref{fig:F8} and ~\ref{fig:F10}). This means that the AGN disc luminosity is an important factor, if not the main, in deciphering AGN types with SED models, and that a probable evolutionary path from intermediate types can be obtained with the AGN disc luminosity \citep{2014MNRAS.438.3340E}.

This result could have an impact in cosmological estimations done in AGNs. For example, QSOs are used to measure the luminosity distance at high redshifts to determine cosmological parameters \citep[e.g.][]{2019A&A...628L...4L}. If there is a dependency in the luminosity distance of QSOs with the viewing angle, it would lead to incorrect cosmological estimates \citep{2021ApJ...909...58P}. Even if the dependency is not entirely on the viewing angle but on the brightness of the AGN, then these estimates should be reformulated.

\subsection{The AGN dust winds}\label{sec:DiscCrack}

As we mentioned before, \textsc{X-CIGALE} follows the simple AGN-unification scheme, i.e. the viewing angle determines the AGN type \citep{2020MNRAS.491..740Y}. If this simplified scheme is correct, the viewing angle should be the best classifier in our sample, but it is not.  Our results for the smooth and clumpy torus models (Fritz and SKIRTOR, respectively) show that the observed AGN disc luminosity is a better classifier than the viewing angle. Perhaps the AGN ``zoo'' of galaxies is too complex to be explained with only the angle-dependent obscuration coming from a toroidal structure \citep{2017A&ARv..25....2P}. 

A possible solution is the AGN disc-wind scenario \citep{1992ApJ...385..460E,2006ApJ...648L.101E,2015ARA&A..53..365N}. For example, \citet{2017ApJ...838L..20H} propose a model where dusty winds can explain the observational results on spatially resolved AGNs \citep[e.g.][]{2021A&A...652A..98G,2021A&A...652A..99A}. Later on, this model was extended by \citet{2019ApJ...884..171H}, following interferometry IR and sub-mm observations \citep[e.g.][]{2016A&A...591A..47L} where the AGN structure is composed of disc, wind and wind launching regions instead of a simple toroidal obscuration structure. In this scenario, the multiphase structure is consistent with the relation between the AGN obscured fraction (covering fraction) and AGN luminosity (Eddington ratio) in X-rays \citep{2014MNRAS.437.3550M,2017Natur.549..488R}. The winds provide additional obscuration traced by covering factors and could separate the obscured and unobscured regions, and thus separate Type-1 and Type-2 AGNs. Similarly, \citet{2021ApJ...906...84O} describe a disc wind model analogous to \citet{2019ApJ...884..171H} using X-CLUMPY \citep{2008ApJ...685..160N,2019ApJ...877...95T}, another SED toroidal model. In that work, X-CLUMPY is used in two different models for obscured and unobscured AGNs. With this approach, \citet{2021ApJ...906...84O} also finds a negative correlation between the Eddington ratio and the torus covering factor. Therefore, it is possible that the estimates of the AGN disc luminosity can be affected by other structures, like the ones proposed by \citet{2019ApJ...884..171H} or \citet{2021ApJ...906...84O}. These structures may explain why the estimated observed AGN disc luminosity works better as a classifier in Sy1 and Sy2 with \textsc{X-CIGALE}.

An advantage of dusty wind structures is that they can also explain the origin of the red QSOs. These red QSOs are expected to be red due to dust in the line of sight \citep{1995Natur.375..469W}, although it has recently been proposed that it is related to an evolutionary phase of QSO \citep{2019MNRAS.488.3109K}.  
\citet{2021A&A...649A.102C} studied the nature of red QSO using \textsc{AGNfitter} \citep{2016ApJ...833...98C}, which resides in SED templates of smooth toroidal models \citep{2004MNRAS.355..973S}, and found no connection between the AGN torus and the QSO reddening. However, they also found high-velocity winds in these red QSOs, suggesting that dusty wind structures may explain their nature.

Interestingly, even the intermediate Seyfert types can be explained with the disc-wind scenario. \citet{2014MNRAS.438.3340E} show that an evolutionary path of intermediate Seyfert types is related to the wind streamlines. When the accretion rate decreases the AGN luminosity decreases because the clouds in the wind streamlines move from high to low altitudes, generating an evolution sequence from 1.0 to 1.9 AGN types. This path is similar to the one presented in Sect.~\ref{subsec:InterSey} and Figs.~\ref{fig:F8} and \ref{fig:F10}, where AGN disc luminosity decreases with Seyfert type. The estimations presented in this work support this idea, as the mean AGN accretion power is lower in Sy2 galaxies by $\sim0.5$ dex than in Sy1 galaxies. However, we need to keep in mind that: i) the estimates of the physical parameters coming from AGN SED models depend on the chosen model \citep[as shown in Fig.~\ref{fig:F10}\, and by][]{2019ApJ...884...10G,2019ApJ...884...11G}, and ii) the classification criteria for the intermediate types are different in \citet{2014MNRAS.438.3340E} and this work. Therefore, we need more information in terms of intermediate type classifications and in-depth observations of AGN regions with multi-wavelength observations to create a robust connection between theoretical models and observations. 

\subsection{AGN fraction and SFR in AGN types}\label{sec:DiscFAGNSFR}

To describe the difference between Type-1 and Type-2 AGN other physical parameters could be used besides AGN luminosity and viewing angle. In Sect.~\ref{subsec:Dichotomy} and Fig.~\ref{fig:F7}, we mentioned that parameters like $f_{\rm{AGN}}$ and SFR can also differentiate Seyfert types. However, we argue that the differences between AGN types will be smaller in these physical parameters. 

In terms of the AGN contribution to the total IR luminosity ($f_{\rm{AGN}}$), \citet{2016MNRAS.458.4297G} found that Sy2 galaxies tend to have a lower $f_{\rm{AGN}}$ compared with Sy1, in a sample of local galaxies. Similar results are found by \citet{2020MNRAS.499.4325R}, where six Type-1 AGN galaxies have higher $f_{\rm{AGN}}$ than AGN types with an estimated viewing angle of 60\degr, generally Type-2 AGNs. They also found that the $f_{\rm{AGN}}$ could increase with IR luminosity, as  \citet{2012ApJ...744....2A} suggested. Therefore, Sy1 could have a higher AGN fraction than Sy2 because of their higher IR luminosity \citep[e.g.][and references therein]{2019ApJ...872..168S}. We find a similar behaviour in Fig.~\ref{fig:F7}, although the values are widely distributed between both types. In addition, in Fig.~\ref{fig:F9}, we find a slight increase of $f_{\rm{AGN}}$ with redshift as found by other works \citep[e.g][]{2020MNRAS.499.4068W}, especially in Sy2 galaxies, but the statistic is small. The discrepancy in the results may also lie in the selection effects when comparing AGN types of galaxies, due to sample selection \citep[e.g][]{2016ApJ...833...98C}, AGN definitions \citep[e.g][]{2020MNRAS.499.4068W}, or incorrect estimates of $f_{\rm{AGN}}$ because of cold-dust emission \citep[][]{2021arXiv210312747M}. Therefore, it is not yet clear how $f_{\rm{AGN}}$, or its relationship to total IR luminosity, will change at higher redshifts ($z>1$) between AGN types. 

For the SFR, the relative importance is small when separating the two Seyfert types, following the feature selection (Sect.~\ref{sec:FeatureSel}). This result agrees with \citet{2019ApJ...872..168S} and \citet{2021A&A...646A.167M}, who found no differences between AGN types in terms of SFR and stellar mass associated with AGN power in X-rays, although large uncertainties are present for Type-1 AGNs. On the other hand, \citet{2019ApJ...878...11Z} show differences in stellar mass but not in SFR between AGN types, even using different SFR calculation methods (SED, H$_\alpha$ and IR luminosities). In this work, stellar mass is estimated with \textsc{X-CIGALE} but its relative importance is even lower than the importance of the SFR. In addition, we do not find a clear separation between the two Seyfert types in terms of SFR with redshift (Fig.~\ref{fig:F9}). The SFR will increase for both types as the SFR increases with the redshift until the cosmic noon, therefore we are only seeing a selection effect.

Thus, looking at $f_{\rm{AGN}}$ and SFR individually may give some clues about the difference between AGN types. However, the differences are small when compared together with other physical parameters, such as the observed AGN disc luminosity. Furthermore, one of the main problems of looking at different physical parameters resides in the definition between obscured (Type-2) and unobscured (Type-1) AGNs \citep{2018ARA&A..56..625H}, as we will see in the following subsection.

\subsection{AGN Classifications}\label{sec:DiscClass}

In this work, we treated the Seyfert types coming from \citetalias{2000A&AS..143....9W} and \citetalias{2010A&A...518A..10V} as true classifications. However, that does not mean that the Seyfert galaxy sample studied here is not populated with other types of AGN galaxies. For example, galaxies that we assume as unclassified Seyfert end up being mostly QSO in the literature (Sect.~\ref{sec:Predictions}). In any case, we have presented evidence from different AGN setups that seem to agree with the general classification of Type-1 and Type-2 AGN. Then, we could use simple relations, like the one presented in Eq.~\ref{eq:LAGNsep}, to separate AGN galaxies using the features estimated from SED modelling of the AGN component available in \textsc{X-CIGALE}. Machine learning ensemble techniques could also be used to classify galaxies with \textsc{X-CIGALE} outputs, achieving even higher accuracy than just one physical parameter (Table~\ref{tab:metrics}). Nevertheless, these methods will require correct classifications of galaxies that sometimes are not available. 

As we discussed in Sect.~\ref{sec:DiscCrack}, the intermediate Seyfert classifications could help solve the problems of the AGN unification model. If these galaxies are an evolutionary stage of AGNs, as proposed by \citet{2014MNRAS.438.3340E} with disc wind models, then classifying AGN galaxies could be crucial to join the theoretical models with observations. In this work, all the intermediate classifications come from \citetalias{2010A&A...518A..10V} with the quantitative approach of \citet{1992MNRAS.257..677W}, as described in Sect.~\ref{subsec:InterSey}. However, other criteria could be applied in the classification of intermediate AGN galaxies \citep[e.g.][]{2012MNRAS.426.2703S}. More spectral information and consistent classifications will be needed in intermediate AGN galaxies to clarify this evolutionary stage, for example using the on-going DEVILS survey. Future observations with the \textit{James Webb Space Telescope}
(JWST) will detect obscured AGNs and calculate AGN fractional contributions through photometry and spectral line features \citep[e.g.][]{2021ApJ...906...35S,2021ApJ...908..144Y}, which will elucidate the path for classifying these type of galaxies.  

\subsection{X-ray information}\label{sec:DiscXray}

The efforts to create catalogues of AGN galaxies with X-ray data are crucial to classify galaxies missed in optical catalogues \citep[e.g.][]{2017ApJ...850...74K}. These missed galaxies are in general obscured AGN galaxies (Type-2), which are difficult to classify, even using ensemble methods in optical wavelengths \citep{2021MNRAS.503.4136G}. However, it is good to keep in mind that Type-2 AGNs could remain undetected in IR and X-ray colour surveys \citep{2020MNRAS.495.1853P}. Therefore, a multi-wavelength study, like the one presented in this work, is ideal to tackle the completeness and classification problem in AGN galaxies. 

Unfortunately, a limitation of this study is the lack of X-ray data when fitting the SED, although there is a correlation between the AGN MIR luminosity and the 2–10 keV  X-ray luminosity in Type-1 and Type-2 AGNs \citep{2009A&A...502..457G,2019ApJ...872..168S}. We conclude that the quantity of photometric data in NED and CDS is not enough nor homogeneous to include the X-ray module in the \textsc{X-CIGALE} setups. None the less, X-rays can help to constrain non-physical parameters in SED tools as \textsc{X-CIGALE} \citep{2020MNRAS.491..740Y}. Recent studies with different SEDs models that include X-ray data show how important the inclusion of these data could be to constrain the AGN models \citep[e.g.][]{2019ApJ...872..168S,2021ApJ...906...84O,2021A&A...646A..29M,2021A&A...646A.167M}. In the future, we expect to understand the probable evolutionary paths discussed in this work for different AGN types, with the AGN  luminosity, $f_{\rm{AGN}}$, and X-ray data. In the next decade, the combination of the \textit{Athena space telescope} with \textsc{X-CIGALE} will help to unambiguously determine the presence of AGN for this type of works \citep{2020MNRAS.491..740Y}.

\section{Conclusions}\label{sec:conclu}

We have used a sample of 13\,173 Seyfert galaxies from \citetalias{2000A&AS..143....9W} and  \citetalias{2010A&A...518A..10V} to assess the importance of the viewing angle in AGN SED models. We have used a data-driven approach by retrieving photometric data from astronomical databases (CDS and NED) to be used in the SED analysis with \textsc{X-CIGALE}. Two AGN SED toroidal models (Fritz and SKIRTOR setups) were used with different viewing angle configurations to verify the effect of viewing angle selection in the estimated physical parameters. These estimates have been validated by comparing our results with those from \citet{2017A&A...597A..51V}, showing good agreement except for the AGN fractions, which can be related to the different assumptions of the grids.

Our main conclusions are the following:
\begin{itemize}
    \item[1.] The estimated viewing angle from \textsc{X-CIGALE} seems to be the second best discriminator when assessing AGN type. This result is supported by different prediction metrics and importance scores in machine learning algorithms which favour the observed AGN disc luminosity as the most important physical parameter. 
    \item[2.] The initial viewing angle assumption in \textsc{X-CIGALE} does not significantly affect the other estimated physical parameters if at least two viewing angles that follow the AGN Type-1 and Type-2 classifications are taken into account. 
    \item[3.] At different redshifts ($z \lesssim 0.5$), the smooth and clumpy torus AGN models seem to favour viewing angles around 20-30\degr and 60-70\degr when looking at Sy1 (Type-1 AGN) and Sy2 (Type-2 AGN), respectively. While in terms of the observed AGN disc luminosity, we propose to use a limit (Eq.~\ref{eq:LAGNsep}) that separates both types. These values may predict the AGN type in unclassified AGN galaxies, as shown in the case of unclassified and discrepant Seyfert galaxies.
    \item[4.] Machine learning ensemble methods can be used for AGN classification tasks but require the use of several parameters from \textsc{X-CIGALE}. These parameters include individual physical parameters that are important for the classification (e.g. viewing angle or observed AGN disc luminosity). Nevertheless, these methods require correct classifications (training data) that often vary based on criteria.
    \item[5.] The observed and intrinsic AGN disc luminosity decreases from Type-1 to Type-2 AGNs in the intermediate Seyfert types. This decrease may be explained by accretion rates within AGN disc wind models, which show an evolutionary path among these AGN types. However, more information is needed to create a robust connection between theoretical models and observations. 
\end{itemize}

In this work, we have demonstrated usefulness of the broad-band SED tool as \textsc{X-CIGALE} to classify AGN galaxies in Type-1 and Type-2. Thus, \textsc{X-CIGALE} could be a powerful tool to characterise AGNs in the upcoming years. Future space telescopes like the JWST and \textit{Athena} will get crucial photometry and spectroscopy to constrain AGN physical parameters like luminosity and $f_{\rm{AGN}}$, which will improve AGN galaxy classifications. These classifications along with other physical parameters will help us understand the real scenario that describes AGN galaxies. 

\section*{Acknowledgements}

K. Ma{\l}ek has been supported by the Polish National Science Centre grant (UMO-2018/30/E/ST9/00082). We would like to thank the Center for Information Technology of the University of Groningen for their support and for providing access to the Peregrine high performance computing cluster. We acknowledge the anonymous referee for a careful reading of the manuscript and very helpful suggestions and comments.

This research made use of Astropy,\footnote{http://www.astropy.org} a community-developed core Python package for Astronomy \citep{2013A&A...558A..33A,2018AJ....156..123A}. This research has made use of the SIMBAD database, operated at CDS, Strasbourg, France. This research has made use of the NASA/IPAC Extragalactic Database (NED), which is funded by the National Aeronautics and Space Administration and operated by the California Institute of Technology. This research has made use of NASA’s Astrophysics Data System Bibliographic Services. Part of the initial exploratory analysis was performed with TOPCAT \citep{2005ASPC..347...29T}. 

\section*{Data Availability}

Most of the data and code underlying this article are available in Zenodo, at \url{https://doi.org/10.5281/zenodo.5227294}. The \textsc{X-CIGALE} estimated data from the AGN setups are available in a companion repository in Zenodo, at \url{https://doi.org/10.5281/zenodo.5221764}, which also includes a script to create similar figures like the one presented in Fig.~\ref{fig:SEDExample} for all galaxies in this work.
 



\bibliographystyle{mnras}
\bibliography{ManualBib,ADS} 

\begin{thebibliography}{}
\makeatletter
\relax
\def\mn@urlcharsother{\let\do\@makeother \do\$\do\&\do\#\do\^\do\_\do\%\do\~}
\def\mn@doi{\begingroup\mn@urlcharsother \@ifnextchar [ {\mn@doi@}
  {\mn@doi@[]}}
\def\mn@doi@[#1]#2{\def\@tempa{#1}\ifx\@tempa\@empty \href
  {http://dx.doi.org/#2} {doi:#2}\else \href {http://dx.doi.org/#2} {#1}\fi
  \endgroup}
\def\mn@eprint#1#2{\mn@eprint@#1:#2::\@nil}
\def\mn@eprint@arXiv#1{\href {http://arxiv.org/abs/#1} {{\tt arXiv:#1}}}
\def\mn@eprint@dblp#1{\href {http://dblp.uni-trier.de/rec/bibtex/#1.xml}
  {dblp:#1}}
\def\mn@eprint@#1:#2:#3:#4\@nil{\def\@tempa {#1}\def\@tempb {#2}\def\@tempc
  {#3}\ifx \@tempc \@empty \let \@tempc \@tempb \let \@tempb \@tempa \fi \ifx
  \@tempb \@empty \def\@tempb {arXiv}\fi \@ifundefined
  {mn@eprint@\@tempb}{\@tempb:\@tempc}{\expandafter \expandafter \csname
  mn@eprint@\@tempb\endcsname \expandafter{\@tempc}}}

\bibitem[\protect\citeauthoryear{{Abazajian} et~al.,}{{Abazajian}
  et~al.}{2009}]{2009ApJS..182..543A}
{Abazajian} K.~N.,  et~al., 2009, \mn@doi [\apjs]
  {10.1088/0067-0049/182/2/543}, \href
  {https://ui.adsabs.harvard.edu/abs/2009ApJS..182..543A} {182, 543}

\bibitem[\protect\citeauthoryear{{Ai} et~al.,}{{Ai}
  et~al.}{2016}]{2016AJ....151...24A}
{Ai} Y.~L.,  et~al., 2016, \mn@doi [\aj] {10.3847/0004-6256/151/2/24}, \href
  {https://ui.adsabs.harvard.edu/abs/2016AJ....151...24A} {151, 24}

\bibitem[\protect\citeauthoryear{{Albareti} et~al.,}{{Albareti}
  et~al.}{2015}]{2015MNRAS.452.4153A}
{Albareti} F.~D.,  et~al., 2015, \mn@doi [\mnras] {10.1093/mnras/stv1406},
  \href {https://ui.adsabs.harvard.edu/abs/2015MNRAS.452.4153A} {452, 4153}

\bibitem[\protect\citeauthoryear{{Alonso-Herrero}, {Pereira-Santaella}, {Rieke}
   \& {Rigopoulou}}{{Alonso-Herrero} et~al.}{2012}]{2012ApJ...744....2A}
{Alonso-Herrero} A.,  {Pereira-Santaella} M.,  {Rieke} G.~H.,   {Rigopoulou}
  D.,  2012, \mn@doi [\apj] {10.1088/0004-637X/744/1/2}, \href
  {https://ui.adsabs.harvard.edu/abs/2012ApJ...744....2A} {744, 2}

\bibitem[\protect\citeauthoryear{{Alonso-Herrero} et~al.,}{{Alonso-Herrero}
  et~al.}{2021}]{2021A&A...652A..99A}
{Alonso-Herrero} A.,  et~al., 2021, \mn@doi [\aap]
  {10.1051/0004-6361/202141219}, \href
  {https://ui.adsabs.harvard.edu/abs/2021A&A...652A..99A} {652, A99}

\bibitem[\protect\citeauthoryear{{Antonucci}}{{Antonucci}}{1993}]{1993ARA&A..31..473A}
{Antonucci} R.,  1993, \mn@doi [\araa] {10.1146/annurev.aa.31.090193.002353},
  \href {https://ui.adsabs.harvard.edu/abs/1993ARA&A..31..473A} {31, 473}

\bibitem[\protect\citeauthoryear{{Antonucci} \& {Miller}}{{Antonucci} \&
  {Miller}}{1985}]{1985ApJ...297..621A}
{Antonucci} R.~R.~J.,  {Miller} J.~S.,  1985, \mn@doi [\apj] {10.1086/163559},
  \href {https://ui.adsabs.harvard.edu/abs/1985ApJ...297..621A} {297, 621}

\bibitem[\protect\citeauthoryear{{Astropy Collaboration} et~al.,}{{Astropy
  Collaboration} et~al.}{2013}]{2013A&A...558A..33A}
{Astropy Collaboration} et~al., 2013, \mn@doi [\aap]
  {10.1051/0004-6361/201322068}, \href
  {https://ui.adsabs.harvard.edu/abs/2013A&A...558A..33A} {558, A33}

\bibitem[\protect\citeauthoryear{{Astropy Collaboration} et~al.,}{{Astropy
  Collaboration} et~al.}{2018}]{2018AJ....156..123A}
{Astropy Collaboration} et~al., 2018, \mn@doi [\aj] {10.3847/1538-3881/aabc4f},
  \href {https://ui.adsabs.harvard.edu/abs/2018AJ....156..123A} {156, 123}

\bibitem[\protect\citeauthoryear{{Baqui} et~al.,}{{Baqui}
  et~al.}{2021}]{2021A&A...645A..87B}
{Baqui} P.~O.,  et~al., 2021, \mn@doi [\aap] {10.1051/0004-6361/202038986},
  \href {https://ui.adsabs.harvard.edu/abs/2021A&A...645A..87B} {645, A87}

\bibitem[\protect\citeauthoryear{{Bernhard}, {Tadhunter}, {Mullaney},
  {Grimmett}, {Rosario}  \& {Alexander}}{{Bernhard}
  et~al.}{2021}]{2021MNRAS.503.2598B}
{Bernhard} E.,  {Tadhunter} C.,  {Mullaney} J.~R.,  {Grimmett} L.~P.,
  {Rosario} D.~J.,   {Alexander} D.~M.,  2021, \mn@doi [\mnras]
  {10.1093/mnras/stab419}, \href
  {https://ui.adsabs.harvard.edu/abs/2021MNRAS.503.2598B} {503, 2598}

\bibitem[\protect\citeauthoryear{{Bluck}, {Maiolino}, {S{\'a}nchez}, {Ellison},
  {Thorp}, {Piotrowska}, {Teimoorinia}  \& {Bundy}}{{Bluck}
  et~al.}{2020}]{2020MNRAS.492...96B}
{Bluck} A. F.~L.,  {Maiolino} R.,  {S{\'a}nchez} S.~F.,  {Ellison} S.~L.,
  {Thorp} M.~D.,  {Piotrowska} J.~M.,  {Teimoorinia} H.,   {Bundy} K.~A.,
  2020, \mn@doi [\mnras] {10.1093/mnras/stz3264}, \href
  {https://ui.adsabs.harvard.edu/abs/2020MNRAS.492...96B} {492, 96}

\bibitem[\protect\citeauthoryear{{Boquien}, {Burgarella}, {Roehlly}, {Buat},
  {Ciesla}, {Corre}, {Inoue}  \& {Salas}}{{Boquien}
  et~al.}{2019}]{2019A&A...622A.103B}
{Boquien} M.,  {Burgarella} D.,  {Roehlly} Y.,  {Buat} V.,  {Ciesla} L.,
  {Corre} D.,  {Inoue} A.~K.,   {Salas} H.,  2019, \mn@doi [\aap]
  {10.1051/0004-6361/201834156}, \href
  {https://ui.adsabs.harvard.edu/abs/2019A&A...622A.103B} {622, A103}

\bibitem[\protect\citeauthoryear{{Braatz}, {Wilson}, {Gezari}, {Varosi}  \&
  {Beichman}}{{Braatz} et~al.}{1993}]{1993ApJ...409L...5B}
{Braatz} J.~A.,  {Wilson} A.~S.,  {Gezari} D.~Y.,  {Varosi} F.,   {Beichman}
  C.~A.,  1993, \mn@doi [\apjl] {10.1086/186846}, \href
  {https://ui.adsabs.harvard.edu/abs/1993ApJ...409L...5B} {409, L5}

\bibitem[\protect\citeauthoryear{{Breiman}}{{Breiman}}{2001}]{2001MachL..45....5B}
{Breiman} L.,  2001, \mn@doi [Machine Learning] {10.1023/A:1010933404324},
  \href {https://ui.adsabs.harvard.edu/abs/2001MachL..45....5B} {45, 5}

\bibitem[\protect\citeauthoryear{{Brunet} et~al.,}{{Brunet}
  et~al.}{2018}]{2018EPJWC.18602004B}
{Brunet} C.,  et~al., 2018, in European Physical Journal Web of Conferences. p.
  02004, \mn@doi{10.1051/epjconf/201818602004}

\bibitem[\protect\citeauthoryear{{Bruzual} \& {Charlot}}{{Bruzual} \&
  {Charlot}}{2003}]{2003MNRAS.344.1000B}
{Bruzual} G.,  {Charlot} S.,  2003, \mn@doi [\mnras]
  {10.1046/j.1365-8711.2003.06897.x}, \href
  {https://ui.adsabs.harvard.edu/abs/2003MNRAS.344.1000B} {344, 1000}

\bibitem[\protect\citeauthoryear{{Buat}, {Ciesla}, {Boquien}, {Ma{\l}ek}  \&
  {Burgarella}}{{Buat} et~al.}{2019}]{2019A&A...632A..79B}
{Buat} V.,  {Ciesla} L.,  {Boquien} M.,  {Ma{\l}ek} K.,   {Burgarella} D.,
  2019, \mn@doi [\aap] {10.1051/0004-6361/201936643}, \href
  {https://ui.adsabs.harvard.edu/abs/2019A&A...632A..79B} {632, A79}

\bibitem[\protect\citeauthoryear{{Buat} et~al.,}{{Buat}
  et~al.}{2021}]{2021A&A...654A..93B}
{Buat} V.,  et~al., 2021, \mn@doi [\aap] {10.1051/0004-6361/202141797}, \href
  {https://ui.adsabs.harvard.edu/abs/2021A&A...654A..93B} {654, A93}

\bibitem[\protect\citeauthoryear{{Burgarella}, {Buat}  \&
  {Iglesias-P{\'a}ramo}}{{Burgarella} et~al.}{2005}]{2005MNRAS.360.1413B}
{Burgarella} D.,  {Buat} V.,   {Iglesias-P{\'a}ramo} J.,  2005, \mn@doi
  [\mnras] {10.1111/j.1365-2966.2005.09131.x}, \href
  {https://ui.adsabs.harvard.edu/abs/2005MNRAS.360.1413B} {360, 1413}

\bibitem[\protect\citeauthoryear{{Calistro Rivera}, {Lusso}, {Hennawi}  \&
  {Hogg}}{{Calistro Rivera} et~al.}{2016}]{2016ApJ...833...98C}
{Calistro Rivera} G.,  {Lusso} E.,  {Hennawi} J.~F.,   {Hogg} D.~W.,  2016,
  \mn@doi [\apj] {10.3847/1538-4357/833/1/98}, \href
  {https://ui.adsabs.harvard.edu/abs/2016ApJ...833...98C} {833, 98}

\bibitem[\protect\citeauthoryear{{Calistro Rivera} et~al.,}{{Calistro Rivera}
  et~al.}{2021}]{2021A&A...649A.102C}
{Calistro Rivera} G.,  et~al., 2021, \mn@doi [\aap]
  {10.1051/0004-6361/202040214}, \href
  {https://ui.adsabs.harvard.edu/abs/2021A&A...649A.102C} {649, A102}

\bibitem[\protect\citeauthoryear{{Calzetti}, {Armus}, {Bohlin}, {Kinney},
  {Koornneef}  \& {Storchi-Bergmann}}{{Calzetti}
  et~al.}{2000}]{2000ApJ...533..682C}
{Calzetti} D.,  {Armus} L.,  {Bohlin} R.~C.,  {Kinney} A.~L.,  {Koornneef} J.,
   {Storchi-Bergmann} T.,  2000, \mn@doi [\apj] {10.1086/308692}, \href
  {https://ui.adsabs.harvard.edu/abs/2000ApJ...533..682C} {533, 682}

\bibitem[\protect\citeauthoryear{{Cameron}, {Storey}, {Rotaciuc}, {Genzel},
  {Verstraete}, {Drapatz}, {Siebenmorgen}  \& {Lee}}{{Cameron}
  et~al.}{1993}]{1993ApJ...419..136C}
{Cameron} M.,  {Storey} J. W.~V.,  {Rotaciuc} V.,  {Genzel} R.,  {Verstraete}
  L.,  {Drapatz} S.,  {Siebenmorgen} R.,   {Lee} T.~J.,  1993, \mn@doi [\apj]
  {10.1086/173467}, \href
  {https://ui.adsabs.harvard.edu/abs/1993ApJ...419..136C} {419, 136}

\bibitem[\protect\citeauthoryear{{Carleo}, {Cirac}, {Cranmer}, {Daudet},
  {Schuld}, {Tishby}, {Vogt-Maranto}  \& {Zdeborov{\'a}}}{{Carleo}
  et~al.}{2019}]{2019RvMP...91d5002C}
{Carleo} G.,  {Cirac} I.,  {Cranmer} K.,  {Daudet} L.,  {Schuld} M.,  {Tishby}
  N.,  {Vogt-Maranto} L.,   {Zdeborov{\'a}} L.,  2019, \mn@doi [Reviews of
  Modern Physics] {10.1103/RevModPhys.91.045002}, \href
  {https://ui.adsabs.harvard.edu/abs/2019RvMP...91d5002C} {91, 045002}

\bibitem[\protect\citeauthoryear{{Casey}}{{Casey}}{2012}]{2012MNRAS.425.3094C}
{Casey} C.~M.,  2012, \mn@doi [\mnras] {10.1111/j.1365-2966.2012.21455.x},
  \href {https://ui.adsabs.harvard.edu/abs/2012MNRAS.425.3094C} {425, 3094}

\bibitem[\protect\citeauthoryear{{Chabrier}}{{Chabrier}}{2003}]{2003PASP..115..763C}
{Chabrier} G.,  2003, \mn@doi [\pasp] {10.1086/376392}, \href
  {https://ui.adsabs.harvard.edu/abs/2003PASP..115..763C} {115, 763}

\bibitem[\protect\citeauthoryear{{Chen} \& {Guestrin}}{{Chen} \&
  {Guestrin}}{2016}]{2016arXiv160302754C}
{Chen} T.,  {Guestrin} C.,  2016, arXiv e-prints, \href
  {https://ui.adsabs.harvard.edu/abs/2016arXiv160302754C} {p. arXiv:1603.02754}

\bibitem[\protect\citeauthoryear{{Chen} et~al.,}{{Chen}
  et~al.}{2018}]{2018A&A...615A.167C}
{Chen} S.,  et~al., 2018, \mn@doi [\aap] {10.1051/0004-6361/201832678}, \href
  {https://ui.adsabs.harvard.edu/abs/2018A&A...615A.167C} {615, A167}

\bibitem[\protect\citeauthoryear{{Ciesla} et~al.,}{{Ciesla}
  et~al.}{2015}]{2015A&A...576A..10C}
{Ciesla} L.,  et~al., 2015, \mn@doi [\aap] {10.1051/0004-6361/201425252}, \href
  {https://ui.adsabs.harvard.edu/abs/2015A&A...576A..10C} {576, A10}

\bibitem[\protect\citeauthoryear{{Ciesla}, {Elbaz}, {Schreiber}, {Daddi}  \&
  {Wang}}{{Ciesla} et~al.}{2018}]{2018A&A...615A..61C}
{Ciesla} L.,  {Elbaz} D.,  {Schreiber} C.,  {Daddi} E.,   {Wang} T.,  2018,
  \mn@doi [\aap] {10.1051/0004-6361/201832715}, \href
  {https://ui.adsabs.harvard.edu/abs/2018A&A...615A..61C} {615, A61}

\bibitem[\protect\citeauthoryear{{Cusumano} et~al.,}{{Cusumano}
  et~al.}{2010}]{2010A&A...524A..64C}
{Cusumano} G.,  et~al., 2010, \mn@doi [\aap] {10.1051/0004-6361/201015249},
  \href {https://ui.adsabs.harvard.edu/abs/2010A&A...524A..64C} {524, A64}

\bibitem[\protect\citeauthoryear{{D'Abrusco}, {Massaro}, {Paggi}, {Smith},
  {Masetti}, {Landoni}  \& {Tosti}}{{D'Abrusco}
  et~al.}{2014}]{2014ApJS..215...14D}
{D'Abrusco} R.,  {Massaro} F.,  {Paggi} A.,  {Smith} H.~A.,  {Masetti} N.,
  {Landoni} M.,   {Tosti} G.,  2014, \mn@doi [\apjs]
  {10.1088/0067-0049/215/1/14}, \href
  {https://ui.adsabs.harvard.edu/abs/2014ApJS..215...14D} {215, 14}

\bibitem[\protect\citeauthoryear{{D'Abrusco} et~al.,}{{D'Abrusco}
  et~al.}{2019}]{2019ApJS..242....4D}
{D'Abrusco} R.,  et~al., 2019, \mn@doi [\apjs] {10.3847/1538-4365/ab16f4},
  \href {https://ui.adsabs.harvard.edu/abs/2019ApJS..242....4D} {242, 4}

\bibitem[\protect\citeauthoryear{{Dale}, {Helou}, {Magdis}, {Armus},
  {D{\'\i}az-Santos}  \& {Shi}}{{Dale} et~al.}{2014}]{2014ApJ...784...83D}
{Dale} D.~A.,  {Helou} G.,  {Magdis} G.~E.,  {Armus} L.,  {D{\'\i}az-Santos}
  T.,   {Shi} Y.,  2014, \mn@doi [\apj] {10.1088/0004-637X/784/1/83}, \href
  {https://ui.adsabs.harvard.edu/abs/2014ApJ...784...83D} {784, 83}

\bibitem[\protect\citeauthoryear{{Dietrich} et~al.,}{{Dietrich}
  et~al.}{2018}]{2018MNRAS.480.3562D}
{Dietrich} J.,  et~al., 2018, \mn@doi [\mnras] {10.1093/mnras/sty2056}, \href
  {https://ui.adsabs.harvard.edu/abs/2018MNRAS.480.3562D} {480, 3562}

\bibitem[\protect\citeauthoryear{{Dong} et~al.,}{{Dong}
  et~al.}{2018}]{2018AJ....155..189D}
{Dong} X.~Y.,  et~al., 2018, \mn@doi [\aj] {10.3847/1538-3881/aab5ae}, \href
  {https://ui.adsabs.harvard.edu/abs/2018AJ....155..189D} {155, 189}

\bibitem[\protect\citeauthoryear{{Duarte Puertas}, {Vilchez},
  {Iglesias-P{\'a}ramo}, {Kehrig}, {P{\'e}rez-Montero}  \&
  {Rosales-Ortega}}{{Duarte Puertas} et~al.}{2017}]{2017A&A...599A..71D}
{Duarte Puertas} S.,  {Vilchez} J.~M.,  {Iglesias-P{\'a}ramo} J.,  {Kehrig} C.,
   {P{\'e}rez-Montero} E.,   {Rosales-Ortega} F.~F.,  2017, \mn@doi [\aap]
  {10.1051/0004-6361/201629044}, \href
  {https://ui.adsabs.harvard.edu/abs/2017A&A...599A..71D} {599, A71}

\bibitem[\protect\citeauthoryear{{Dullemond} \& {van Bemmel}}{{Dullemond} \&
  {van Bemmel}}{2005}]{2005A&A...436...47D}
{Dullemond} C.~P.,  {van Bemmel} I.~M.,  2005, \mn@doi [\aap]
  {10.1051/0004-6361:20041763}, \href
  {https://ui.adsabs.harvard.edu/abs/2005A&A...436...47D} {436, 47}

\bibitem[\protect\citeauthoryear{{Efstathiou}}{{Efstathiou}}{2006}]{2006MNRAS.371L..70E}
{Efstathiou} A.,  2006, \mn@doi [\mnras] {10.1111/j.1745-3933.2006.00210.x},
  \href {https://ui.adsabs.harvard.edu/abs/2006MNRAS.371L..70E} {371, L70}

\bibitem[\protect\citeauthoryear{{Efstathiou} \& {Rowan-Robinson}}{{Efstathiou}
  \& {Rowan-Robinson}}{1995}]{1995MNRAS.273..649E}
{Efstathiou} A.,  {Rowan-Robinson} M.,  1995, \mn@doi [\mnras]
  {10.1093/mnras/273.3.649}, \href
  {https://ui.adsabs.harvard.edu/abs/1995MNRAS.273..649E} {273, 649}

\bibitem[\protect\citeauthoryear{{Efstathiou}, {Hough}  \&
  {Young}}{{Efstathiou} et~al.}{1995}]{1995MNRAS.277.1134E}
{Efstathiou} A.,  {Hough} J.~H.,   {Young} S.,  1995, \mn@doi [\mnras]
  {10.1093/mnras/277.3.1134}, \href
  {https://ui.adsabs.harvard.edu/abs/1995MNRAS.277.1134E} {277, 1134}

\bibitem[\protect\citeauthoryear{{Elitzur} \& {Shlosman}}{{Elitzur} \&
  {Shlosman}}{2006}]{2006ApJ...648L.101E}
{Elitzur} M.,  {Shlosman} I.,  2006, \mn@doi [\apjl] {10.1086/508158}, \href
  {https://ui.adsabs.harvard.edu/abs/2006ApJ...648L.101E} {648, L101}

\bibitem[\protect\citeauthoryear{{Elitzur}, {Ho}  \& {Trump}}{{Elitzur}
  et~al.}{2014}]{2014MNRAS.438.3340E}
{Elitzur} M.,  {Ho} L.~C.,   {Trump} J.~R.,  2014, \mn@doi [\mnras]
  {10.1093/mnras/stt2445}, \href
  {https://ui.adsabs.harvard.edu/abs/2014MNRAS.438.3340E} {438, 3340}

\bibitem[\protect\citeauthoryear{{Emmering}, {Blandford}  \&
  {Shlosman}}{{Emmering} et~al.}{1992}]{1992ApJ...385..460E}
{Emmering} R.~T.,  {Blandford} R.~D.,   {Shlosman} I.,  1992, \mn@doi [\apj]
  {10.1086/170955}, \href
  {https://ui.adsabs.harvard.edu/abs/1992ApJ...385..460E} {385, 460}

\bibitem[\protect\citeauthoryear{{Fabian}, {Vasudevan}  \& {Gandhi}}{{Fabian}
  et~al.}{2008}]{2008MNRAS.385L..43F}
{Fabian} A.~C.,  {Vasudevan} R.~V.,   {Gandhi} P.,  2008, \mn@doi [\mnras]
  {10.1111/j.1745-3933.2008.00430.x}, \href
  {https://ui.adsabs.harvard.edu/abs/2008MNRAS.385L..43F} {385, L43}

\bibitem[\protect\citeauthoryear{{Fischer}, {Crenshaw}, {Kraemer}  \&
  {Schmitt}}{{Fischer} et~al.}{2013}]{2013ApJS..209....1F}
{Fischer} T.~C.,  {Crenshaw} D.~M.,  {Kraemer} S.~B.,   {Schmitt} H.~R.,  2013,
  \mn@doi [\apjs] {10.1088/0067-0049/209/1/1}, \href
  {https://ui.adsabs.harvard.edu/abs/2013ApJS..209....1F} {209, 1}

\bibitem[\protect\citeauthoryear{{Flesch}}{{Flesch}}{2015}]{2015PASA...32...10F}
{Flesch} E.~W.,  2015, \mn@doi [\pasa] {10.1017/pasa.2015.10}, \href
  {https://ui.adsabs.harvard.edu/abs/2015PASA...32...10F} {32, e010}

\bibitem[\protect\citeauthoryear{{Flesch}}{{Flesch}}{2021}]{2021yCat.7290....0F}
{Flesch} E.~W.,  2021, VizieR Online Data Catalog, \href
  {https://ui.adsabs.harvard.edu/abs/2021yCat.7290....0F} {p. VII/290}

\bibitem[\protect\citeauthoryear{{Florez} et~al.,}{{Florez}
  et~al.}{2020}]{2020MNRAS.497.3273F}
{Florez} J.,  et~al., 2020, \mn@doi [\mnras] {10.1093/mnras/staa2200}, \href
  {https://ui.adsabs.harvard.edu/abs/2020MNRAS.497.3273F} {497, 3273}

\bibitem[\protect\citeauthoryear{Friedman}{Friedman}{2001}]{friedman2001greedy}
Friedman J.~H.,  2001, Annals of statistics, pp 1189--1232

\bibitem[\protect\citeauthoryear{{Fritz}, {Franceschini}  \&
  {Hatziminaoglou}}{{Fritz} et~al.}{2006}]{2006MNRAS.366..767F}
{Fritz} J.,  {Franceschini} A.,   {Hatziminaoglou} E.,  2006, \mn@doi [\mnras]
  {10.1111/j.1365-2966.2006.09866.x}, \href
  {https://ui.adsabs.harvard.edu/abs/2006MNRAS.366..767F} {366, 767}

\bibitem[\protect\citeauthoryear{{Gandhi}, {Horst}, {Smette}, {H{\"o}nig},
  {Comastri}, {Gilli}, {Vignali}  \& {Duschl}}{{Gandhi}
  et~al.}{2009}]{2009A&A...502..457G}
{Gandhi} P.,  {Horst} H.,  {Smette} A.,  {H{\"o}nig} S.,  {Comastri} A.,
  {Gilli} R.,  {Vignali} C.,   {Duschl} W.,  2009, \mn@doi [\aap]
  {10.1051/0004-6361/200811368}, \href
  {https://ui.adsabs.harvard.edu/abs/2009A&A...502..457G} {502, 457}

\bibitem[\protect\citeauthoryear{{Garc{\'\i}a-Burillo}
  et~al.,}{{Garc{\'\i}a-Burillo} et~al.}{2021}]{2021A&A...652A..98G}
{Garc{\'\i}a-Burillo} S.,  et~al., 2021, \mn@doi [\aap]
  {10.1051/0004-6361/202141075}, \href
  {https://ui.adsabs.harvard.edu/abs/2021A&A...652A..98G} {652, A98}

\bibitem[\protect\citeauthoryear{{Gentile Fusillo}, {G{\"a}nsicke}  \&
  {Greiss}}{{Gentile Fusillo} et~al.}{2015}]{2015MNRAS.448.2260G}
{Gentile Fusillo} N.~P.,  {G{\"a}nsicke} B.~T.,   {Greiss} S.,  2015, \mn@doi
  [\mnras] {10.1093/mnras/stv120}, \href
  {https://ui.adsabs.harvard.edu/abs/2015MNRAS.448.2260G} {448, 2260}

\bibitem[\protect\citeauthoryear{{Gkini}, {Plionis}, {Chira}  \&
  {Koulouridis}}{{Gkini} et~al.}{2021}]{2021A&A...650A..75G}
{Gkini} A.,  {Plionis} M.,  {Chira} M.,   {Koulouridis} E.,  2021, \mn@doi
  [\aap] {10.1051/0004-6361/202140278}, \href
  {https://ui.adsabs.harvard.edu/abs/2021A&A...650A..75G} {650, A75}

\bibitem[\protect\citeauthoryear{{Golob}, {Sawicki}, {Goulding}  \&
  {Coupon}}{{Golob} et~al.}{2021}]{2021MNRAS.503.4136G}
{Golob} A.,  {Sawicki} M.,  {Goulding} A.~D.,   {Coupon} J.,  2021, \mn@doi
  [\mnras] {10.1093/mnras/stab719}, \href
  {https://ui.adsabs.harvard.edu/abs/2021MNRAS.503.4136G} {503, 4136}

\bibitem[\protect\citeauthoryear{{Gonz{\'a}lez-Mart{\'\i}n}
  et~al.,}{{Gonz{\'a}lez-Mart{\'\i}n} et~al.}{2019a}]{2019ApJ...884...10G}
{Gonz{\'a}lez-Mart{\'\i}n} O.,  et~al., 2019a, \mn@doi [\apj]
  {10.3847/1538-4357/ab3e6b}, \href
  {https://ui.adsabs.harvard.edu/abs/2019ApJ...884...10G} {884, 10}

\bibitem[\protect\citeauthoryear{{Gonz{\'a}lez-Mart{\'\i}n}
  et~al.,}{{Gonz{\'a}lez-Mart{\'\i}n} et~al.}{2019b}]{2019ApJ...884...11G}
{Gonz{\'a}lez-Mart{\'\i}n} O.,  et~al., 2019b, \mn@doi [\apj]
  {10.3847/1538-4357/ab3e4f}, \href
  {https://ui.adsabs.harvard.edu/abs/2019ApJ...884...11G} {884, 11}

\bibitem[\protect\citeauthoryear{{Granato} \& {Danese}}{{Granato} \&
  {Danese}}{1994}]{1994MNRAS.268..235G}
{Granato} G.~L.,  {Danese} L.,  1994, \mn@doi [\mnras]
  {10.1093/mnras/268.1.235}, \href
  {https://ui.adsabs.harvard.edu/abs/1994MNRAS.268..235G} {268, 235}

\bibitem[\protect\citeauthoryear{{Gruppioni} et~al.,}{{Gruppioni}
  et~al.}{2016}]{2016MNRAS.458.4297G}
{Gruppioni} C.,  et~al., 2016, \mn@doi [\mnras] {10.1093/mnras/stw577}, \href
  {https://ui.adsabs.harvard.edu/abs/2016MNRAS.458.4297G} {458, 4297}

\bibitem[\protect\citeauthoryear{{Gupta}, {Sikora}  \& {Nalewajko}}{{Gupta}
  et~al.}{2016}]{2016MNRAS.461.2346G}
{Gupta} M.,  {Sikora} M.,   {Nalewajko} K.,  2016, \mn@doi [\mnras]
  {10.1093/mnras/stw1473}, \href
  {https://ui.adsabs.harvard.edu/abs/2016MNRAS.461.2346G} {461, 2346}

\bibitem[\protect\citeauthoryear{{Hickox} \& {Alexander}}{{Hickox} \&
  {Alexander}}{2018}]{2018ARA&A..56..625H}
{Hickox} R.~C.,  {Alexander} D.~M.,  2018, \mn@doi [\araa]
  {10.1146/annurev-astro-081817-051803}, \href
  {https://ui.adsabs.harvard.edu/abs/2018ARA&A..56..625H} {56, 625}

\bibitem[\protect\citeauthoryear{{H{\"o}nig}}{{H{\"o}nig}}{2019}]{2019ApJ...884..171H}
{H{\"o}nig} S.~F.,  2019, \mn@doi [\apj] {10.3847/1538-4357/ab4591}, \href
  {https://ui.adsabs.harvard.edu/abs/2019ApJ...884..171H} {884, 171}

\bibitem[\protect\citeauthoryear{{H{\"o}nig} \& {Kishimoto}}{{H{\"o}nig} \&
  {Kishimoto}}{2017}]{2017ApJ...838L..20H}
{H{\"o}nig} S.~F.,  {Kishimoto} M.,  2017, \mn@doi [\apjl]
  {10.3847/2041-8213/aa6838}, \href
  {https://ui.adsabs.harvard.edu/abs/2017ApJ...838L..20H} {838, L20}

\bibitem[\protect\citeauthoryear{{J{\"a}rvel{\"a}}, {L{\"a}hteenm{\"a}ki}  \&
  {Le{\'o}n-Tavares}}{{J{\"a}rvel{\"a}} et~al.}{2015}]{2015A&A...573A..76J}
{J{\"a}rvel{\"a}} E.,  {L{\"a}hteenm{\"a}ki} A.,   {Le{\'o}n-Tavares} J.,
  2015, \mn@doi [\aap] {10.1051/0004-6361/201424694}, \href
  {https://ui.adsabs.harvard.edu/abs/2015A&A...573A..76J} {573, A76}

\bibitem[\protect\citeauthoryear{{Jayasinghe} et~al.,}{{Jayasinghe}
  et~al.}{2018}]{2018MNRAS.477.3145J}
{Jayasinghe} T.,  et~al., 2018, \mn@doi [\mnras] {10.1093/mnras/sty838}, \href
  {https://ui.adsabs.harvard.edu/abs/2018MNRAS.477.3145J} {477, 3145}

\bibitem[\protect\citeauthoryear{Kass \& Raftery}{Kass \&
  Raftery}{1995}]{kass1995bayes}
Kass R.~E.,  Raftery A.~E.,  1995, Journal of the american statistical
  association, 90, 773

\bibitem[\protect\citeauthoryear{{Kauffmann} et~al.,}{{Kauffmann}
  et~al.}{2003}]{2003MNRAS.346.1055K}
{Kauffmann} G.,  et~al., 2003, \mn@doi [\mnras]
  {10.1111/j.1365-2966.2003.07154.x}, \href
  {https://ui.adsabs.harvard.edu/abs/2003MNRAS.346.1055K} {346, 1055}

\bibitem[\protect\citeauthoryear{{Klindt}, {Alexander}, {Rosario}, {Lusso}  \&
  {Fotopoulou}}{{Klindt} et~al.}{2019}]{2019MNRAS.488.3109K}
{Klindt} L.,  {Alexander} D.~M.,  {Rosario} D.~J.,  {Lusso} E.,   {Fotopoulou}
  S.,  2019, \mn@doi [\mnras] {10.1093/mnras/stz1771}, \href
  {https://ui.adsabs.harvard.edu/abs/2019MNRAS.488.3109K} {488, 3109}

\bibitem[\protect\citeauthoryear{{Koss} et~al.,}{{Koss}
  et~al.}{2017}]{2017ApJ...850...74K}
{Koss} M.,  et~al., 2017, \mn@doi [\apj] {10.3847/1538-4357/aa8ec9}, \href
  {https://ui.adsabs.harvard.edu/abs/2017ApJ...850...74K} {850, 74}

\bibitem[\protect\citeauthoryear{{Krawczyk}, {Richards}, {Mehta}, {Vogeley},
  {Gallagher}, {Leighly}, {Ross}  \& {Schneider}}{{Krawczyk}
  et~al.}{2013}]{2013ApJS..206....4K}
{Krawczyk} C.~M.,  {Richards} G.~T.,  {Mehta} S.~S.,  {Vogeley} M.~S.,
  {Gallagher} S.~C.,  {Leighly} K.~M.,  {Ross} N.~P.,   {Schneider} D.~P.,
  2013, \mn@doi [\apjs] {10.1088/0067-0049/206/1/4}, \href
  {https://ui.adsabs.harvard.edu/abs/2013ApJS..206....4K} {206, 4}

\bibitem[\protect\citeauthoryear{{Krawczyk}, {Richards}, {Gallagher},
  {Leighly}, {Hewett}, {Ross}  \& {Hall}}{{Krawczyk}
  et~al.}{2015}]{2015AJ....149..203K}
{Krawczyk} C.~M.,  {Richards} G.~T.,  {Gallagher} S.~C.,  {Leighly} K.~M.,
  {Hewett} P.~C.,  {Ross} N.~P.,   {Hall} P.~B.,  2015, \mn@doi [\aj]
  {10.1088/0004-6256/149/6/203}, \href
  {https://ui.adsabs.harvard.edu/abs/2015AJ....149..203K} {149, 203}

\bibitem[\protect\citeauthoryear{{Krolik} \& {Begelman}}{{Krolik} \&
  {Begelman}}{1988}]{1988ApJ...329..702K}
{Krolik} J.~H.,  {Begelman} M.~C.,  1988, \mn@doi [\apj] {10.1086/166414},
  \href {https://ui.adsabs.harvard.edu/abs/1988ApJ...329..702K} {329, 702}

\bibitem[\protect\citeauthoryear{{LaMassa} et~al.,}{{LaMassa}
  et~al.}{2015}]{2015ApJ...800..144L}
{LaMassa} S.~M.,  et~al., 2015, \mn@doi [\apj] {10.1088/0004-637X/800/2/144},
  \href {https://ui.adsabs.harvard.edu/abs/2015ApJ...800..144L} {800, 144}

\bibitem[\protect\citeauthoryear{{Lebouteiller}, {Barry}, {Goes}, {Sloan},
  {Spoon}, {Weedman}, {Bernard-Salas}  \& {Houck}}{{Lebouteiller}
  et~al.}{2015}]{2015ApJS..218...21L}
{Lebouteiller} V.,  {Barry} D.~J.,  {Goes} C.,  {Sloan} G.~C.,  {Spoon}
  H.~W.~W.,  {Weedman} D.~W.,  {Bernard-Salas} J.,   {Houck} J.~R.,  2015,
  \mn@doi [\apjs] {10.1088/0067-0049/218/2/21}, \href
  {https://ui.adsabs.harvard.edu/abs/2015ApJS..218...21L} {218, 21}

\bibitem[\protect\citeauthoryear{{Leja}, {Johnson}, {Conroy}  \& {van
  Dokkum}}{{Leja} et~al.}{2018}]{2018ApJ...854...62L}
{Leja} J.,  {Johnson} B.~D.,  {Conroy} C.,   {van Dokkum} P.,  2018, \mn@doi
  [\apj] {10.3847/1538-4357/aaa8db}, \href
  {https://ui.adsabs.harvard.edu/abs/2018ApJ...854...62L} {854, 62}

\bibitem[\protect\citeauthoryear{{Liu}, {Yuan}, {Dong}, {Zhou}  \& {Liu}}{{Liu}
  et~al.}{2018}]{2018ApJS..235...40L}
{Liu} H.-Y.,  {Yuan} W.,  {Dong} X.-B.,  {Zhou} H.,   {Liu} W.-J.,  2018,
  \mn@doi [\apjs] {10.3847/1538-4365/aab88e}, \href
  {https://ui.adsabs.harvard.edu/abs/2018ApJS..235...40L} {235, 40}

\bibitem[\protect\citeauthoryear{{L{\'o}pez-Gonzaga}, {Burtscher}, {Tristram},
  {Meisenheimer}  \& {Schartmann}}{{L{\'o}pez-Gonzaga}
  et~al.}{2016}]{2016A&A...591A..47L}
{L{\'o}pez-Gonzaga} N.,  {Burtscher} L.,  {Tristram} K.~R.~W.,  {Meisenheimer}
  K.,   {Schartmann} M.,  2016, \mn@doi [\aap] {10.1051/0004-6361/201527590},
  \href {https://ui.adsabs.harvard.edu/abs/2016A&A...591A..47L} {591, A47}

\bibitem[\protect\citeauthoryear{{Lusso}, {Piedipalumbo}, {Risaliti},
  {Paolillo}, {Bisogni}, {Nardini}  \& {Amati}}{{Lusso}
  et~al.}{2019}]{2019A&A...628L...4L}
{Lusso} E.,  {Piedipalumbo} E.,  {Risaliti} G.,  {Paolillo} M.,  {Bisogni} S.,
  {Nardini} E.,   {Amati} L.,  2019, \mn@doi [\aap]
  {10.1051/0004-6361/201936223}, \href
  {https://ui.adsabs.harvard.edu/abs/2019A&A...628L...4L} {628, L4}

\bibitem[\protect\citeauthoryear{{Lyu} \& {Rieke}}{{Lyu} \&
  {Rieke}}{2018}]{2018ApJ...866...92L}
{Lyu} J.,  {Rieke} G.~H.,  2018, \mn@doi [\apj] {10.3847/1538-4357/aae075},
  \href {https://ui.adsabs.harvard.edu/abs/2018ApJ...866...92L} {866, 92}

\bibitem[\protect\citeauthoryear{{Ma{\l}ek} et~al.,}{{Ma{\l}ek}
  et~al.}{2018}]{2018A&A...620A..50M}
{Ma{\l}ek} K.,  et~al., 2018, \mn@doi [\aap] {10.1051/0004-6361/201833131},
  \href {https://ui.adsabs.harvard.edu/abs/2018A&A...620A..50M} {620, A50}

\bibitem[\protect\citeauthoryear{{Marchesi} et~al.,}{{Marchesi}
  et~al.}{2016}]{2016ApJ...817...34M}
{Marchesi} S.,  et~al., 2016, \mn@doi [\apj] {10.3847/0004-637X/817/1/34},
  \href {https://ui.adsabs.harvard.edu/abs/2016ApJ...817...34M} {817, 34}

\bibitem[\protect\citeauthoryear{{Marin}}{{Marin}}{2016}]{2016MNRAS.460.3679M}
{Marin} F.,  2016, \mn@doi [\mnras] {10.1093/mnras/stw1131}, \href
  {https://ui.adsabs.harvard.edu/abs/2016MNRAS.460.3679M} {460, 3679}

\bibitem[\protect\citeauthoryear{{Masoura}, {Mountrichas}, {Georgantopoulos}
  \& {Plionis}}{{Masoura} et~al.}{2021}]{2021A&A...646A.167M}
{Masoura} V.~A.,  {Mountrichas} G.,  {Georgantopoulos} I.,   {Plionis} M.,
  2021, \mn@doi [\aap] {10.1051/0004-6361/202039238}, \href
  {https://ui.adsabs.harvard.edu/abs/2021A&A...646A.167M} {646, A167}

\bibitem[\protect\citeauthoryear{Matthews}{Matthews}{1975}]{matthews1975comparison}
Matthews B.~W.,  1975, Biochimica et Biophysica Acta (BBA)-Protein Structure,
  405, 442

\bibitem[\protect\citeauthoryear{{McKinney}, {Hayward}, {Rosenthal},
  {Martinez-Galarza}, {Pope}, {Sajina}  \& {Smith}}{{McKinney}
  et~al.}{2021}]{2021arXiv210312747M}
{McKinney} J.,  {Hayward} C.~C.,  {Rosenthal} L.~J.,  {Martinez-Galarza} J.~R.,
   {Pope} A.,  {Sajina} A.,   {Smith} H.~A.,  2021, arXiv e-prints, \href
  {https://ui.adsabs.harvard.edu/abs/2021arXiv210312747M} {p. arXiv:2103.12747}

\bibitem[\protect\citeauthoryear{{Merloni} et~al.,}{{Merloni}
  et~al.}{2014}]{2014MNRAS.437.3550M}
{Merloni} A.,  et~al., 2014, \mn@doi [\mnras] {10.1093/mnras/stt2149}, \href
  {https://ui.adsabs.harvard.edu/abs/2014MNRAS.437.3550M} {437, 3550}

\bibitem[\protect\citeauthoryear{{Meusinger}, {Hinze}  \& {de
  Hoon}}{{Meusinger} et~al.}{2011}]{2011A&A...525A..37M}
{Meusinger} H.,  {Hinze} A.,   {de Hoon} A.,  2011, \mn@doi [\aap]
  {10.1051/0004-6361/201015520}, \href
  {https://ui.adsabs.harvard.edu/abs/2011A&A...525A..37M} {525, A37}

\bibitem[\protect\citeauthoryear{{Miettinen}}{{Miettinen}}{2018}]{2018Ap&SS.363..197M}
{Miettinen} O.,  2018, \mn@doi [\apss] {10.1007/s10509-018-3418-7}, \href
  {https://ui.adsabs.harvard.edu/abs/2018Ap&SS.363..197M} {363, 197}

\bibitem[\protect\citeauthoryear{{Mountrichas}, {Buat}, {Yang}, {Boquien},
  {Burgarella}  \& {Ciesla}}{{Mountrichas} et~al.}{2021}]{2021A&A...646A..29M}
{Mountrichas} G.,  {Buat} V.,  {Yang} G.,  {Boquien} M.,  {Burgarella} D.,
  {Ciesla} L.,  2021, \mn@doi [\aap] {10.1051/0004-6361/202039401}, \href
  {https://ui.adsabs.harvard.edu/abs/2021A&A...646A..29M} {646, A29}

\bibitem[\protect\citeauthoryear{{Mullaney}, {Alexander}, {Goulding}  \&
  {Hickox}}{{Mullaney} et~al.}{2011}]{2011MNRAS.414.1082M}
{Mullaney} J.~R.,  {Alexander} D.~M.,  {Goulding} A.~D.,   {Hickox} R.~C.,
  2011, \mn@doi [\mnras] {10.1111/j.1365-2966.2011.18448.x}, \href
  {https://ui.adsabs.harvard.edu/abs/2011MNRAS.414.1082M} {414, 1082}

\bibitem[\protect\citeauthoryear{{Nenkova}, {Ivezi{\'c}}  \&
  {Elitzur}}{{Nenkova} et~al.}{2002}]{2002ApJ...570L...9N}
{Nenkova} M.,  {Ivezi{\'c}} {\v{Z}}.,   {Elitzur} M.,  2002, \mn@doi [\apjl]
  {10.1086/340857}, \href
  {https://ui.adsabs.harvard.edu/abs/2002ApJ...570L...9N} {570, L9}

\bibitem[\protect\citeauthoryear{{Nenkova}, {Sirocky}, {Nikutta}, {Ivezi{\'c}}
  \& {Elitzur}}{{Nenkova} et~al.}{2008}]{2008ApJ...685..160N}
{Nenkova} M.,  {Sirocky} M.~M.,  {Nikutta} R.,  {Ivezi{\'c}} {\v{Z}}.,
  {Elitzur} M.,  2008, \mn@doi [\apj] {10.1086/590483}, \href
  {https://ui.adsabs.harvard.edu/abs/2008ApJ...685..160N} {685, 160}

\bibitem[\protect\citeauthoryear{{Netzer}}{{Netzer}}{2015}]{2015ARA&A..53..365N}
{Netzer} H.,  2015, \mn@doi [\araa] {10.1146/annurev-astro-082214-122302},
  \href {https://ui.adsabs.harvard.edu/abs/2015ARA&A..53..365N} {53, 365}

\bibitem[\protect\citeauthoryear{{Noll}, {Burgarella}, {Giovannoli}, {Buat},
  {Marcillac}  \& {Mu{\~n}oz-Mateos}}{{Noll}
  et~al.}{2009}]{2009A&A...507.1793N}
{Noll} S.,  {Burgarella} D.,  {Giovannoli} E.,  {Buat} V.,  {Marcillac} D.,
  {Mu{\~n}oz-Mateos} J.~C.,  2009, \mn@doi [\aap]
  {10.1051/0004-6361/200912497}, \href
  {https://ui.adsabs.harvard.edu/abs/2009A&A...507.1793N} {507, 1793}

\bibitem[\protect\citeauthoryear{{Oberto} et~al.,}{{Oberto}
  et~al.}{2020}]{2020ASPC..522..105O}
{Oberto} A.,  et~al., 2020, in {Ballester} P.,  {Ibsen} J.,  {Solar} M.,
  {Shortridge} K.,  eds,  Astronomical Society of the Pacific Conference Series
  Vol. 522, Astronomical Data Analysis Software and Systems XXVII. p.~105

\bibitem[\protect\citeauthoryear{{Ogawa}, {Ueda}, {Tanimoto}  \&
  {Yamada}}{{Ogawa} et~al.}{2021}]{2021ApJ...906...84O}
{Ogawa} S.,  {Ueda} Y.,  {Tanimoto} A.,   {Yamada} S.,  2021, \mn@doi [\apj]
  {10.3847/1538-4357/abccce}, \href
  {https://ui.adsabs.harvard.edu/abs/2021ApJ...906...84O} {906, 84}

\bibitem[\protect\citeauthoryear{{Oh}, {Yi}, {Schawinski}, {Koss},
  {Trakhtenbrot}  \& {Soto}}{{Oh} et~al.}{2015}]{2015ApJS..219....1O}
{Oh} K.,  {Yi} S.~K.,  {Schawinski} K.,  {Koss} M.,  {Trakhtenbrot} B.,
  {Soto} K.,  2015, \mn@doi [\apjs] {10.1088/0067-0049/219/1/1}, \href
  {https://ui.adsabs.harvard.edu/abs/2015ApJS..219....1O} {219, 1}

\bibitem[\protect\citeauthoryear{{Osterbrock}}{{Osterbrock}}{1977}]{1977ApJ...215..733O}
{Osterbrock} D.~E.,  1977, \mn@doi [\apj] {10.1086/155407}, \href
  {https://ui.adsabs.harvard.edu/abs/1977ApJ...215..733O} {215, 733}

\bibitem[\protect\citeauthoryear{{Osterbrock}}{{Osterbrock}}{1981}]{1981ApJ...249..462O}
{Osterbrock} D.~E.,  1981, \mn@doi [\apj] {10.1086/159306}, \href
  {https://ui.adsabs.harvard.edu/abs/1981ApJ...249..462O} {249, 462}

\bibitem[\protect\citeauthoryear{{Osterbrock} \& {Pogge}}{{Osterbrock} \&
  {Pogge}}{1985}]{1985ApJ...297..166O}
{Osterbrock} D.~E.,  {Pogge} R.~W.,  1985, \mn@doi [\apj] {10.1086/163513},
  \href {https://ui.adsabs.harvard.edu/abs/1985ApJ...297..166O} {297, 166}

\bibitem[\protect\citeauthoryear{{Padovani} et~al.,}{{Padovani}
  et~al.}{2017}]{2017A&ARv..25....2P}
{Padovani} P.,  et~al., 2017, \mn@doi [\aapr] {10.1007/s00159-017-0102-9},
  \href {https://ui.adsabs.harvard.edu/abs/2017A&ARv..25....2P} {25, 2}

\bibitem[\protect\citeauthoryear{{Panessa}, {Castangia}, {Malizia}, {Bassani},
  {Tarchi}, {Bazzano}  \& {Ubertini}}{{Panessa}
  et~al.}{2020}]{2020A&A...641A.162P}
{Panessa} F.,  {Castangia} P.,  {Malizia} A.,  {Bassani} L.,  {Tarchi} A.,
  {Bazzano} A.,   {Ubertini} P.,  2020, \mn@doi [\aap]
  {10.1051/0004-6361/201937407}, \href
  {https://ui.adsabs.harvard.edu/abs/2020A&A...641A.162P} {641, A162}

\bibitem[\protect\citeauthoryear{{P{\^a}ris} et~al.,}{{P{\^a}ris}
  et~al.}{2018}]{2018A&A...613A..51P}
{P{\^a}ris} I.,  et~al., 2018, \mn@doi [\aap] {10.1051/0004-6361/201732445},
  \href {https://ui.adsabs.harvard.edu/abs/2018A&A...613A..51P} {613, A51}

\bibitem[\protect\citeauthoryear{{Pedregosa} et~al.,}{{Pedregosa}
  et~al.}{2012}]{2012arXiv1201.0490P}
{Pedregosa} F.,  et~al., 2012, arXiv e-prints, \href
  {https://ui.adsabs.harvard.edu/abs/2012arXiv1201.0490P} {p. arXiv:1201.0490}

\bibitem[\protect\citeauthoryear{{P{\'e}rez-Torres}, {Mattila},
  {Alonso-Herrero}, {Aalto}  \& {Efstathiou}}{{P{\'e}rez-Torres}
  et~al.}{2021}]{2021A&ARv..29....2P}
{P{\'e}rez-Torres} M.,  {Mattila} S.,  {Alonso-Herrero} A.,  {Aalto} S.,
  {Efstathiou} A.,  2021, \mn@doi [\aapr] {10.1007/s00159-020-00128-x}, \href
  {https://ui.adsabs.harvard.edu/abs/2021A&ARv..29....2P} {29, 2}

\bibitem[\protect\citeauthoryear{{Pier} \& {Krolik}}{{Pier} \&
  {Krolik}}{1992}]{1992ApJ...401...99P}
{Pier} E.~A.,  {Krolik} J.~H.,  1992, \mn@doi [\apj] {10.1086/172042}, \href
  {https://ui.adsabs.harvard.edu/abs/1992ApJ...401...99P} {401, 99}

\bibitem[\protect\citeauthoryear{{Pouliasis}, {Mountrichas}, {Georgantopoulos},
  {Ruiz}, {Yang}  \& {Bonanos}}{{Pouliasis} et~al.}{2020}]{2020MNRAS.495.1853P}
{Pouliasis} E.,  {Mountrichas} G.,  {Georgantopoulos} I.,  {Ruiz} A.,  {Yang}
  M.,   {Bonanos} A.~Z.,  2020, \mn@doi [\mnras] {10.1093/mnras/staa1263},
  \href {https://ui.adsabs.harvard.edu/abs/2020MNRAS.495.1853P} {495, 1853}

\bibitem[\protect\citeauthoryear{{Prevot}, {Lequeux}, {Maurice}, {Prevot}  \&
  {Rocca-Volmerange}}{{Prevot} et~al.}{1984}]{1984A&A...132..389P}
{Prevot} M.~L.,  {Lequeux} J.,  {Maurice} E.,  {Prevot} L.,
  {Rocca-Volmerange} B.,  1984, \aap, \href
  {https://ui.adsabs.harvard.edu/abs/1984A&A...132..389P} {132, 389}

\bibitem[\protect\citeauthoryear{{Prince}, {Czerny}  \& {Pollo}}{{Prince}
  et~al.}{2021}]{2021ApJ...909...58P}
{Prince} R.,  {Czerny} B.,   {Pollo} A.,  2021, \mn@doi [\apj]
  {10.3847/1538-4357/abd775}, \href
  {https://ui.adsabs.harvard.edu/abs/2021ApJ...909...58P} {909, 58}

\bibitem[\protect\citeauthoryear{{Rakshit}, {Stalin}, {Chand}  \&
  {Zhang}}{{Rakshit} et~al.}{2017}]{2017ApJS..229...39R}
{Rakshit} S.,  {Stalin} C.~S.,  {Chand} H.,   {Zhang} X.-G.,  2017, \mn@doi
  [\apjs] {10.3847/1538-4365/aa6971}, \href
  {https://ui.adsabs.harvard.edu/abs/2017ApJS..229...39R} {229, 39}

\bibitem[\protect\citeauthoryear{{Ramos Almeida} \& {Ricci}}{{Ramos Almeida} \&
  {Ricci}}{2017}]{2017NatAs...1..679R}
{Ramos Almeida} C.,  {Ricci} C.,  2017, \mn@doi [Nature Astronomy]
  {10.1038/s41550-017-0232-z}, \href
  {https://ui.adsabs.harvard.edu/abs/2017NatAs...1..679R} {1, 679}

\bibitem[\protect\citeauthoryear{{Ramos Padilla}, {Ashby}, {Smith},
  {Mart{\'\i}nez-Galarza}, {Beverage}, {Dietrich}, {Higuera-G.}  \&
  {Weiner}}{{Ramos Padilla} et~al.}{2020}]{2020MNRAS.499.4325R}
{Ramos Padilla} A.~F.,  {Ashby} M.~L.~N.,  {Smith} H.~A.,
  {Mart{\'\i}nez-Galarza} J.~R.,  {Beverage} A.~G.,  {Dietrich} J.,
  {Higuera-G.} M.-A.,   {Weiner} A.~S.,  2020, \mn@doi [\mnras]
  {10.1093/mnras/staa2813}, \href
  {https://ui.adsabs.harvard.edu/abs/2020MNRAS.499.4325R} {499, 4325}

\bibitem[\protect\citeauthoryear{{Ricci} et~al.,}{{Ricci}
  et~al.}{2017}]{2017Natur.549..488R}
{Ricci} C.,  et~al., 2017, \mn@doi [\nat] {10.1038/nature23906}, \href
  {https://ui.adsabs.harvard.edu/abs/2017Natur.549..488R} {549, 488}

\bibitem[\protect\citeauthoryear{{Salmon} et~al.,}{{Salmon}
  et~al.}{2016}]{2016ApJ...827...20S}
{Salmon} B.,  et~al., 2016, \mn@doi [\apj] {10.3847/0004-637X/827/1/20}, \href
  {https://ui.adsabs.harvard.edu/abs/2016ApJ...827...20S} {827, 20}

\bibitem[\protect\citeauthoryear{{Salpeter}}{{Salpeter}}{1955}]{1955ApJ...121..161S}
{Salpeter} E.~E.,  1955, \mn@doi [\apj] {10.1086/145971}, \href
  {https://ui.adsabs.harvard.edu/abs/1955ApJ...121..161S} {121, 161}

\bibitem[\protect\citeauthoryear{{Satyapal}, {Kamal}, {Cann}, {Secrest}  \&
  {Abel}}{{Satyapal} et~al.}{2021}]{2021ApJ...906...35S}
{Satyapal} S.,  {Kamal} L.,  {Cann} J.~M.,  {Secrest} N.~J.,   {Abel} N.~P.,
  2021, \mn@doi [\apj] {10.3847/1538-4357/abbfaf}, \href
  {https://ui.adsabs.harvard.edu/abs/2021ApJ...906...35S} {906, 35}

\bibitem[\protect\citeauthoryear{{Schmidt}, {Ferreiro}, {Vega Neme}  \&
  {Oio}}{{Schmidt} et~al.}{2016}]{2016A&A...596A..95S}
{Schmidt} E.~O.,  {Ferreiro} D.,  {Vega Neme} L.,   {Oio} G.~A.,  2016, \mn@doi
  [\aap] {10.1051/0004-6361/201629343}, \href
  {https://ui.adsabs.harvard.edu/abs/2016A&A...596A..95S} {596, A95}

\bibitem[\protect\citeauthoryear{{Scott}}{{Scott}}{2015}]{2015mdet.book.....S}
{Scott} D.~W.,  2015, {Multivariate Density Estimation: Theory, Practice, and
  Visualization}

\bibitem[\protect\citeauthoryear{{Scott} \& {Stewart}}{{Scott} \&
  {Stewart}}{2014}]{2014MNRAS.438.2253S}
{Scott} A.~E.,  {Stewart} G.~C.,  2014, \mn@doi [\mnras]
  {10.1093/mnras/stt2341}, \href
  {https://ui.adsabs.harvard.edu/abs/2014MNRAS.438.2253S} {438, 2253}

\bibitem[\protect\citeauthoryear{{Serra}, {Amblard}, {Temi}, {Burgarella},
  {Giovannoli}, {Buat}, {Noll}  \& {Im}}{{Serra}
  et~al.}{2011}]{2011ApJ...740...22S}
{Serra} P.,  {Amblard} A.,  {Temi} P.,  {Burgarella} D.,  {Giovannoli} E.,
  {Buat} V.,  {Noll} S.,   {Im} S.,  2011, \mn@doi [\apj]
  {10.1088/0004-637X/740/1/22}, \href
  {https://ui.adsabs.harvard.edu/abs/2011ApJ...740...22S} {740, 22}

\bibitem[\protect\citeauthoryear{{Siebenmorgen}, {Heymann}  \&
  {Efstathiou}}{{Siebenmorgen} et~al.}{2015}]{2015A&A...583A.120S}
{Siebenmorgen} R.,  {Heymann} F.,   {Efstathiou} A.,  2015, \mn@doi [\aap]
  {10.1051/0004-6361/201526034}, \href
  {https://ui.adsabs.harvard.edu/abs/2015A&A...583A.120S} {583, A120}

\bibitem[\protect\citeauthoryear{{Silva}, {Maiolino}  \& {Granato}}{{Silva}
  et~al.}{2004}]{2004MNRAS.355..973S}
{Silva} L.,  {Maiolino} R.,   {Granato} G.~L.,  2004, \mn@doi [\mnras]
  {10.1111/j.1365-2966.2004.08380.x}, \href
  {https://ui.adsabs.harvard.edu/abs/2004MNRAS.355..973S} {355, 973}

\bibitem[\protect\citeauthoryear{{Souchay} et~al.,}{{Souchay}
  et~al.}{2015}]{2015A&A...583A..75S}
{Souchay} J.,  et~al., 2015, \mn@doi [\aap] {10.1051/0004-6361/201526092},
  \href {https://ui.adsabs.harvard.edu/abs/2015A&A...583A..75S} {583, A75}

\bibitem[\protect\citeauthoryear{{Stalevski}, {Fritz}, {Baes}, {Nakos}  \&
  {Popovi{\'c}}}{{Stalevski} et~al.}{2012}]{2012MNRAS.420.2756S}
{Stalevski} M.,  {Fritz} J.,  {Baes} M.,  {Nakos} T.,   {Popovi{\'c}}
  L.~{\v{C}}.,  2012, \mn@doi [\mnras] {10.1111/j.1365-2966.2011.19775.x},
  \href {https://ui.adsabs.harvard.edu/abs/2012MNRAS.420.2756S} {420, 2756}

\bibitem[\protect\citeauthoryear{{Stalevski}, {Ricci}, {Ueda}, {Lira}, {Fritz}
  \& {Baes}}{{Stalevski} et~al.}{2016}]{2016MNRAS.458.2288S}
{Stalevski} M.,  {Ricci} C.,  {Ueda} Y.,  {Lira} P.,  {Fritz} J.,   {Baes} M.,
  2016, \mn@doi [\mnras] {10.1093/mnras/stw444}, \href
  {https://ui.adsabs.harvard.edu/abs/2016MNRAS.458.2288S} {458, 2288}

\bibitem[\protect\citeauthoryear{{Stern} \& {Laor}}{{Stern} \&
  {Laor}}{2012}]{2012MNRAS.426.2703S}
{Stern} J.,  {Laor} A.,  2012, \mn@doi [\mnras]
  {10.1111/j.1365-2966.2012.21772.x}, \href
  {https://ui.adsabs.harvard.edu/abs/2012MNRAS.426.2703S} {426, 2703}

\bibitem[\protect\citeauthoryear{{Suh} et~al.,}{{Suh}
  et~al.}{2019}]{2019ApJ...872..168S}
{Suh} H.,  et~al., 2019, \mn@doi [\apj] {10.3847/1538-4357/ab01fb}, \href
  {https://ui.adsabs.harvard.edu/abs/2019ApJ...872..168S} {872, 168}

\bibitem[\protect\citeauthoryear{{Sun} \& {Shen}}{{Sun} \&
  {Shen}}{2015}]{2015ApJ...804L..15S}
{Sun} J.,  {Shen} Y.,  2015, \mn@doi [\apjl] {10.1088/2041-8205/804/1/L15},
  \href {https://ui.adsabs.harvard.edu/abs/2015ApJ...804L..15S} {804, L15}

\bibitem[\protect\citeauthoryear{{Tamayo} et~al.,}{{Tamayo}
  et~al.}{2016}]{2016ApJ...832L..22T}
{Tamayo} D.,  et~al., 2016, \mn@doi [\apjl] {10.3847/2041-8205/832/2/L22},
  \href {https://ui.adsabs.harvard.edu/abs/2016ApJ...832L..22T} {832, L22}

\bibitem[\protect\citeauthoryear{{Tanimoto}, {Ueda}, {Odaka}, {Kawaguchi},
  {Fukazawa}  \& {Kawamuro}}{{Tanimoto} et~al.}{2019}]{2019ApJ...877...95T}
{Tanimoto} A.,  {Ueda} Y.,  {Odaka} H.,  {Kawaguchi} T.,  {Fukazawa} Y.,
  {Kawamuro} T.,  2019, \mn@doi [\apj] {10.3847/1538-4357/ab1b20}, \href
  {https://ui.adsabs.harvard.edu/abs/2019ApJ...877...95T} {877, 95}

\bibitem[\protect\citeauthoryear{{Taylor}}{{Taylor}}{2005}]{2005ASPC..347...29T}
{Taylor} M.~B.,  2005, in {Shopbell} P.,  {Britton} M.,   {Ebert} R.,  eds,
  Astronomical Society of the Pacific Conference Series Vol. 347, Astronomical
  Data Analysis Software and Systems XIV. p.~29

\bibitem[\protect\citeauthoryear{Tharwat}{Tharwat}{2020}]{tharwat2020classification}
Tharwat A.,  2020, Applied Computing and Informatics

\bibitem[\protect\citeauthoryear{{Thorne} et~al.,}{{Thorne}
  et~al.}{2021}]{2021MNRAS.505..540T}
{Thorne} J.~E.,  et~al., 2021, \mn@doi [\mnras] {10.1093/mnras/stab1294}, \href
  {https://ui.adsabs.harvard.edu/abs/2021MNRAS.505..540T} {505, 540}

\bibitem[\protect\citeauthoryear{{Toba} et~al.,}{{Toba}
  et~al.}{2014}]{2014ApJ...788...45T}
{Toba} Y.,  et~al., 2014, \mn@doi [\apj] {10.1088/0004-637X/788/1/45}, \href
  {https://ui.adsabs.harvard.edu/abs/2014ApJ...788...45T} {788, 45}

\bibitem[\protect\citeauthoryear{{Toba} et~al.,}{{Toba}
  et~al.}{2021}]{2021ApJ...912...91T}
{Toba} Y.,  et~al., 2021, \mn@doi [\apj] {10.3847/1538-4357/abe94a}, \href
  {https://ui.adsabs.harvard.edu/abs/2021ApJ...912...91T} {912, 91}

\bibitem[\protect\citeauthoryear{{Urry} \& {Padovani}}{{Urry} \&
  {Padovani}}{1995}]{1995PASP..107..803U}
{Urry} C.~M.,  {Padovani} P.,  1995, \mn@doi [\pasp] {10.1086/133630}, \href
  {https://ui.adsabs.harvard.edu/abs/1995PASP..107..803U} {107, 803}

\bibitem[\protect\citeauthoryear{{V{\'e}ron-Cetty} \&
  {V{\'e}ron}}{{V{\'e}ron-Cetty} \& {V{\'e}ron}}{2010}]{2010A&A...518A..10V}
{V{\'e}ron-Cetty} M.~P.,  {V{\'e}ron} P.,  2010, \mn@doi [\aap]
  {10.1051/0004-6361/201014188}, \href
  {https://ui.adsabs.harvard.edu/abs/2010A&A...518A..10V} {518, A10}

\bibitem[\protect\citeauthoryear{{Vika}, {Ciesla}, {Charmandaris}, {Xilouris}
  \& {Lebouteiller}}{{Vika} et~al.}{2017}]{2017A&A...597A..51V}
{Vika} M.,  {Ciesla} L.,  {Charmandaris} V.,  {Xilouris} E.~M.,
  {Lebouteiller} V.,  2017, \mn@doi [\aap] {10.1051/0004-6361/201629031}, \href
  {https://ui.adsabs.harvard.edu/abs/2017A&A...597A..51V} {597, A51}

\bibitem[\protect\citeauthoryear{{Villarroel}, {Nyholm}, {Karlsson},
  {Comer{\'o}n}, {Korn}, {Sollerman}  \& {Zackrisson}}{{Villarroel}
  et~al.}{2017}]{2017ApJ...837..110V}
{Villarroel} B.,  {Nyholm} A.,  {Karlsson} T.,  {Comer{\'o}n} S.,  {Korn}
  A.~J.,  {Sollerman} J.,   {Zackrisson} E.,  2017, \mn@doi [\apj]
  {10.3847/1538-4357/aa5d5a}, \href
  {https://ui.adsabs.harvard.edu/abs/2017ApJ...837..110V} {837, 110}

\bibitem[\protect\citeauthoryear{{Virtanen} et~al.,}{{Virtanen}
  et~al.}{2020}]{2020NatMe..17..261V}
{Virtanen} P.,  et~al., 2020, \mn@doi [Nature Methods]
  {10.1038/s41592-019-0686-2}, \href
  {https://ui.adsabs.harvard.edu/abs/2020NatMe..17..261V} {17, 261}

\bibitem[\protect\citeauthoryear{{Wada}}{{Wada}}{2015}]{2015ApJ...812...82W}
{Wada} K.,  2015, \mn@doi [\apj] {10.1088/0004-637X/812/1/82}, \href
  {https://ui.adsabs.harvard.edu/abs/2015ApJ...812...82W} {812, 82}

\bibitem[\protect\citeauthoryear{{Wang} et~al.,}{{Wang}
  et~al.}{2020}]{2020MNRAS.499.4068W}
{Wang} T.-W.,  et~al., 2020, \mn@doi [\mnras] {10.1093/mnras/staa2988}, \href
  {https://ui.adsabs.harvard.edu/abs/2020MNRAS.499.4068W} {499, 4068}

\bibitem[\protect\citeauthoryear{{Webster}, {Francis}, {Petersont},
  {Drinkwater}  \& {Masci}}{{Webster} et~al.}{1995}]{1995Natur.375..469W}
{Webster} R.~L.,  {Francis} P.~J.,  {Petersont} B.~A.,  {Drinkwater} M.~J.,
  {Masci} F.~J.,  1995, \mn@doi [\nat] {10.1038/375469a0}, \href
  {https://ui.adsabs.harvard.edu/abs/1995Natur.375..469W} {375, 469}

\bibitem[\protect\citeauthoryear{{Wenger} et~al.,}{{Wenger}
  et~al.}{2000}]{2000A&AS..143....9W}
{Wenger} M.,  et~al., 2000, \mn@doi [\aaps] {10.1051/aas:2000332}, \href
  {https://ui.adsabs.harvard.edu/abs/2000A&AS..143....9W} {143, 9}

\bibitem[\protect\citeauthoryear{{Winkler}}{{Winkler}}{1992}]{1992MNRAS.257..677W}
{Winkler} H.,  1992, \mn@doi [\mnras] {10.1093/mnras/257.4.677}, \href
  {https://ui.adsabs.harvard.edu/abs/1992MNRAS.257..677W} {257, 677}

\bibitem[\protect\citeauthoryear{{Yang} et~al.,}{{Yang}
  et~al.}{2020}]{2020MNRAS.491..740Y}
{Yang} G.,  et~al., 2020, \mn@doi [\mnras] {10.1093/mnras/stz3001}, \href
  {https://ui.adsabs.harvard.edu/abs/2020MNRAS.491..740Y} {491, 740}

\bibitem[\protect\citeauthoryear{{Yang} et~al.,}{{Yang}
  et~al.}{2021}]{2021ApJ...908..144Y}
{Yang} G.,  et~al., 2021, \mn@doi [\apj] {10.3847/1538-4357/abd6c1}, \href
  {https://ui.adsabs.harvard.edu/abs/2021ApJ...908..144Y} {908, 144}

\bibitem[\protect\citeauthoryear{{Zhou}, {Wang}, {Yuan}, {Lu}, {Dong}, {Wang}
  \& {Lu}}{{Zhou} et~al.}{2006}]{2006ApJS..166..128Z}
{Zhou} H.,  {Wang} T.,  {Yuan} W.,  {Lu} H.,  {Dong} X.,  {Wang} J.,   {Lu} Y.,
   2006, \mn@doi [\apjs] {10.1086/504869}, \href
  {https://ui.adsabs.harvard.edu/abs/2006ApJS..166..128Z} {166, 128}

\bibitem[\protect\citeauthoryear{{Zou}, {Yang}, {Brandt}  \& {Xue}}{{Zou}
  et~al.}{2019}]{2019ApJ...878...11Z}
{Zou} F.,  {Yang} G.,  {Brandt} W.~N.,   {Xue} Y.,  2019, \mn@doi [\apj]
  {10.3847/1538-4357/ab1eb1}, \href
  {https://ui.adsabs.harvard.edu/abs/2019ApJ...878...11Z} {878, 11}

\makeatother
\end{thebibliography}




\appendix

\section{Narrow-line Sy1 galaxies}\label{App:S1n}

The narrow-line Sy1 (NLSy1) are AGN galaxies that share similarities with Sy1 and Sy2, but cannot be identified as an intermediate Seyfert type \citep[][]{1985ApJ...297..166O}. These galaxies are classified as NLSy1 due to their i) H$\beta$ line emission profile (broad and narrow components), ii) ratio between H$\beta$ and [\ion{O}{III}] fluxes below 3 ($R<3$), iii) narrow FWHM, and iv) the presence of [\ion{Fe}{II}] multiplets \citep[e.g.][]{2006ApJS..166..128Z,2017ApJS..229...39R,2018A&A...615A.167C}. These characteristics indicate that NLSy1 are similar to Sy1. None the less, NLSy1 seem to show higher Eddington rates, lower black hole masses and different viewing angles \citep{2017ApJS..229...39R}. 

\begin{figure*}
	\includegraphics[width=\textwidth]{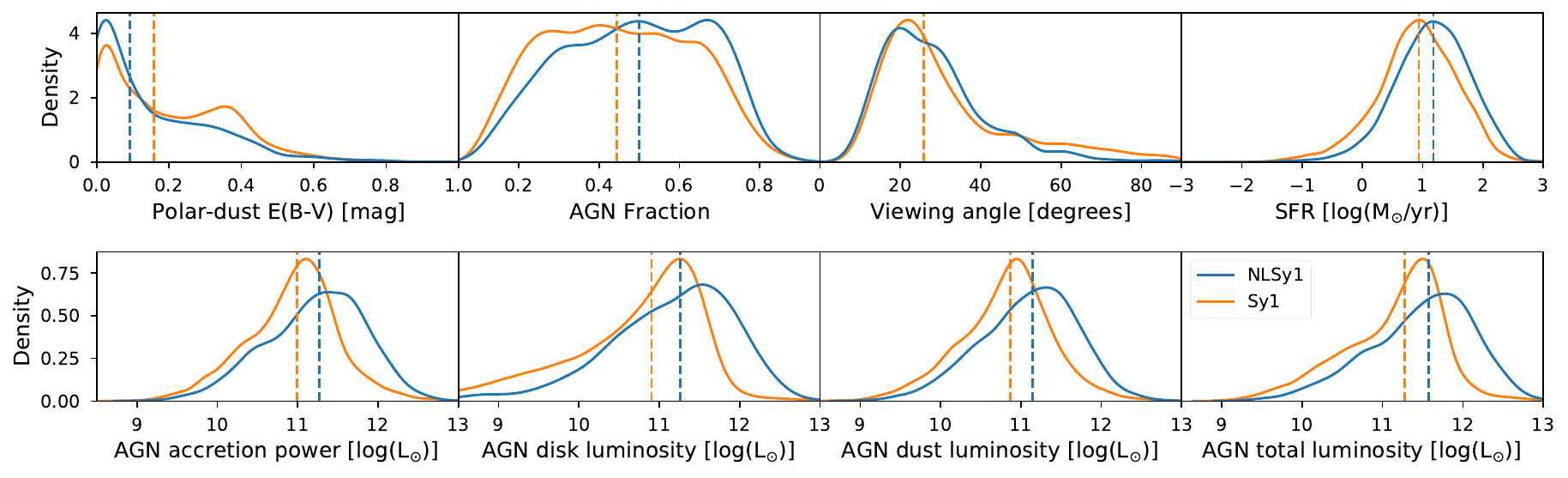}
    \caption{Probability density functions for the AGN estimated parameters and SFR comparing NLSy1 (blue) and Sy1 (orange) galaxies for the SKIRTOR setup with the respective median values (vertical dashed lines). No difference is observed in the viewing angle median estimates. However, NLSy1 show a higher luminosity in the intrinsic component (accretion power), the emitted disc, dust luminosities and the total AGN luminosity.}
    \label{fig:A1}
\end{figure*}

We assumed that  NLSy1 are Sy1 galaxies in the catalogue of \citetalias{2010A&A...518A..10V}, as most of these NLSy1 were classified as Sy1 in \citetalias{2000A&AS..143....9W}. However, the results from the AGN setups in this work show small differences in the AGN parameters between these two AGN types. In Figure~\ref{fig:A1}, we present the density functions of the AGN physical parameters for NLSy1 and Sy1 as classified by \citetalias{2010A&A...518A..10V}. For almost all physical parameters the median values for these AGN types are different, although it does not show any difference for the viewing angle. The null hypothesis of the KS test is always rejected when comparing NLSy1 and Sy1 galaxies. All the parameters have a higher $D$ value than the critical value, $D_{\rm{crit}}=0.04$ in the SKIRTOR setup. This means that both samples originate from different distributions. Nevertheless, in terms of the viewing angle, we notice the smallest difference between these two types ($D=0.08$). This small difference supports our assumption that Sy1 and NLSy1 can be treated as the same type in this work, as our focus resides in the AGN viewing angle.

Conversely, there are larger differences in the total AGN luminosity, as well in terms of their components (disc and re-emitted dust) and intrinsic accretion power ($D>0.24$). These differences may be related to the higher accretion of the AGN in this type of galaxy. Nevertheless, the difference in the AGN physical parameters for Sy1 and NLSy1 may require the X-ray bands, which were not obtained in this work (Sect.~\ref{sec:DiscXray}), as they seem to show a steeper X-ray slope in NLSy1 \citep{2014MNRAS.438.2253S}. Then, a more specific study of these two types of galaxies will be required to understand their nature. 

\section{Mock results}\label{App:mock}
We perform a mock analysis on the main physical parameters to verify the quality of the fits inside \textsc{X-CIGALE}. We use the estimations from the SKIRTOR and Fritz setups. The mock analysis is performed inside \textsc{X-CIGALE} by creating mock values from the original photometry flux and errors and the best fit of the object. Then, the same method used in the original estimation is applied to obtain mock estimations. With this analysis, we can estimate the reliability of the obtained estimations for the physical parameters \citep[][]{2019A&A...622A.103B,2020MNRAS.491..740Y}. 

\begin{figure*}
	\includegraphics[width=\textwidth]{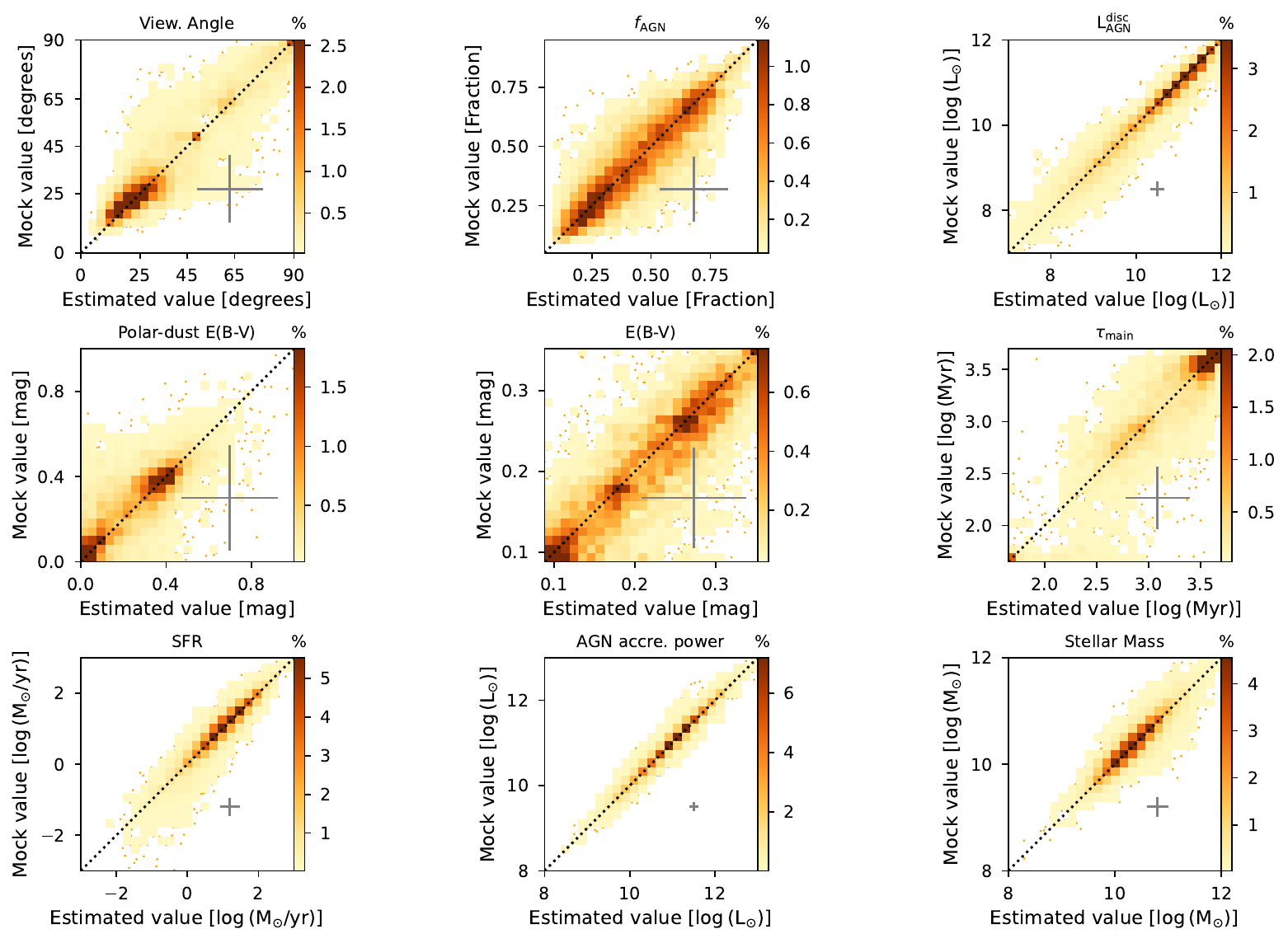}
    \caption{Mock versus estimated values from SKIRTOR setup for nine parameters studied in this work. The 2-dimensional histograms show the percentage of galaxies in the sample that fall in the parameter space sampled by 25 cells in each dimension. The pointed line represents the 1:1 relation and the grey crosses represent the median estimated error for each of the parameters. Cells with only one galaxy are not coloured, instead we draw these individual galaxies as orange dots.}
    \label{fig:A2}
\end{figure*}

Figure~\ref{fig:A2} shows the mock analysis for the SKIRTOR setup in the seven selected parameters used for this study (Sect.~\ref{sec:FeatureSel}) with the addition of the accretion power (intrinsic disc luminosity) and stellar mass. In general, all the physical parameters are well correlated, although some parameters tend to have larger uncertainties such as polar dust and viewing angle. The e-folding time of the main stellar population ($\tau_{\rm{main}}$) is the most affected parameter when comparing with mock values. This result is expected as the age estimates inside \textsc{CIGALE} are not well constrained, as noted by \citet[][]{2017A&A...597A..51V} and shown in Sect.~\ref{sec:verif}. The mock analysis for the Fritz setup (found in the online repository) shows similar results as the Fritz setup. Therefore, the quality of the fits allows us to analyse the estimated physical parameters in this work.

\section{Classifications of individual objects in the literature}

For the 14 unclassified and 45 discrepant Seyfert classifications discussed in Sect.~\ref{sec:Predictions}, we searched the literature for other classifications of activity type. In Table~\ref{tab:Liter}, we present the activity type information and the reference for these classifications. We did not find another classification for some of these galaxies, therefore these galaxies are not listed in Table~\ref{tab:Liter}.

\begin{table*}
\centering
\caption{Activity types found in the literature for unclassified and discrepant Seyfert galaxies.}
\label{tab:Liter}
\begin{tabular}{lcc}
\hline
\hline
Object ID & Activity classification types & References \\
\hline
\multicolumn{3}{c}{No Type in VCV \& SMB}\\
\hline
2MASX J12140343-1921428 & Mixed blazar; QSO; Sy1 & 2,19; 16; 18\\
2MASX J23032790+1443491 & Composite; Sy1 & 20; 18\\      
CADIS 16-505716 & QSO & 18 \\
LEDA 1485346 & Sy1 & 18 \\
LEDA 3095610 & QSO; Sy1 & 16; 18\\
LEDA 3096762 & QSO & 18\\
MCG+00-11-002 & Sy1 & 18\\
MCG+03-45-003 & Sy1; Sy2 & 18; 23\\
QSO B1238+6232 & QSO & 18\\
{[HB93] 0248+011A}& QSO & 18\\
\hline
\multicolumn{3}{c}{Type in VCV or SMB}\\
\hline
2E 2294 & QSO; Sy1 & 6,9,12,15,16,21; 18 \\
2E 2628 & QSO & 6,9,13,15,16,18,21 \\
2E 3786 & QSO & 6,9,13,15,16,18,21 \\
2MASS J00423990+3017514 & Sy1 & 18,23\\    
2MASS J01341936+0146479 & QSO; Sy1 & 6,16,18\\
2MASS J02500703+0025251 & QSO; SF; Sy1 & 6,10,14,16,21,22; 8; 18,20\\
2MASS J08171856+5201477 & Sy1; NLSy1 & 3,18,20; 7\\
2MASS J09393182+5449092 & QSO; Sy1 & 16,9,4,18,21; 20\\
2MASS J09455439+4238399 & QSO; Sy1; NLSy1 & 14; 20; 4,7,17,18\\
2MASS J09470326+4640425 & QSO; NLSy1 & 6,9,16,18,21; 7\\
2MASS J09594856+5942505 & QSO; Sy1 & 6,9,10,12,14,16,21; 18 \\
2MASS J10102753+4132389 & QSO & 6,9,10,13,15,18,21 \\
2MASS J10470514+5444060 & QSO; Sy1 & 4,6,9,15,16,21; 18,20 \\
2MASS J12002696+3317286 & QSO; Sy1 & 16,9,6,14,21; 18 \\
2MASS J15142051+4244453 & QSO; Sy1 & 4,6,12,14,15,16,21; 18,20 \\
2MASSI J0930176+470720 & QSO; Sy1 & 6,9,23,16,18,21; 20 \\
2MASX J02522087+0043307 & QSO; Sy1; SF & 4,14,16,22; 18,20; 8 \\
2MASX J02593816+0042167 & QSO; Sy1; SF & 6,12,14,16,21,22; 18,20; 8\\
2MASX J06374318-7538458 & Sy1 & 18\\
2MASX J09420770+0228053 & Sy2; LINER & 18; 20 \\
2MASX J09443702-2633554 & QSO; Sy1; Sy1.5 & 16; 18; 1\\
2MASX J09483841+4030436 & QSO; Sy1 & 14; 18,20 \\
2MASX J10155660-2002268 & Sy1 & 18\\
2MASX J10194946+3322041 & Sy1; NLSy1 & 18; 7\\
2MASX J15085291+6814074 & Sy1 & 18\\
2MASX J16383091-2055246 & Sy1; NLSy1 & 18; 1,5,11 \\
2MASX J21033788-0455396 & Sy1 & 18\\
2MASX J22024516-1304538 & Sy1 & 18\\
2dFGRS TGN357Z241 & QSO; Sy1 & 4,16; 18\\
3C 286 & Mixed Blazar; QSO & 2,19; 6,12,13,18,21\\ 
6dFGS gJ034205.4-370322 & Mixed Blazar; QSO; Sy1 & 2,19; 16; 18\\
6dFGS gJ043944.9-454043 & QSO & 16,18 \\
6dFGS gJ084628.7-121409 & Sy1; NLSy1 & 18; 5\\
CTS 11 & Sy1; NLSy1 & 18; 11 \\
HE 0226-4110 & QSO; NLSy1 & 16,18; 5\\
ICRF J025937.6+423549 & Mixed Blazar; QSO & 19; 18 \\
ICRF J081100.6+571412 & QSO & 6,12,13,14,15,18,21 \\
ICRF J100646.4-215920 & Mixed Blazar; Sy1 & 2,19; 18 \\
ICRF J110153.4+624150 & BZQ; QSO; SF & 2,18; 6,9,12,13,18,21; 8 \\
ICRF J135704.4+191907 & BZQ; QSO & 2,18,23; 6,9,12,13,14,18,21 \\
IRAS 10295-1831 & Sy1 & 18\\
Mrk 1361 & QSO; Sy1; Sy2 & 14; 20; 18\\
PB 162 & QSO; Sy1; NLSy1; SF & 14,16; 18,20; 7; 8 \\
UGC 10683 & Sy1 & 18\\
\hline
\end{tabular}
\begin{flushleft}
\textbf{References:} 1: \citet{2020A&A...641A.162P}; 2: \citet{2019ApJS..242....4D}; 3: \citet{2018ApJS..235...40L}; 4: \citet{2018AJ....155..189D}; 5: \citet{2018A&A...615A.167C}; 6: \citet{2018A&A...613A..51P}; 7: \citet{2017ApJS..229...39R}; 8: \citet{2017A&A...599A..71D}; 9: \citet{2016MNRAS.461.2346G}; 10: \citet{2016AJ....151...24A}; 11: \citet{2016A&A...596A..95S}; 12: \citet{2015MNRAS.452.4153A}; 13: \citet{2015MNRAS.448.2260G}; 14: \citet{2015ApJ...804L..15S}; 15: \citet{2015AJ....149..203K}; 16: \citet{2015A&A...583A..75S}; 17: \citet{2015A&A...573A..76J}; 18: \citet{2015PASA...32...10F}, version 7.2 \citet{2021yCat.7290....0F}; 19: \citet{2014ApJS..215...14D}; 20: \citet{2014ApJ...788...45T}; 21: \citet{2013ApJS..206....4K}; 22: \citet{2011A&A...525A..37M}; 23: \citet{2010A&A...524A..64C}, coordinates match within 2\arcsec, distance varies between 0.56\arcsec and 1.77\arcsec.\\
\end{flushleft}
\end{table*}


\bsp	
\label{lastpage}
\end{document}